\documentclass[noshowpacs,peprintnumbers,amssymb,nofootinbib,aps]{revtex4}
\usepackage{epsfig,color,wrapfig}
\usepackage{amsmath}
\usepackage{graphicx}
\usepackage{amssymb}
\usepackage{slashbox}
\usepackage{bm}
\usepackage{array,multirow}
\usepackage[toc]{appendix}
\usepackage{etoolbox}
\usepackage{scalerel}
\usepackage{slashed}
\usepackage{color}
\usepackage{caption}

\DeclareCaptionStyle{mystyle}
  {format=plain,%
    textformat=period,
    justification=RaggedRight,
    singlelinecheck=true,
  }

\DeclareCaptionStyle{singlelinecentered}
  [justification=Centering]
  {style=mystyle}

\DeclareCaptionStyle{singlelineraggedleft}
  [justification=RaggedLeft]
  {style=mystyle}

\usepackage{amsmath}
\usepackage{amssymb}
\usepackage{slashbox}
\usepackage{bm}
\usepackage{dcolumn,multirow,subfig}
\usepackage[normalem]{ulem}

\usepackage{array}

\usepackage{hyperref}

{\newcommand{\lsim}{\mbox{\raisebox{-.6ex}{~$\stackrel{<}{\sim}$~}}}
	{\newcommand{\gsim}{\mbox{\raisebox{-.6ex}{~$\stackrel{>}{\sim}$~}}}

		\newcommand{\bmt}{\begin{pmatrix}}
			\newcommand{\emt}{\end{pmatrix}}
		\newcommand{\ba}{\begin{array}{c}}
			\newcommand{\ea}{\end{array}}
		\newcommand{\be}{\begin{equation}}
		\newcommand{\ee}{\end{equation}}
		\newcommand{\bea}{\begin{eqnarray}}
		\newcommand{\eea}{\end{eqnarray}}
		\newcommand{\nn}{\nonumber}
		\newcommand{\bi}{\begin{itemize}}
			\newcommand{\ei}{\end{itemize}}
		
		\newcommand{\baz}{\begin{array}{cc}}
			
			\newcommand{\mathsym}[1]{{}}

			\newcommand{\bt}{\begin{tabular}}
				\newcommand{\et}{\end{tabular}}

			\newcommand{\benu}{\begin{enumerate}}
				\newcommand{\eenu}{\end{enumerate}}
			
			\newcommand{\bav}{\begin{array}{cccc}}

\begin{document} 
	
\title{\boldmath Correlating the anomalous results in $b \to s$ decays with inert Higgs doublet dark matter and muon $(g-2)$}

\author{Basabendu Barman}
\email{bb1988@iitg.ac.in }
\affiliation{Indian Institute of Technology, North Guwahati, Guwahati 781039, Assam, India }

\author{Debasish Borah}
\email{dborah@iitg.ac.in}
\affiliation{Indian Institute of Technology, North Guwahati, Guwahati 781039, Assam, India }

\author{Lopamudra Mukherjee}
\email{mukherjeelopa@iitg.ac.in }
\affiliation{Indian Institute of Technology, North Guwahati, Guwahati 781039, Assam, India }

\author{Soumitra Nandi}
\email{soumitra.nandi@iitg.ac.in}
\affiliation{Indian Institute of Technology, North Guwahati, Guwahati 781039, Assam, India }


\begin{abstract}
In this article, we have considered an extension of the inert Higgs doublet model with $SU(2)_L$ singlet vector like fermions. Our model is capable of addressing some interesting anomalous results in $b\to s\ell^+\ell^-$ decays (like $R(K^{(*)})$) and in muon $(g-2)$. Apart from explaining these anomalies, and being consistent with other flavour data, the model satisfies relevant constraints in the dark matter sector, while remaining within the reach of ongoing direct detection experiments. The model also produces signatures at the large hadron collider (LHC) with final states comprised of dilepton, dijet and missing energy, providing signals to be probed at higher luminosity.  
\end{abstract}

\maketitle

\section{Introduction}
\label{sec:intro}

The low energy observables in $B$ decays and $B_q-\bar{B_q}$ (q = d,s) mixings play an important role in the indirect detection of new physics (NP). In this regard, the flavour changing neutral current (FCNC) processes, such as $b\to s $, are unique in a sense that in the standard model (SM) they contribute at the loop level thereby keeping their contributions suppressed, in general. For the last couple of years, the semileptonic decays $b\to s \ell^+\ell^-$ ($\ell = \mu, e$) have got lot of attention. The observed ratios of the exclusive branching fractions such as $R(K^{(*)}) = \mathcal{B}(B \to K^{(*)} \mu^+\mu^-)/\mathcal{B}(B \to K^{(*)} e^+ e^-)$ have shown anomalous behaviours with the measured values deviating from their respective SM expectations. The LHCb collaboration has measured~\cite{Aaij:2019wad,Aaij:2017}

\begin{equation}
 R(K) = 0.846^{+0.060 \> + 0.016}_{-0.054 \> -0.014}, \ \ \text{in the bin with dilepton mass squared $q^2 \in [1.1,6]$ {\rm {GeV}$^2$}}, 
\end{equation}

and 

\begin{equation}
 R(K^*) = 
 \begin{cases}
   0.660^{+0.110}_{- 0.070} \pm 0.024,\ \ \text{$q^2 \in [0.045,1.1]$ {\it {\rm GeV}$^2$}},\\
   0.685^{+0.113}_{- 0.069} \pm 0.047,\ \ \text{$q^2 \in [1.1,6]$ {\it {\rm GeV}$^2$}}.
 \end{cases}
\end{equation}

The corresponding SM predictions are, respectively, $R(K) = 1.0004 (8) $, and  

\begin{equation}
 R(K^*) = 
 \begin{cases}
   0.920 \pm 0.007,\ \ \text{$q^2 \in [0.045,1.1]$ {\rm {GeV}$^2$}},\\
   0.996 \pm 0.002,\ \ \text{$q^2 \in [1.1,6]$ {\rm {GeV}$^2$}},
 \end{cases}
\end{equation}

the details of which can be found in~\cite{Hiller:2003,Bordone:2016}. In our analysis, we have not included the very recent results on $R(K^{(*)})$ by Belle collaboration \cite{Abdesselam:2019wac,Abdesselam:2019lab}. This is because the data has large error bars, making a meaningful comparison of the results will be difficult. Therefore, the observed data indicate a possible violation of lepton universality. There have been plenty of analysis on the NP explanations of the observed discrepancies, which we are not going to elaborate here. In order to explain the observed discrepancies, one needs to develop a new mechanism that will generate lepton universality violation (LUV) either at the tree level or via loops.  

Amongst the other important observables, anomalous magnetic moment of muon shows deviation between theory and experiment. Particle magnetic moments are good probes of physics beyond the SM, and the similar study could shed light on our understanding of quantum electrodynamics (QED) and the SM. The anomalous magnetic moment of muon has been measured very precisely while it has also been predicted in the SM to a great accuracy. The muon anomalous magnetic moment is defined as

\begin{equation}
 a_{\mu} \equiv \frac{g_{\mu}-2}{2},
\end{equation}

which includes the quantum loop effects, and parametrizes the small calculable deviation from $g_{\mu} = 2$ (Lande's $g$ factor). The SM contributions to $a_{\mu}$
can be expressed as 

\begin{equation}
 a_{\mu}^{SM} = a_{\mu}^{\rm QED} + a_{\mu}^{\rm EW} + a_{\mu}^{\rm Had},
\end{equation}

where $a_{\mu}^{\rm QED}$, $a_{\mu}^{\rm EW}$ and $a_{\mu}^{\rm Had}$ are the contributions from QED loops, electroweak loops and hadronic loops respectively. This quantity has been measured very accurately and at present the difference between the predicted and the measured value is given by 

\begin{equation}
 \Delta a_{\mu} = a_{\mu}^{exp} - a_{\mu}^{SM} = 26.8 (7.6)\times 10^{-10}, 
\end{equation}

which shows there is still room for NP beyond the SM (for details see~\cite{pdg2018}). In this study, we will look for a NP model which is capable of addressing simultaneously both the above mentioned excesses.  

On the other hand, dark matter (DM) has been understood to be present in significant amount in the present Universe, roughly five times the abundance of ordinary baryonic matter~\cite{Ade:2015xua}. The present dark matter abundance, measured by the Planck~\cite{Ade:2015xua} is often quoted as

\begin{equation}
\Omega_{\text{DM}} h^2 =\frac{\rho_{\text{DM}}}{\rho_{\text{c}}}h^2= 0.1198 \pm 0.0015
\label{dm_relic}
\end{equation}
where $h = H_0/(100 \;\text{km} \text{s}^{-1} \text{Mpc}^{-1})$, $\rho_{DM}$, and $\rho_{\text{c}}=\frac{3H^2_0}{8\pi G}$ are, respectively, the present day normalized 
Hubble expansion rate ($H_0$), DM density, and the critical density of the universe, whereas $G$ is the universal constant of gravity. Such cosmological evidences are also complemented by astrophysical evidences suggesting the presence of non-luminous and non-baryonic matter component in the universe~\cite{Zwicky:1933gu,Rubin:1970zza,Clowe:2006eq}.

In the SM, we do not have a suitable DM candidate which satisfies the requirements as given in~\cite{Taoso:2007qk}. This has led to several beyond the standard model (BSM) proposals which can successfully explain DM in the Universe. Amongst different BSM prescriptions, the paradigm with a generic weakly interacting massive particle (WIMP) is well motivated. In such scenarios, the DM particle has mass and interactions typically around the electroweak ballpark and can give rise to the correct dark matter relic abundance, a remarkable coincidence often referred to as the \textit{WIMP Miracle} (see, for example,~\cite{Arcadi:2017kky}). Since WIMP dark matter scenarios involve additional physics around the electroweak scale, it is tempting to speculate if the same new physics can have plausible explanations for the observed flavour anomalies like $R(K^{(*)})$, $a_{\mu}$ mentioned earlier. Within such unified framework, one needs to find out the allowed NP parameter space consistent with flavour data as well as the requirements for a DM candidate. Also, it is necessary to check that the required NP parameter spaces are consistent with all the other relevant measurements which are not anomalous. There have been several attempts along this direction, some of which can be found in~\cite{Kawamura:2017ecz, Sierra:2015fma, Baek:2017sew, Cline:2017aed,Cline:2017qqu} and references therein. Apart from being consistent with all these observations, it is also important for such a scenario to be predictive at different experiments like direct detection of dark matter, collider searches and so on.

In a model independent analysis~\cite{Geng:2017svp}, by considering an effective theory framework, it has been shown that the deficit in the lepton universality ratio $R(K^{(*)})$ can be best explained by the set of the operators
$ {\cal O}^{\ell}_9 = [\bar{b} \gamma_{\mu} P_L s][\bar{l}\gamma^{\mu} l ]$ and 
${\cal O}^{\ell}_{10}=[\bar{b} \gamma_{\mu} P_L s ][\bar{l}\gamma^{\mu}\gamma_5 l ]$. Therefore, the NP models under considerations should give rise to these four-fermi interactions either via tree or loop level diagrams for the process $b\to s\ell\ell$. Here, we consider the inert Higgs doublet model (IDM), which is a simple extension of the SM by an additional scalar field $\Phi_2$ transforming as doublet under $SU(2)_L$ gauge symmetry and has hypercharge $Y=1$. The model has been introduced in~\cite{Deshpande:1977rw}, and later studied extensively by several groups in the context of DM phenomenology~\cite{Ma:2006km,Barbieri:2006dq,Cirelli:2005uq, LopezHonorez:2006gr,Honorez:2010re,LopezHonorez:2010tb, Arhrib:2013ela,Dasgupta:2014hha,Borah:2017dqx, Borah:2017dfn}. 

In this model, an additional discrete $Z_2$ symmetry is introduced in order to prevent the coupling of this scalar field to the SM fermions. Under this $Z_2$ symmetry, the additional scalar field transforms as $\Phi_2 \rightarrow -\Phi_2$ whereas all SM fields are even. If the lightest component of $\Phi_2$ is electromagnetically neutral, it can be stable and hence a good DM candidate. Being inert in nature, IDM will not contribute to the decay $b\to s \ell\ell$. Hence, we have extended this model by considering three generations of vector like $SU(2)_L$ singlet down type quarks and charged leptons, odd under the $Z_2$ symmetry so that they can couple to the SM quarks and leptons only through the inert scalar doublet. The lightest component of $\Phi_2$ remains the lightest $Z_2$ odd particle of the model and hence the DM candidate. We have shown that apart from explaining the DM abundance of the Universe, the model can also explain the observed pattern in $R(K^{(*)})$. This model has family non-universal Yukawa couplings between $\Phi_2$, vector like fermions and the SM fermions. Hence, it will contribute to $b\to s\ell\ell$ at one loop level. Now, if we consider a hierarchical structure between the Yukawa couplings of electron and muon with the new vector like fermions, then we can expect to get $R(K^{(*)}) \ne 1$. The additional vector like fermions can also contribute to the relic abundance, as well as direct detection scattering rates of DM in this model, giving us a complementary probe of the model parameters in both DM and flavour experiments.

In the pure IDM there exists two mass ranges where DM relic abundance can be satisfied: one in the low mass regime below the $W$ boson mass threshold $(M_{\text{DM}} < M_W)$ and the other around 550 GeV or above. In our extended IDM, there will be additional annihilation channels of DM. Therefore, it is important to rescan the parameter space for both the pure and the extended IDM. The direct detection scattering in pure IDM is primarily mediated by the SM Higgs and faces the strongest constraints from the direct detection experiments in the low mass regime. For example, the latest data from the LUX experiment rules out DM-nucleon spin independent cross section above around $2.2 \times 10^{-46} \; \text{cm}^2$ for DM mass of around 50 GeV \cite{Akerib:2016vxi}. On the other hand, the recently released results from the XENON-1T experiment rules out spin independent WIMP-nucleon interaction cross section above $7.7\times 10^{-47}\; \text{cm}^2$ for DM mass of 35 GeV \cite{Aprile:2017iyp}. These strong bounds reduce the allowed DM masses in the low mass regime to a very narrow region near the SM like Higgs resonance $M_{\text{DM}} \approx m_h/2$. Although the direct detection limits can be somewhat relaxed in the high mass regime $(M_{\text{DM}}  \gtrapprox 550 \; \text{GeV})$, the production of DM at colliders will be suppressed compared to the low mass regime. In the presence of additional vector like quarks, there are additional diagrams which will contribute to the spin independent direct detection cross section. We in fact find that, compared to the pure IDM, the presence of new vector like fermions can keep the dark matter direct detection rates closer to the experimental upper bound for some choices of parameters.

The mediators of our model couples to SM quarks and leptons, therefore interesting collider signature are expected with leptons and/or jets in the final state with missing energy. We study the final states containing $(\ell^+\ell^- + \slashed{E_T})$, $(jj  + \slashed{E_T})$ and $(\ell^+\ell^- + jj + \slashed{E_T})$ to unravel the model in the large hadron collider (LHC). These final states are already explored in supersymmetry (SUSY) searches, and important constraints have been obtained on the parameter space~\cite{Khachatryan:2016fll,Aaboud:2017hrg}. There have also been some studies on collider signatures of pure IDM, for example see \cite{Gustafsson:2012aj, Poulose:2016lvz, Datta:2016nfz}. In our model, we prepare few benchmark scenarios by choosing points from the new parameter spaces which are allowed by flavour data and overcome bounds from the DM searches. We have predicted the kinematical distributions of our signal events and compared them with the respective SM backgrounds. We find that at the high luminosity LHC the model may be observed for a few benchmark scenarios at more than 5$\sigma$ significance. We also check the perturbative unitarity of the model and find that for the chosen benchmark points the model can remain perturbative up to an energy scale $10^5-10^7$ GeV.         

The paper is organised as follows: in Sec.~\ref{sec1} we discuss the particle content and possible interactions, followed by the dark matter phenomenology of the model in Sec.~\ref{sec:dmpheno}; ~constraints from muon $(g-2)$ and lepton flavour violating decays are discussed in Sec.~\ref{sec:muong2}; contributions in $b\to s $ transitions are studied in Sec.~\ref{sec:bs}; results from DM and flavour analysis are discussed Sec.~\ref{sec:resul} and some benchmark points are also chosen for further collider study; we then discuss the fate of this model at the LHC in Sec.~\ref{sec:colliderpheno}, pointing out the possibility of probing it in future higher luminosity; the RGE runnings are discussed in Sec.~\ref{sec:rge} and finally we summarize in Sec.~\ref{sec:concl}. 

\section{IDM with Vector Like Fermions}
\label{sec1}

\begin{table}[htbp]
\begin{center}
\renewcommand{\arraystretch}{1.5}
\begin{tabular}{|c|c|c|}
\hline
Particles & $SU(3)_c \times SU(2)_L \times U(1)_Y$ & $Z_2 $   \\
\hline
$Q_L=\begin{pmatrix}u_{L}\\
d_{L}\end{pmatrix}$ & $(3, 2, \frac{1}{6})$ & +  \\
$u_R$ & $(3, 1, \frac{2}{3})$ & +  \\
$d_R$ & $(3, 1, -\frac{1}{3})$ & +  \\
$L_L=\begin{pmatrix}\nu_{L}\\
e_{L}\end{pmatrix}$ & $(1, 2, -\frac{1}{2})$ & +  \\
$e_R$ & $(1, 1, -1)$ & + \\
$\Phi_1$ & $(1,2,\frac{1}{2})$ & + \\
\hline
$\Phi_2$ & $(1,2,\frac{1}{2})$ & - \\
${D}_{L,R}$ & $(3, 1, -\frac{1}{3})$ & - \\
$E_{L,R}$ & $(1, 1, -1)$ & - \\
\hline
\end{tabular}
\end{center}
\captionsetup{style=singlelinecentered}
\caption{Particle content of the extension of IDM by vector like fermions}
\label{table1}
\end{table}

As mentioned earlier, the IDM is an extension of the SM by an additional global discrete $Z_2$ symmetry under which a newly incorporated scalar doublet $\Phi_2$ 
transforms as $\Phi_2 \rightarrow -\Phi_2$, while the usual SM fields are even under $Z_2$. 
The requirement of keeping the $Z_2$ symmetry unbroken prevents the neutral component of the second Higgs doublet from acquiring a non-zero vacuum expectation 
value (vev). Since the same discrete symmetry prevents any coupling of $\Phi_2$ with the SM fermions, it automatically makes the lightest component of $\Phi_2$ stable 
and hence a good DM candidate. The scalar potential of the model involving the SM Higgs doublet $\Phi_1$ and the inert doublet $\Phi_2$ can be written as
\begin{equation}
\begin{aligned}
V(\Phi_1,\Phi_2) &=  \mu_1^2|\Phi_1|^2 +\mu_2^2|\Phi_2|^2+\frac{\lambda_1}{2}|\Phi_1|^4+\frac{\lambda_2}{2}|\Phi_2|^4+\lambda_3|\Phi_1|^2|\Phi_2|^2 \nonumber \\
& +\lambda_4|\Phi_1^\dag \Phi_2|^2 + \bigg\{\frac{\lambda_5}{2}(\Phi_1^\dag \Phi_2)^2 + \text{h.c.}\bigg\}.
\end{aligned}
\label {scalarpot}
\end{equation}
As the electroweak symmetry has to be broken by the vev of $\Phi_1$, we assume $\mu^2_1 <0$. Also, $\mu^2_2 >0$ is assumed so that $\Phi_2$ does not acquire a vev. 
Writing the scalar fields in terms of components and expanding the field $\Phi_1$ about the non-zero vev, we have
\begin{equation}
\Phi_1=\begin{pmatrix} 0 \\  \frac{ v +h }{\sqrt 2} \end{pmatrix} , \Phi_2=\begin{pmatrix} H^\pm\\  \frac{H^0+iA^0}{\sqrt 2} \end{pmatrix}
\end{equation}
in unitary gauge. Here $v$ is the vev of the neutral component of $\Phi_1$. After electroweak symmetry breaking (EWSB), the masses of the physical scalars, 
at tree level, can be written as
\begin{eqnarray}
m_h^2 &=& \lambda_1 v^2 ,\nonumber\\
M_{H^\pm}^2 &=& \mu_2^2 + \frac{1}{2}\lambda_3 v^2 , \nonumber\\
M_{H^0}^2 &=& \mu_2^2 + \frac{1}{2}(\lambda_3+\lambda_4+\lambda_5)v^2=M^2_{H^\pm}+
\frac{1}{2}\left(\lambda_4+\lambda_5\right)v^2  , \nonumber\\
M_{A^0}^2 &=& \mu_2^2 + \frac{1}{2}(\lambda_3+\lambda_4-\lambda_5)v^2=M^2_{H^\pm}+
\frac{1}{2}\left(\lambda_4-\lambda_5\right)v^2.
\label{mass_relation}
\end{eqnarray}
Here $m_h \approx 125$ GeV is the mass of the SM Higgs, $M_{H^0}, M_{A^0}$ are the masses of the CP even and CP odd scalars of the inert doublet while $M_{H^\pm}$ being 
the mass of the charged scalar. Without any loss of generality, we consider $\lambda_5 <0, \lambda_4+\lambda_5 <0$ so that the CP even scalar is the lightest $Z_2$ 
odd particle and hence a stable dark matter candidate.

Apart from the $Z_2$ odd scalar doublet $\Phi_2$, we consider additional vector like charged fermions too, which are odd under the same $Z_2$ symmetry. The particle content of the model is shown in Table \ref{table1}. Here $D$ is the down-type vector like quark and $E$ is the vector like lepton. This allows the coupling of the inert doublet scalar with the SM fermions through the vector like fermion portal. The relevant Yukawa Lagrangian can be written as

\begin{align}\label{newlag}
{\cal L}  &= (y_u)_{ij} \bar{Q}_i \tilde{\Phi}_1 u_{Rj} + (y_d)_{ij} \bar{Q}_i \Phi_1 d_{Rj} + (y_e)_{ij} \bar{L}_i \Phi_1 e_{Rj} 
+ (\lambda^D)_{ij} \bar{Q}_i \Phi_2 D_{Rj}  + (\lambda^{E})_{ij} \bar{L}_i \Phi_2 E_{Rj} \nonumber \\
& + M_{D_i} \bar{D}_{Li} D_{Ri} + M_{E_i} \bar{E}_{Li} E_{Ri} + \text{h.c.} 
\end{align}
where $\tilde{\Phi}_{1,2} = i \tau_2 \Phi^*_{1,2}$ and $\lambda^{E,D}$ are the Yukawa couplings associated with the vector fermion interactions. Also, $i$ and $j$ are the generation indices. 

We are working in a basis where the vector like fermion fields are diagonal. Also, the SM fields can be rotated to the corresponding mass basis.  In our analysis, in the quark sector the phenomenologically relevant new interactions involve only SM down-type quarks. Therefore, while rotating the SM quark fields, in principle, we can choose a basis where down-type quarks are diagonal. In such case, there won't be any changes in the definitions of our new Yukawas. On the other hand, if we work in a basis where SM up-type quarks are diagonal, then there will be additional flavour mixing of CKM type in the relevant interactions. These mixings will redefine our Yukawas ($\lambda^D_{ij}$) in Eq.\ref{newlag} to say ${\lambda^D_{ij}}^{\prime}$. However, phenomenologically this will not add any new information in our analysis. The similar analysis could be done with ${\lambda^D_{ij}}^{\prime}$.

In the lepton sector, the new charge current interactions involving neutrinos are not phenomenologically relevant in our analysis. The phenomenologically relevant new interactions involve only charged lepton Yukawas. Therefore, the additional vector like charged fermions do not contribute to the generation of light neutrino mass. Now, even if neutrino masses are generated by some new physics mechanism, the corresponding mixing angles will not enter in any of the observables  or physical processes discussed in this work.

\section{Dark Matter Phenomenology}
\label{sec:dmpheno}

In this section we discuss the DM phenomenology of this model in terms of relic density and direct search bounds. We divide the discussion into the following two subsections.

\subsection{Relic abundance of DM}
\label{sec2}

For a single component DM, the relic abundance can be obtained by solving the Boltzmann equation (BEQ):

\begin{equation}
\frac{dn_{\rm DM}}{dt}+3Hn_{\rm DM} = -\langle \sigma v \rangle (n^2_{\rm DM} -(n^{\rm eq}_{\rm DM})^2),
\end{equation}

where $n_{\rm DM}$ is the number density of the $\rm DM$ particle, $n^{\rm eq}_{\rm DM}$ is the equilibrium number density and $H$ is the Hubble expansion rate. The thermally averaged annihilation cross section $ \langle \sigma v \rangle $ can be expanded in powers of (non-relativistic) velocity as: $ \langle \sigma v \rangle = a +b v^2+...$, where the first term corresponds to $s$-wave, the second terms corresponds to $p$-wave and so on. Under this approximation, BEQ can be solved numerically to find the present day relic density of the DM~\cite{Kolb:1990vq,Scherrer:1985zt}:

\begin{equation}
\Omega_{\rm DM} h^2 \approx \frac{1.04 \times 10^9 x_F}{M_{\text{Pl}} \sqrt{g_*} (a+3b/x_F)}~,
\end{equation}

where $x_F = M_{\rm DM}/T_F$, $T_F$ is the freeze-out temperature, $M_{\rm DM}$ is the mass of dark matter, $g_*$ is the total number of relativistic degrees of freedom (DOF) at the time of freeze-out ($\sim 106$) and and $M_{\text{Pl}} \approx 2.4\times 10^{18}$ GeV is the reduced Planck mass. WIMPs generally freeze out at: $x_F \approx \{20-30\}$. Generically, $x_F$ can be obtained from the relation:

\begin{equation}
x_F = \ln \frac{0.038gM_{\text{Pl}}M_{\rm DM}<\sigma v>}{g_*^{1/2}x_F^{1/2}}~,
\label{xf}
\end{equation}

which is derived from the equality condition of DM interaction rate $\Gamma = n_{\rm DM} \langle \sigma v \rangle$ with the rate of expansion of the Universe $H \approx g^{1/2}_*\frac{T^2}{M_{Pl}}$ (i.e, the freeze-out condition). 

For all practical purposes, one can obtain the approximate analytical solution for relic density as~\cite{Jungman:1995df} :

\begin{equation}
\Omega_{\rm DM} h^2 \approx \frac{3 \times 10^{-27} cm^3 s^{-1}}{\langle \sigma v \rangle}.
\label{eq:relic}
\end{equation}

The thermally averaged annihilation cross section $\langle \sigma v \rangle$ is given by \cite{Gondolo:1990dk}
\begin{equation}
\langle \sigma v \rangle = \frac{1}{8m^4T K^2_2(M_{DM}/T)} \int^{\infty}_{4M_{DM}^2}\sigma (s-4M_{DM}^2)\surd{s}K_1(\surd{s}/T) ds~,
\end{equation}

where $K_i$'s are modified Bessel functions of order $i$.


In presence of co-annihilation, the effective cross section can be expressed as~\cite{Griest:1990kh}: 
\begin{align}
\sigma_{eff} &= \sum_{i,j}^{N}\langle \sigma_{ij} v\rangle r_ir_j \nonumber \\
&= \sum_{i,j}^{N}\langle \sigma_{ij}v\rangle \frac{g_ig_j}{g^2_{eff}}(1+\Delta_i)^{3/2}(1+\Delta_j)^{3/2}e^{\big(-x_F(\Delta_i + \Delta_j)\big)}, \nonumber \\
\end{align}

where $x_F = \frac{M_{DM}}{T_F}$ and $\Delta_i = \frac{m_i-M_{\text{DM}}}{M_{\text{DM}}}$, where the masses of the heavier components of the inert Higgs doublet are denoted by $m_{i}$. Total number of effective DOF is given by: 

\begin{align}
g_{eff} &= \sum_{i=1}^{N}g_i(1+\Delta_i)^{3/2}e^{-x_F\Delta_i}.
\end{align}

Thermally averaged cross section then reads:

\begin{align}
\langle \sigma_{ij} v \rangle &= \frac{x_F}{8m^2_im^2_jM_{\text{DM}}K_2((m_i/M_{\text{DM}})x_F)K_2((m_j/M_{\text{DM}})x_F)} \times \nonumber \\
& \int^{\infty}_{(m_i+m_j)^2}ds \sigma_{ij}(s-2(m_i^2+m_j^2)) \sqrt{s}K_1(\sqrt{s}x_F/M_{\text{DM}}). \nonumber \\
\label{eq:thcs}
\end{align}

The relic density can be again computed by approximate analytical solution:

\bea
\Omega_{\rm DM} h^2 = \frac{2.4\times 10^{-10}}{\sigma_{eff}}~\rm GeV^{-2}.
\eea

In the present model, discussed in the previous section, we consider one of the neutral component of the scalar doublet $\Phi_2$ namely $H^0$, as the DM candidate for our analysis. This is similar to the inert doublet model of dark matter discussed extensively in the literature~\cite{Ma:2006km,Barbieri:2006dq,Cirelli:2005uq, LopezHonorez:2006gr,Honorez:2010re,LopezHonorez:2010tb,Arhrib:2013ela,Dasgupta:2014hha}. In the low mass regime ($M_{H^0} \equiv M_{\text{DM}}\leq M_W$), the annihilation of DM to the SM fermions (through s-channel Higgs mediation) dominates over other channels. As pointed out in~\cite{Honorez:2010re}, the annihilation $H^0 H^0 \rightarrow W W^* \rightarrow W f \bar{f^{\prime}}$ also plays a role in the $M_{\text{DM}} \leq M_W$ region. Depending on the mass differences $M_{H^\pm}-M_{H^0}\,(\equiv \Delta M_{H^\pm}), M_{A^0}-M_{H^0}\,( \equiv \Delta M_{A^0})$, co-annihilation of $H^0, H^\pm$ and $H^0, A^0$ become important in determining the relic abundance of the DM. Typically, when the heavier components of the inert scalar doublet have masses close to the DM mass, they can be thermally accessible at the epoch of DM freeze-out. Therefore, the annihilation cross section of DM in such a case gets additional contributions from co-annihilations between the DM and the heavier components of the scalar doublet $\Phi_2$.

\subsection{Dark matter direct search}
\label{sec:dd}

\begin{figure}[htb!]
$$
\includegraphics[scale=0.56]{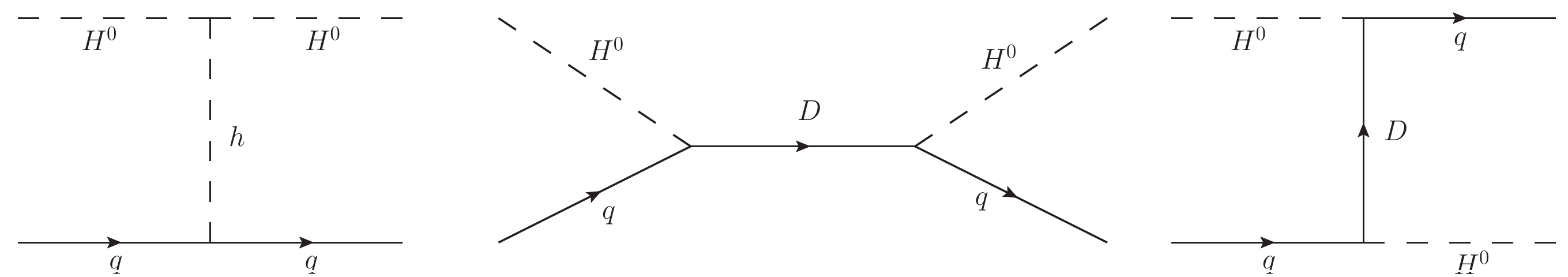}
$$
\captionsetup{style=singlelineraggedleft}
\caption{The Feynman diagrams contributing to the direct search of $H^{0}$. Here, $D$ indicates the contributions from down type vector like quarks (all the three generations)}
\label{fig:dirsearch}
\end{figure}

As mentioned earlier, there are severe constraints on spin independent DM-nucleon scattering rates from ongoing experiments \cite{Akerib:2016vxi, Aprile:2017iyp}. In the pure IDM, the tree level DM-nucleon elastic scattering can arise through the SM Higgs mediation and the current bounds on direct detection cross section can rule out some portion of the parameter space satisfying relic specially in the low mass regime $M_{\text{DM}} \approx m_h/2$ where bounds are stronger. 
The elastic DM nucleon scattering in the present model gets additional contributions from exotic quark $D$, as depicted in Fig.~\ref{fig:dirsearch} where the first diagram corresponds to the usual SM Higgs mediated one. The additional contributions will come from the rest of the two diagrams. There is another possible diagram mediated by $Z$-boson, even in the pure IDM, but that has already been excluded by recent direct search data. Therefore, in order to forbid the $Z$-mediated channel, the mass of $A^{0}$ has to be kept higher than that of $H^{0}$ by a non-zero value, higher than typical kinetic energy $(\mathcal{O}(100 \; \text{keV}))$ of a DM particle so that $H^0$ can not scatter inelastically into $A^0$. The chosen mass splitting in our analysis satisfy this bound as well as the ones from LEP II data~\cite{Lundstrom:2008ai}. Hence, in this model we have three direct search graphs corresponding to $t$-channel Higgs and exotic quark mediation and another $s$-channel diagram mediated by the vector like quark. Due to these additional diagrams, the direct detection rates of the extended IDM can be more promising compare to the pure IDM, as we will discuss later. In the limit of very large exotic quark masses or very small couplings of exotic quarks to DM, the direct detection rates will converge towards the ones known for pure IDM.

\section{Muon $(g-2)$ and the lepton flavour violation (LFV) decays}
\label{sec:muong2}

The effective vertex of photon with any charged particle is given by: 

\begin{equation}
 {\bar u}(p^{\prime}) e \Gamma_\mu u(p) = {\bar u}(p^{\prime})\left[e \gamma_{\mu} F_1(q^2) + \frac{i e \sigma_{\mu\nu} q^{\nu}}{2 m_f} F_2(q^2)+ ...\right] u(p).
\end{equation}

The factor $g_{\mu} \equiv 2 (F_1(0) + F_2(0))$, and the anomalous magnetic moment is given as $a_{\mu} \equiv F_2(0) \ne 0$ (since $F_1(0) = 1$ at all order). Similarly, the amplitude for the LFV decays $\ell_i \to \ell_j \gamma$ can be written as: 

\begin{equation}
 M_{\gamma} = {\bar u}_{\ell_j}(p^{\prime})\left[A_L q^2 \gamma_{\mu} P_L  + i A_R  m_{\ell_i} \sigma_{\mu\nu} q^{\nu} P_R \right] u_{\ell_i}(p). 
\end{equation}

The associated branching fraction can be expressed as:

\begin{equation}
 \mathcal{B}(\ell_i \to \ell_j\gamma) = \frac{\alpha \tau_{\ell_i}}{4 } m_{\ell_i}^5 A_R^2, 
\end{equation}
where $\tau_{\ell_i}$ is the life time of the lepton $\ell_i$ and $\alpha = 1/137$ is the fine structure constant.

\begin{figure}[htb!]
\centering
\includegraphics[scale=0.5]{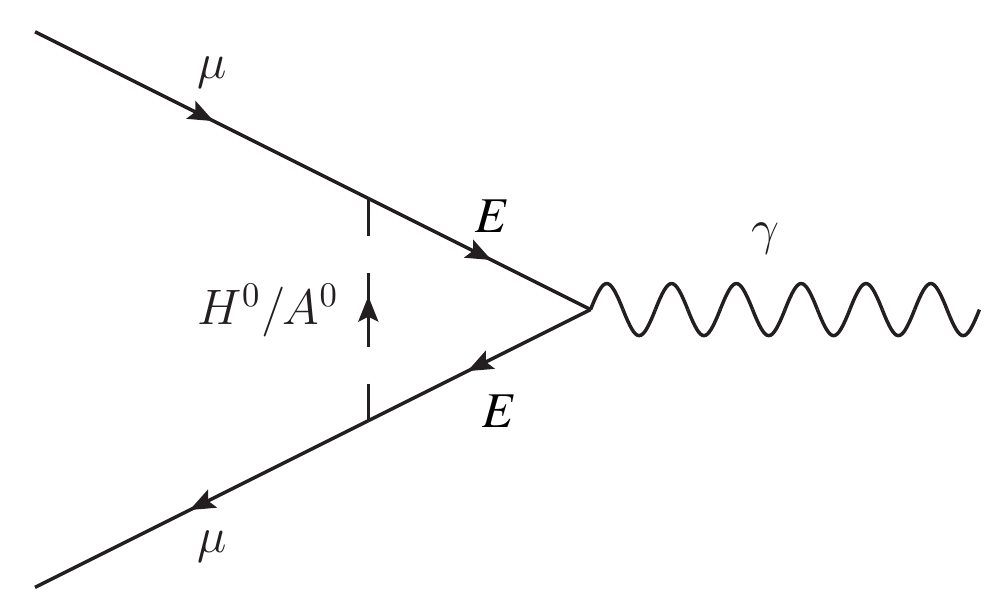}
\includegraphics[scale=0.45]{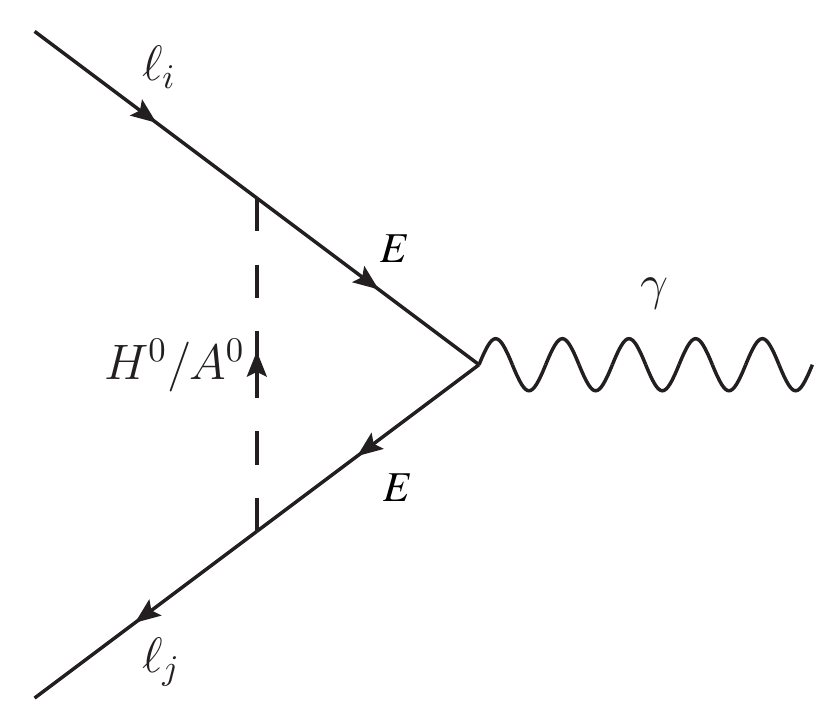}
\captionsetup{style=singlelineraggedleft}
\caption{Feynman diagrams contributing to muon anomalous magnetic moment $a_\mu$ (left) and lepton flavour violating decays (right). Here, $E \ (= E_1/E_2/E_3)$ is the vector like lepton}
\label{fig:muong2}
\end{figure}

In our model, the leading contributions in $a_\mu$ and the LFV decays like $\tau\to \mu\gamma$, $\mu\to e\gamma$ and $\tau \to e\gamma$ are obtained from the diagrams in Fig.~\ref{fig:muong2}. In the loop, we have either $H^0$ or $A^0$ and the vector like lepton $E$ (which could be either of $E_1$, $E_2$ or $E_3$). The diagram on the left hand side will contribute to $a_{\mu}$, which is given by \cite{Leveille:1977rc,Bhattacharyya:2009,Lindner:2016bgg}
\begin{equation}
a_{\mu} = \sum_{i} \frac{\lambda^E_{2i}{\lambda^E_{2i}}^* m_{\mu}^2}{16\pi^2}\left[  \frac{1}{M_{H^0}^2} \bigg(\xi_1(r^{H^0}_{E_i}) - 
 \xi_2(r^{H^0}_{E_i})\bigg) + \frac{1}{M_{A^0}^2} \bigg(\xi_1(r^{A^0}_{E_i} - \xi_2(r^{A^0}_{E_i})\bigg) \right],
\end{equation}

with $r^X_{E_i} = m_{E_i}^2/M_{X}^2$ ($X = H^0$ or $A^0$). Here, $m_{E_i}$ is the mass of i-th generation vector like lepton and $\lambda^E_{2i}$ is the corresponding coupling to muon as expressed in Eq. \ref{newlag}. Note that in the Yukawas of the vector like fermions, the first index corresponds to the SM fermion generation while the second index represents that for vector like fermions. The functions $\xi_1$ and $\xi_2$ are given by: 

\begin{align}
\xi_1(r) &=  \frac{-3 + 4 r - r^2 }{2 (1-r)^3} - \frac{\ln r}{(1-r)^3} \nn \\ 
\xi_2(r) &=  \frac{1}{6(1-r)^4}[-11 + 18 r - 9 r^2 + 2 r^3 - 6 \ln r].
\end{align}

The contributions to the decay $\ell_i \to \ell_j \gamma$ will be obtained from the RHS diagram of Fig.~\ref{fig:muong2}, which is given as \cite{Bhattacharyya:2009,Lindner:2016bgg} :
\begin{equation}
 A_R = \sum_{k} \frac{ \lambda^E_{ik} {\lambda^E_{jk}}^* m_{\ell_i}}{{16 \pi^2}}\left[  \frac{1}{M_{H^0}^2} \bigg(\xi_1(r^{H^0}_{E_k}) - \xi_2(r^{H^0}_{E_k})\bigg) + \frac{1}{M_{A^0}^2} \bigg(\xi_1(r^{A^0}_{E_k}) - \xi_2(r^{A^0}_{E_k})\bigg) \right].
\end{equation}

In this section we have only shown the analytical expressions of various contributions in $a_{\mu}$ and $\mathcal{B}(\ell_i \to \ell_j \gamma)$, the numerical results are presented in section \ref{sec:resul}.

\section{NP contributions in $b\to s$ decays}
\label{sec:bs}
\subsection{ $b\to s \ell^+\ell^-$ decays ($\ell = \mu$, $e$)}
\label{sec:bsll}

As mentioned earlier, the FCNC transitions such as $b \rightarrow s$  are important probes of flavour physics and are highly sensitive to NP contributions. The effective Hamiltonian for the $b \rightarrow s$ transitions at low energy can be written as \cite{Altmannshofer:2008dz,Becirevic:2012fy}:

\begin{equation}
\mathcal{H}_{\text{eff}} = - \frac{4 G_F}{\sqrt{2}}V_{tb} V_{ts}^{\ast} \left(\sum_{i=1...6} C_i \mathcal{O}_i + 
\sum_{i=7,8,9,10,S,P} ( C_i \mathcal{O}_i + C_i^{\prime} \mathcal{O}_i^{\prime})\right) + h.c.
\label{Heff}
\end{equation}

where $O_i$ and $O_i^{\prime}$'s are the dimension six effective operators which are given as below, 

\begin{eqnarray}
\mathcal{O}_1 &=& (\bar{s}_{\alpha} c_{\beta})_{V-A} (\bar{c}_{\beta}b_\alpha)_{V-A}, \hskip 1cm \mathcal{O}_2 = (\bar{s}c)_{V-A}(\bar{c}b)_{V-A}, \nonumber \\
\mathcal{O}_3 &=& (\bar{s}b)_{V-A}\sum_{q}(\bar{q}q)_{V-A}, \hskip 1.2cm \mathcal{O}_4 = (\bar{s}_\alpha b_\beta)_{V-A} \sum_{q}(\bar{q}_{\beta} q_{\alpha})_{V-A}, \nonumber \\
\mathcal{O}_5 &=& (\bar{s}b)_{V-A} \sum_{q}(\bar{q}q)_{V+A}, \hskip 1.2cm \mathcal{O}_6 = (\bar{s}_{\alpha}b_{\beta})_{V-A} \sum_{q}(\bar{q}_{\beta}q_{\alpha})_{V+A}, \nonumber \\
\mathcal{O}_7 &=& \frac{e}{g^2}m_b (\bar{s}\sigma_{\mu \nu} P_R b)F^{\mu \nu}, \hskip 1.2cm \mathcal{O}_7^{'} = \frac{e}{g^2}m_b (\bar{s}\sigma_{\mu \nu} 
P_L b)F^{\mu \nu}, \nonumber \\
\mathcal{O}_8 &=& \frac{1}{g}m_b (\bar{s}\sigma_{\mu \nu} T^a P_R b)G^{\mu \nu},\hskip 1cm \mathcal{O}_8^{'} = \frac{1}{g}m_b (\bar{s}\sigma_{\mu \nu}
T^a P_L b)G^{\mu \nu}, \nonumber \\
\mathcal{O}_9 &=& \frac{e^2}{g^2} (\bar{s}\gamma_{\mu} P_L b)(\bar{l}\gamma^{\mu}l),\hskip 1.6cm \mathcal{O}_9^{'} = \frac{e^2}{g^2} (\bar{s}\gamma_{\mu} 
P_R b)(\bar{l}\gamma^{\mu}l), \nonumber \\
\mathcal{O}_{10} &=& \frac{e^2}{g^2} (\bar{s}\gamma_{\mu} P_L b)(\bar{l}\gamma^{\mu}\gamma_5 l),\hskip 1.23cm \mathcal{O}_{10}^{'} = \frac{e^2}{g^2} 
(\bar{s}\gamma_{\mu} P_R b)(\bar{l}\gamma^{\mu}\gamma_5 l), \nonumber \\
\mathcal{O}_{S} &=& \frac{e^2}{16\pi^2} (\bar{s} P_L b)(\bar{l} l), \hskip 1.9cm \mathcal{O}_{S}^{'} = \frac{e^2}{16\pi^2} (\bar{s} P_R b)(\bar{l} l), \nonumber \\
\mathcal{O}_{P} &=& \frac{e^2}{16\pi^2} (\bar{s} P_L b)(\bar{l}\gamma_5 l),\hskip 1.5cm \mathcal{O}_{P}^{'} = \frac{e^2}{16\pi^2} (\bar{s} P_R b)(\bar{l}\gamma_5 l),
\label{effoperators}
\end{eqnarray}

where $\alpha$ and $\beta$ denote the color indices and the labels $(V \pm A)$ refer to $\gamma_{\mu} ( 1 \pm  \gamma_5)$, and $P_{L,R} = \big(\frac{1 \mp \gamma_5}{2}\big)$. The operators ${\cal O}_1$ to ${\cal O}_{10}$ appear in the SM effective theory, as well as in specific BSM scenarios, while the rest will appear only in NP models. The Wilson coefficients ($C_i$s) corresponding to the SM effective operators can be found in~\cite{Buras:1994dj}. The operators relevant for the decay $b\to s \ell^+\ell^-$ are given by $\mathcal{O}_{9,10}^{(\prime)}$. However, only $\mathcal{O}_{9,10}$ can explain the observed pattern in $R(K^{(*)})$~\cite{Geng:2017svp}. The expression for the decay rate corresponding to the operator basis given in Eq.~\ref{effoperators} are taken from \cite{Altmannshofer:2008dz}. 

Another $b \rightarrow s \mu^+\mu^-$ transition that plays a major role in constraining the NP parameter spaces is the rare decay $B_s \rightarrow \mu^+ \mu^-$. In the SM, this decay occurs via the penguin and the box diagrams, and is helicity suppressed. In the operator basis mentioned in Eq.~\ref{effoperators}, only $\mathcal{O}_{10}$ contributes to this process within SM. Corresponding expression for the branching fraction is given by:

\begin{equation}
	\mathcal{B}(B_s \rightarrow l^+ l^-)^{\text{SM}} = \tau_{B_s}\frac{G_F^2 \alpha^2}{16 \pi^3}|V_{tb}V_{ts}^*|m_{B_s}m_{\mu}^2 
	\sqrt{1-\frac{4  m_{\mu}^2}{m_{B_s}^2}}f_{B_s}^2 |C_{10}|^2.
\end{equation}

The SM prediction \cite{Bobeth:2013uxa} and the measured value \cite{pdg2018, Aaij:2017vad} of the branching fraction for this particular rare decay are respectively given by:

\begin{align}
\mathcal{B}(B_s \rightarrow \mu^+ \mu^-)^{\text{SM}} &= (3.65 \pm 0.23) \times 10^{-9}, \\
\mathcal{B}(B_s \rightarrow \mu^+ \mu^-)^{\text{Expt}}& = (2.7^{+0.6}_{-0.5}) \times 10^{-9}.
\label{Bsnumbers}
\end{align}

We note that the measured value and the SM prediction are consistent with each other within the error bars. This, in turn, will be helpful to constrain new physics parameters.

In the BSM framework, there are several dimension six effective operators which may contribute to the process $B_s \rightarrow \mu^+ \mu^-$. In the operator basis of Eq.~\ref{effoperators}, the expression for the branching fraction will then be modified to:

\begin{equation}
	\begin{split}
	\mathcal{B}(B_s \rightarrow \mu^+ \mu^-)^{\text{BSM}} = \tau_{B_s}f_{B_s}^2m_{B_s}^3\frac{G_F^2 \alpha^2}{64 \pi^3} |V_{tb}V_{ts}^*| & 
	\sqrt{1-\frac{4  m_{\mu}^2}{m_{B_s}^2}} \hspace{0.3cm}        
	\bigg[\frac{m_{B_s}^2}{m_b^2} \hspace{0.1cm} \bigg(1-\frac{4m_{\mu}^2}{m_{B_s}^2}\bigg) \hspace{0.1cm} \left |C_S-C_S^{'}\right |^2 \\ & +
	\left  |\frac{m_{B_s}}{m_b}(C_P-C_P^{'})+2\frac{m_{\mu}}{m_{B_s}}(C_{10}-C_{10}^{'})\right |^2\bigg].
	\label{Bs2mumu}
	\end{split}
\end{equation}

Here, $C_S^{(')}$ and $C_P^{(')}$ are the Wilson coefficients associated with the scalar and pseudoscalar operators. It has been shown that these operators are tightly constrained by the data on $\mathcal{B}(B_s \rightarrow \mu^+ \mu^-)$. Therefore, the contributions from the scalar and pseudoscalar operators can not explain the observed anomalous results in $R(K^{(*)})$ \cite{Hiller:2014yaa,Alonso:2014csa}.



\begin{figure*}[t]
\centering
{\includegraphics[height=3cm]{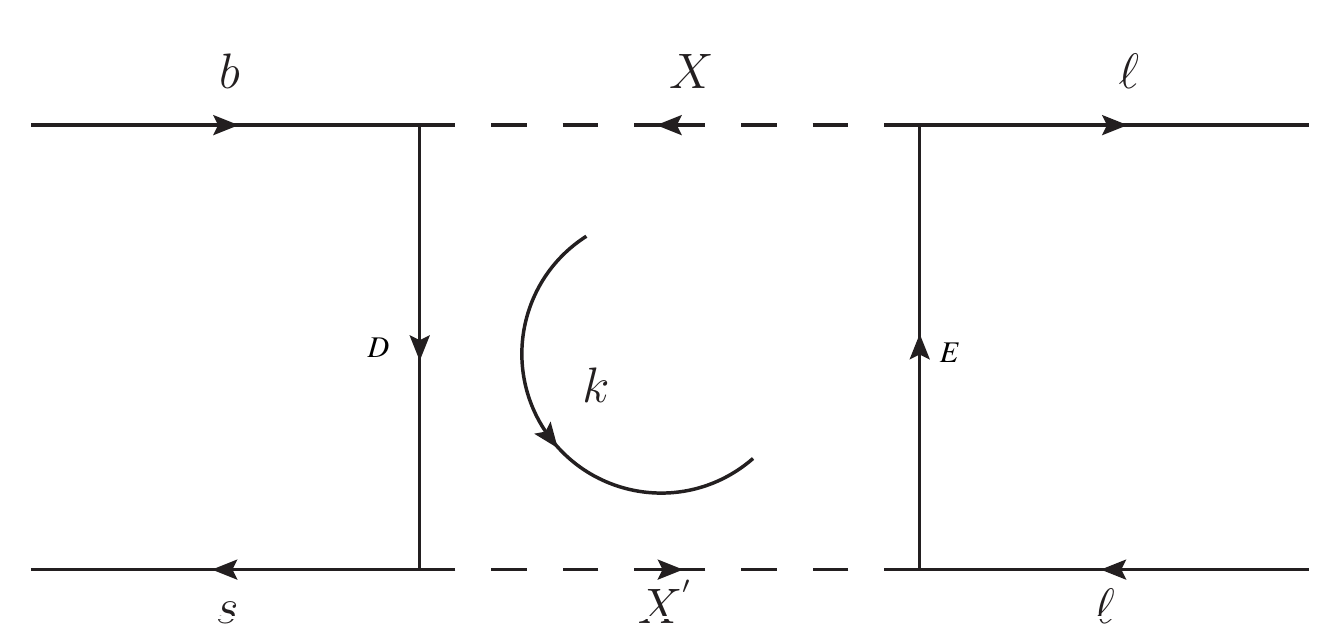}~~~~~~
\includegraphics[height=3cm]{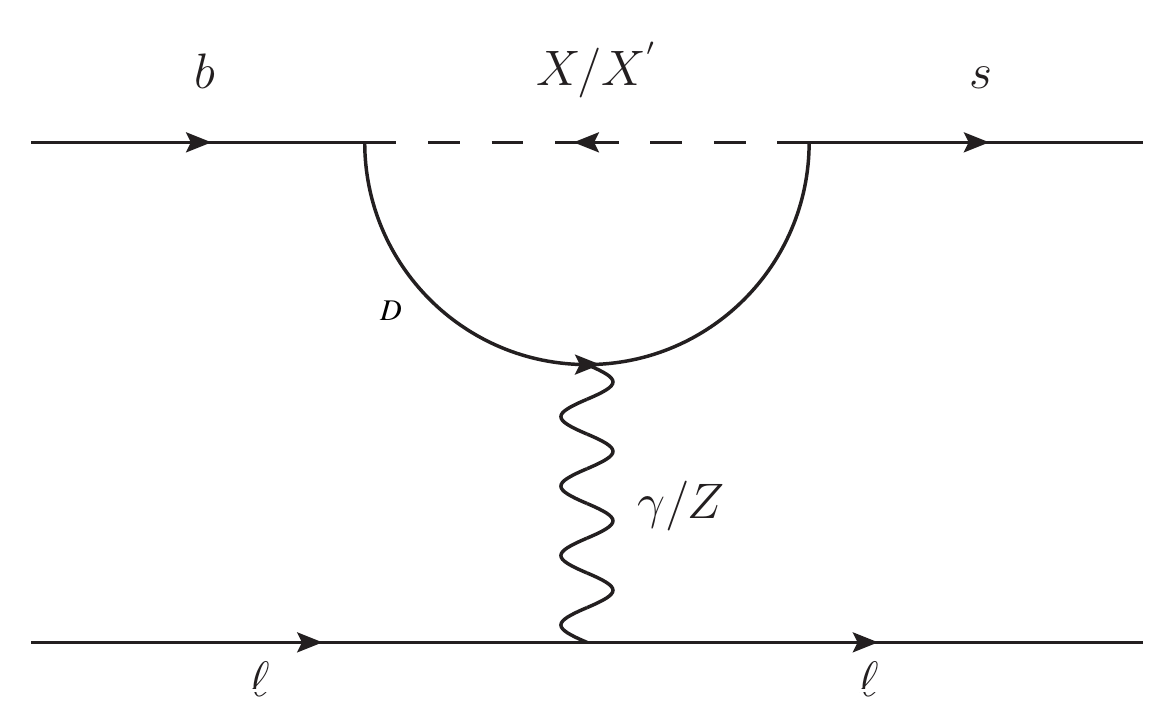}}\\
\includegraphics[height=3cm]{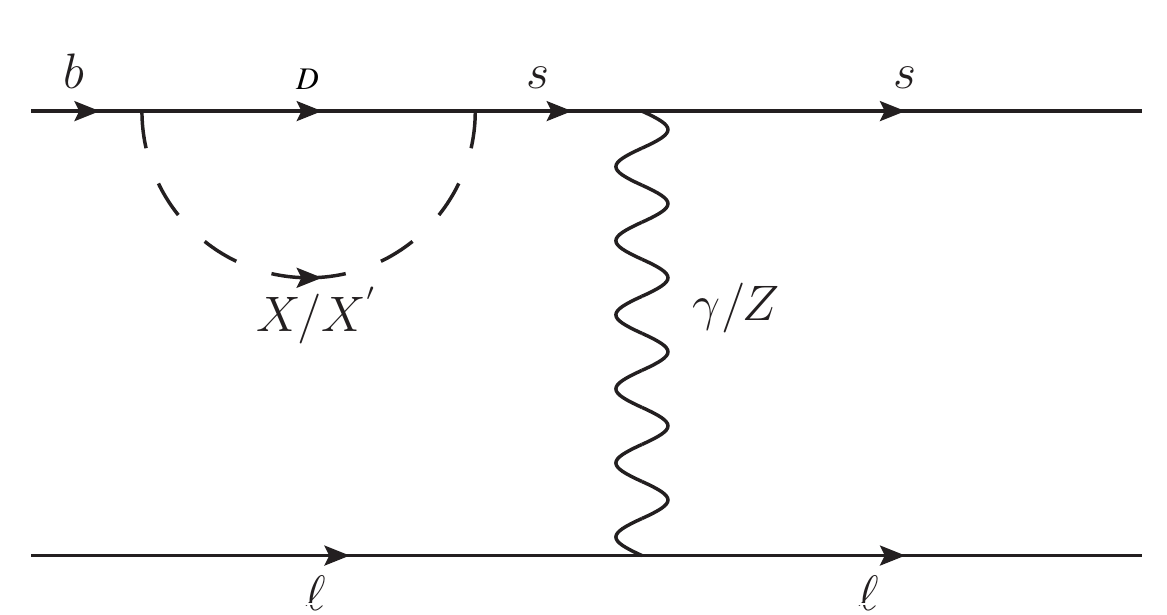}~~~~~~
\includegraphics[height=3cm]{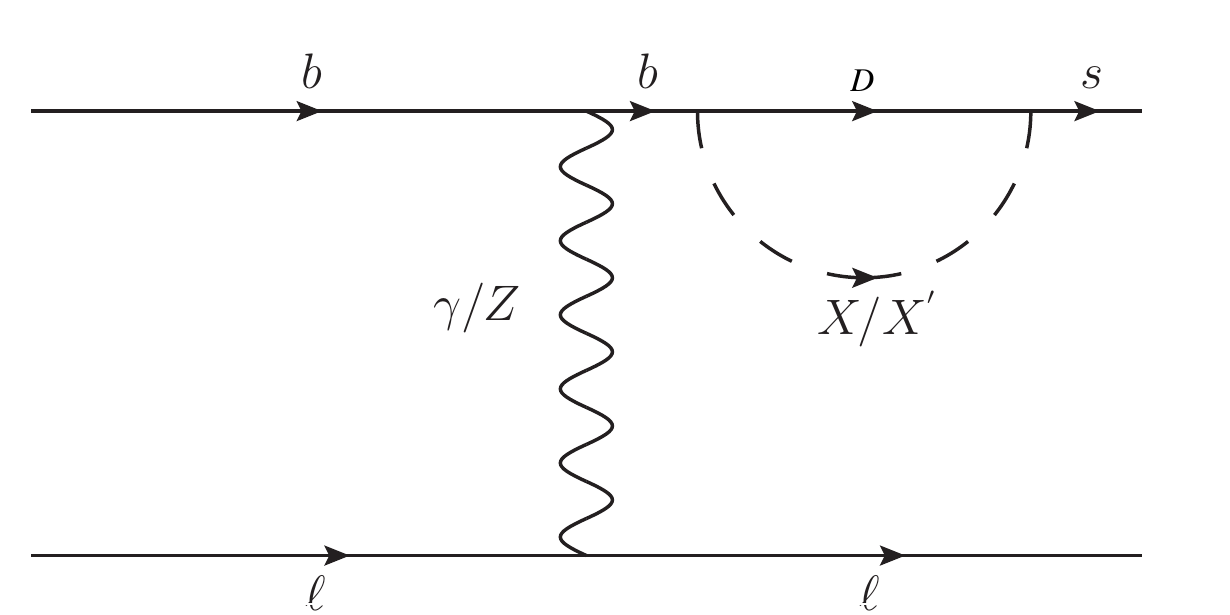}\\
\captionsetup{style=singlelineraggedleft}
\caption{Feynman diagrams contributing to $b \to s \ell \ell$ process. Here $X$ and $X^{'}$ can be either of $H^0$ or $A^0$. The box diagrams with $X = X^{\prime} = H^0/A^0$ will also contribute to $b\to s \ell\ell$ processes}
\label{btosdiag}
\end{figure*}

In our model, the diagrams that will contribute to the process $b\to s\ell\ell$ are shown in Fig.~\ref{btosdiag}, where $X/X^{\prime}$ can be either of $H^0$ or $A^0$. As one can see from Eq.~\ref{newlag}, the new couplings ($\lambda^{E,D}_{ij}$) carry the generation indices of the SM fields (first index) as well as that of the vector like fermions (second index).  Therefore, depending on the type of vector like fermion in the loop, there will be several contributions to the decay amplitude. This will be function of the new Yukawa couplings and the masses of the new particles. However, for the simplicity of the analysis, we have followed the hierarchy: $\lambda^{E,D}_{ij} << \lambda^{E,D}_{ii}$ (i, j = 1, 2 and 3), i.e, the off diagonal Yukawas are suppressed with respect to the diagonal terms. Also, since one of our goals is to explain the $R(K^{(*)})$ anomaly, which requires lepton universality violation, we have further assumed $\lambda^{E,D}_{33} >> \lambda^{E,D}_{22} >> \lambda^{E,D}_{11}$. In this simplified picture, the box diagram with $D_3$ and $E_2$ (in the loop) will have the dominant contribution to the process $b\to s \mu^+\mu^-$. Since the dominant contribution to all the observables mentioned above occur via the third generation of down-type vector like fermion $D_3$ (due the to hierarchy in the couplings), we will from now on talk only about the mass of $D_3$. In general, the contributions from the penguin diagrams are dominant over that of the box diagrams. However, the penguin diagrams alone can not explain $R(K^{(\ast)})$ anomaly, as they contribute equally to the decay rates of $B\to K^{(*)} \mu \mu$ and $B\to K^{(*)} e e$. Perhaps it is possible to explain the observed data by considering contributions from the new box diagrams alone. In such cases, the interference of the SM Wilson coefficients (WC) with that obtained from the box diagrams will play the leading role in explaining the observed pattern in $R(K^{(*)})$ data. If we add the contributions from the penguin diagrams, then there will be interference of the WC obtained from the box and the penguin diagrams. Hence, depending on the size of the individual contributions, the interference of the new box and penguin diagrams could also play an important role in the explanation of the observed data. For completeness, in our analysis we have considered the contributions from all types of diagrams which are shown in Fig.~\ref{btosdiag}.
 
The most general expression for the box diagram with two different scalars $X$ and $X^{'}$ in the loop is given by :
\begin{equation}
	\begin{split}
	i \mathcal{M}_{Box} =\frac{i \pi^2 {\lambda^D_{3i}}^* \lambda^D_{2i} {\lambda^E_{2j}}^* \lambda^E_{2 j}} {(2\pi)^4} \ {\cal A}\ \mathcal{O}_{eff} ,
	\end{split}
	\label{iM}
\end{equation}
 the loop factor is given by \cite{Bhattacharyya:2009}   
\begin{equation}
\begin{split}
{\cal A} = \bigg[\frac{M_{X^{'}}^4}{(M_X^2-M_{X^{'}}^2)(M_{D_i}^2-M_{X^{'}}^2)(M_{E_j}^2-M_{X^{'}}^2)}\text{ln}\bigg(\frac{M_X^2}{M_{X^{'}}^2}\bigg) 
& \\ + \frac{M_{D_i}^4}{(M_{D_i}^2-M_{E_j}^2)(M_{D_i}^2-M_{X^{'}}^2)(M_{D_i}^2-M_{X}^2)}\text{ln}\bigg(\frac{M_{D_i}^2}{M_{X}^2}\bigg)
& \\ + \frac{M_{E_j}^4}{(M_{E_j}^2-M_{D_i}^2)(M_{E_j}^2-M_{X^{'}}^2)(M_{E_j}^2-M_{X}^2)}
\text{ln}\bigg(\frac{M_{E_j}^2}{M_{X}^2}\bigg)\bigg] .
\end{split}
\label{loopfactor}
\end{equation}

Also, we have added the contributions of the box diagrams with only one type of scalar ($H^0$, $A^0$) in the loop. The expression for this can be found from Eq.\ref{loopfactor} by taking the limiting case $X \to X^{'}$.

The effective operator is given by

\begin{eqnarray}
	\mathcal{O}_{eff} &=& [\bar{b} \gamma_{\mu}(1- \gamma_5) s ][\bar{l}\gamma^{\mu}(1- \gamma_5) l ] \nonumber\\
	&=& [\bar{b} \gamma_{\mu}(1- \gamma_5) s ][\bar{l}\gamma^{\mu} l ] - [\bar{b} \gamma_{\mu}(1- \gamma_5) s ][\bar{l}\gamma^{\mu}\gamma_5 l ] \nonumber \\
	&=& \mathcal{O}_9 - \mathcal{O}_{10}.
	\label{Oeff}
\end{eqnarray}

For simplicity, from now on we will rewrite the couplings $\lambda^D_{33} \equiv \lambda_{b}$, $\lambda^D_{23} \equiv \lambda_{s}$ and $\lambda^D_{13} \equiv \lambda_{d}$ . On the other hand, we write $\lambda^E_{11}\equiv \lambda_{e}$, $\lambda^E_{22} \equiv \lambda_{\mu}$ and $\lambda^E_{33} \equiv \lambda_{\tau}$ to simplify our notations. Thus Eq.~\ref{iM} can be written as 

\begin{equation}
	i \mathcal{M}_{Box} \sim i[C_9^{\text{NP}}\mathcal{O}_9 + C_{10}^{\text{NP}}\mathcal{O}_{10}],
\end{equation}

where,

\begin{equation}
	\boxed{C_9^{\text{NP}} = -C_{10}^{\text{NP}} = - \bigg(\frac{ \lambda_{b} \lambda_{s} \lambda_{\mu}^2  \mathcal{A} }{32 \pi^2 }\bigg)},
	\label{CoeffNP}
\end{equation}

which has to be normalized with a factor $\mathcal{N} = -\bigg(\frac{\sqrt{2}}{4 G_F V_{tb}^{*} V_{ts}} \times \frac{4 \pi}{\alpha}\bigg )$ so that the operators are at par with those given in Eq.~\ref{effoperators}.

The amplitude of the photon exchanged penguin diagrams can be written as 
  
\begin{equation}
 \mathcal{M}_{\gamma} = [\bar{b}(A_{L} q^2 \gamma_{\mu} P_{L} + i A_{R} m_{\tau} \sigma_{\mu\nu} q^{\nu} P_R )s] \frac{e^2}{q^2} [\bar{\ell}\gamma^{\mu} \ell ],
\end{equation}

where $q$ is the photon momentum. The dominant contributions will come from $D_3$, therefore, the form-factors $A_{L}$ and $A_{R}$ are induced by the product $\lambda_{s} \lambda_{b}$ coupling. The contribution to $C_{9}$ will come only from $A_{L}$, whose approximate form is given by: 
\begin{equation}
 A_L = \frac{\lambda_{s}\lambda_{b}^{\ast}}{32 \pi^2 M_{X^{(\prime)}}^2} \frac{\xi(r_{D_3})}{3},  
\end{equation}
with  
\begin{equation}
\xi(r_{D_3}) =  \frac{1}{6(1-r_{D_3})^4}[-11 + 18 r_{D_3} - 9 r_{D_3}^2 + 2 r_{D_3}^3 - 6 \ln r_{D_3}].
\end{equation}
and $r_{D_3} = M_{D_3}^2/M_{X^{(\prime)}}^2 $. 

The $Z$-mediated penguin amplitude for the process $b\to s \ell\ell$ can be written as  
\begin{equation}
 \mathcal{M}_Z = [\bar{b} F_{L} \gamma_{\mu} P_{L} s] \frac{1}{M_Z^2} [\bar{\ell}\gamma^{\mu} (a_L^\ell P_L + a_R^\ell P_R) \ell ],
\end{equation}
where 
\begin{equation}
 a_{L}^f = \frac{g}{\cos\theta_W}(t_3^f - Q_f \sin^2\theta_W) ,  \ \ a_{R}^f = \frac{g}{\cos\theta_W}(- Q_f \sin^2\theta_W).
\end{equation}
From the diagrams of Fig.~\ref{btosdiag} we obtain  

\begin{equation}
 F_L = \frac{g}{\cos\theta_W} \frac{\lambda_{s}\lambda_{b}^{\ast}}{32\pi^2} \bigg[ a_R^{D_3} \bigg(\frac{1}{2} - 2C \bigg) + a_L^{D_3} r_{D_3} \xi_0(r_{D_3}) + a_L^s B \bigg]. 
\end{equation}
The finite parts of $C$, $\xi_0$ and $B$ are given by 
\begin{align}
 C &= \frac{1}{2}\int_0^1{dx (1-x)\ \text{ln}[x M_{X^{(')}}^2 + (1-x) M_{D_3}^2] }, \nonumber \\
 \xi_0(r_{D}) &= \int_{0}^{1}dx \frac{1-x}{x+ (1-x)r_{D_3}} \nonumber \\
 B &= \frac{1}{2}\int_0^1{dx x\ \text{ln}[x M_{X^{(')}}^2 + (1-x) M_{D_3}^2] }.
 \end{align}
 
The details of the above calculations can be seen from \cite{Bhattacharyya:2009}. The $Z$-mediated penguin diagrams will contribute to both $C_9$ and $C_{10}$. Therefore, the total contributions to $C_9$ and $C_{10}$ can be extracted from  

\begin{equation}
 \mathcal{M} = \mathcal{M}_{Box} + \mathcal{M}_Z + \mathcal{M}_{\gamma}. 
\end{equation}
The numerical analysis are done the next section (\ref{sec:resul}).  

\subsection{$B_s - \overline{B}_s$ Mixing}
\label{sec:mixing}

\begin{figure}[t]
\centering
\includegraphics[scale=0.8]{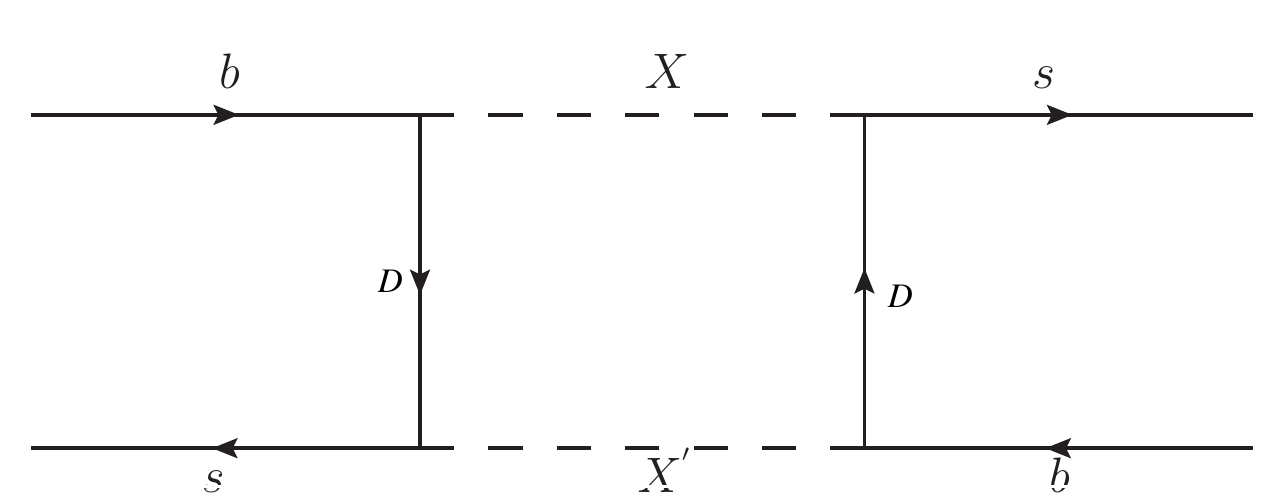} 
\captionsetup{style=singlelineraggedleft}
\caption{The Feynman Diagram contributing to the $B_s$-meson mixing in our model with $X, X^{\prime}$ denoting either $H^0$ or $A^0$. All other possible symmetric diagrams have also been considered during computation. The dominant contributions will come from $D_3$}
\label{fig:BBmix}
\end{figure}

The $\Delta F = 2$ process $B_s - \overline{B}_s$ mixing may play a crucial role in constraining the parameters of our model relevant for $b\to s $ transitions. In this case the important observable is the mass difference $\Delta M_{B_s}$, which is defined as 
\begin{equation}
\Delta M_{B_s} = 2 |M_{12}^{B_s}|.
\end{equation}
In the SM, the dominating contributions to $M_{12}^{B_s}$ will come from the dispersive part of the box diagram amplitude with $W$ boson and the top quark in the loop. 
The mathematical expression for it is given by 
\begin{equation}
M_{12}^{B_s}\bigg\rvert_{SM} = \frac{G_F^2}{12\pi^2} f_{B_s}^2 \hat{B}_{B_s} M_{B_s} M_W^2 (V_{tb}^* V_{ts})^2 \eta_B S_0\bigg(\frac{\overline{m}_t^2}{M_W^2}\bigg),
\end{equation}
$S_0$ is the Inami-Lim function:
\begin{equation}
S_0(x) = \frac{4x - 11 x^2 + x^3}{4(1-x)^2}- \frac{3x^3 \text{ log } x}{2(1-x)^3}.
\end{equation}
The detail of the SM calculations can be seen from \cite{Lenz:2006hd}. 

In the presence of NP there will be additional contributions to $M_{12}^{B_s}$. In our model the dominant contributions will come from the box diagram as shown in Fig.~\ref{fig:BBmix}, which is given by
\begin{equation}
M_{12}^{B_s}\bigg\rvert_{NP} = \bigg(\frac{4\lambda_b^2 \lambda_s^2}{32\pi^2}\bigg) f_{B_s}^2 \hat{B}_{B_s} M_{B_s} S_{NP}
\end{equation}
where,
\begin{equation}
\begin{split}
S_{NP} = \bigg[\frac{M_{D_3}^2}{(M_{X^{'}}^2-M_{D_3}^2) (M_{D_3}^2 - M_X^2)} + \frac{M_{D_3}^4 (M_X^2 + M_{X^{'}}^2) - 2 M_{D_3}^2 M_X^2 M_{X^{'}}^2}{(M_{X^{'}}^2-M_{D_3}^2)^2 (M_{D_3}^2 - M_X^2)^2} \text{Log} \bigg(\frac{M_{D_3}^2}{M_X^2}\bigg) & \\ - \frac{M_{X^{'}}^4}{(M_{X^{'}}^2-M_{D_3}^2)^2 (M_{X^{'}}^2 - M_{X}^2)} \text{Log} \bigg(\frac{M_{X^{'}}^2}{M_{X}^2}\bigg) \bigg]
\end{split}
\end{equation}
There will be several such diagrams with $H^0$ and/or $A^0$ in the loop. Our loop factor can be compared with that given in \cite{Abbott:1979dt}.

In the SM, the $B_s$ mixing phase is negligibly small, also in our analysis we are assuming real Yukawa couplings. Hence, we can express the mixing amplitude as 
\begin{equation}
\Delta M_{B_s} = (\Delta M_{B_s})_{\text{SM}} + (\Delta M_{B_s})_{\text{NP}} =  (\Delta M_{B_s})_{\text{SM}}\ \ \big( 1 + \Delta_{\text{Mix}} \big), 
\label{delta}
\end{equation}
with $\Delta_{\text{Mix}} = \frac{\big(\Delta M_{B_s}\big)_{\text{NP}}}{\big(\Delta M_{B_s}\big)_{\text{SM}}}$. In this ratio, the bag factor and the decay constant will cancel which are the major sources of uncertainties in the SM predictions of the oscillation frequency. $\Delta_{\text{Mix}}$ is sensitive to the NP parameters and using Eq.
\ref{delta} we can find out the maximum allowed ranges of this observable. Using the latest data on $\Delta {M_{B_s}}$ \cite{pdg2018} and the following inputs for decay constant and the bag factor \cite{Charles:2013aka,FLAG2019} 
\begin{equation}
f_{B_s} = 0.2284 \pm 0.0037 \rm~ GeV, \text{and} \ \ B_{B_s} = 1.327 \pm 0.034,  
\label{decaycons}
\end{equation}
we find that $\Delta_{\text{Mix}}$ could be as big as 15\% if we consider the 1$\sigma$ allowed ranges of all the relevant inputs. This could be even 20\% if one uses the projected lattice results as given in ref. \cite{Charles:2013aka}. The $B_d - \overline{B}_d$ mixing data allows sizeable NP contributions ($\approx 30\%$)\cite{Charles:2015gya}. In our model, the contributions to $B_d - \overline{B}_d$ mixing will come from the diagram in Fig.~\ref{fig:BBmix}, with the strange quarks ($s$) in the external legs replaced by down ($d$) quarks. The dominant contribution will thus be proportional to $(\lambda_b\lambda_d)^2$. Therefore, following our assumption of the hierarchical structures of the new Yukawas, the contributions will be highly suppressed. The same argument is also true for $K$-$\overline{K}$ mixing in which the NP contribution due to our model is proportional to $(\lambda_d\lambda_s)^2$. Actually, for all practical purposes, we can set $\lambda_d \approx 0$. This assumption does not have any impact on our final results since we are mostly interested in the observables which do not have dominant contributions from $\lambda_d$.

\section{Results: DM and flavour}
\label{sec:resul}

In this section we discuss the results obtained from the analysis of the DM and flavour sector of our model. We scan the NP parameter space using the constraints from flavour data, relic density and direct detection bounds. In the context of our model, the relevant free parameters are:  $\lambda_{b}$, $\lambda_{s}$, $\lambda_\mu$, $\lambda_{\tau}$,  $M_{H^0}$, $M_{A^0}$, $M_{H^{\pm}}$, $M_{E_3}$, $ M_{E_2}$, $ M_{E_1}$, and  $ M_{D_1} = M_{D_2} = M_{D_3} = M_{D}$
\footnote{Though in our case the significant contributions will come from $M_{D_3} = M_D$.} . 

In order to simplify the analysis, amongst them we have fixed few of the couplings, such as $\lambda_s = 0.01$ and $\lambda_e = 0.001$.
\begin{figure}[t]
\centering
\subfloat[]
{\includegraphics[width=7.5cm]{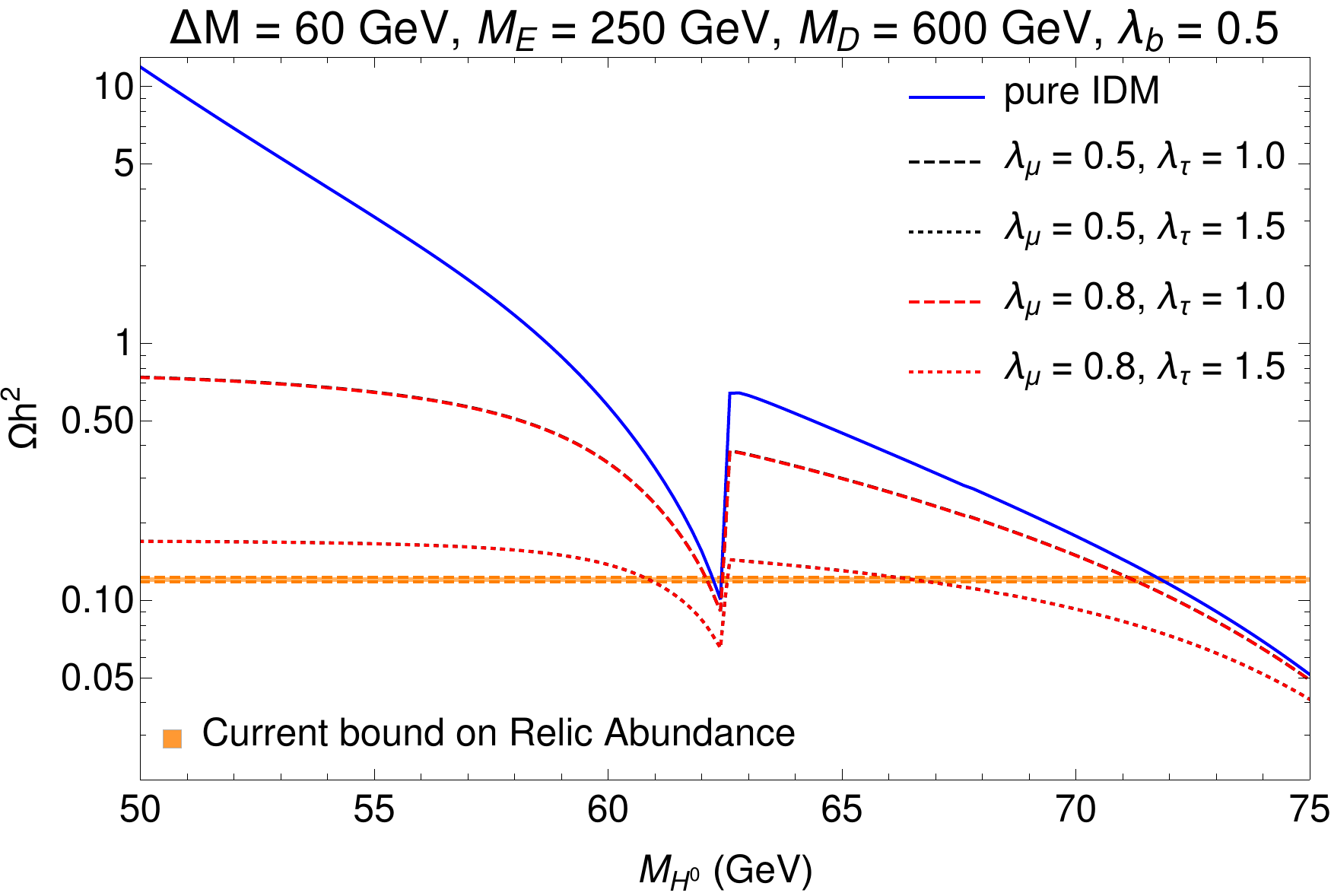}\label{mh0low}}~~~
\subfloat[ ]
{\includegraphics[width=7.5cm]{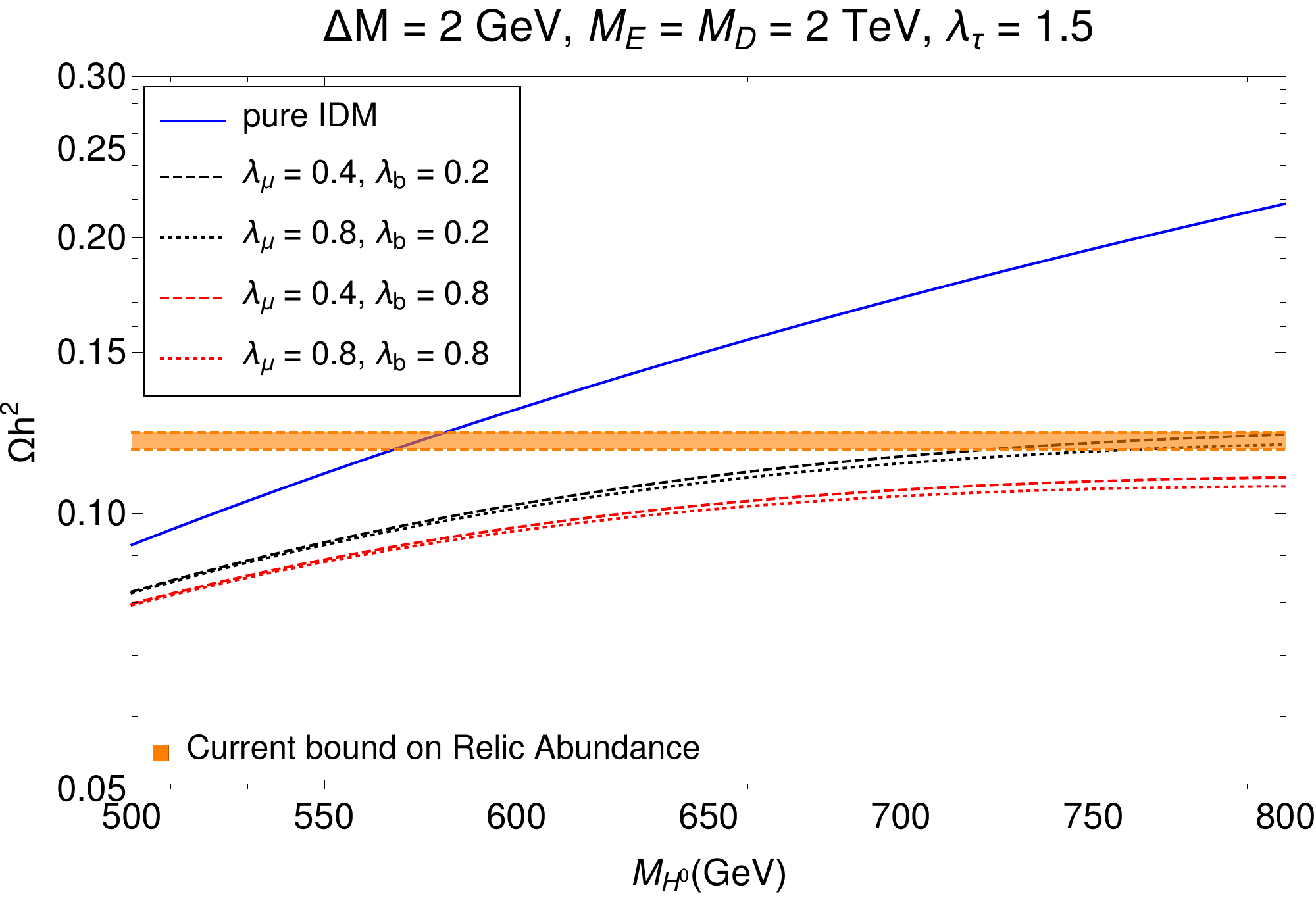}\label{mh0high}}\\
\captionsetup{style=singlelineraggedleft}
\caption{Variations of relic abundance with the DM mass $M_{H^0}$ for different values of the couplings in the (a) low mass and (b) high mass regions of the DM. The plots show that as we switch on the leptonic and/or quark portal couplings, new annihilation channels open up, lowering the relic abundance for a fixed DM mass. In Fig.\ref{mh0low}, the black-dashed and red-dashed lines overlap and hence not visible. This is due to the fact that when $\lambda_\tau$ is large, small variations in $\lambda_\mu$ does not affect the relic abundance. Similarly, the black and red dotted lines are also overlapped. See text for more details}
\label{mh0}
\end{figure}
Also, we choose $\lambda_{b} < 1$, $ \lambda_{\mu} < 1.5$ and $\lambda_\tau \approx 1.5$. The rest of the free parameters are constrained from the $R(K^{(*)})$, $\mathcal{B}(B_s \to \mu\mu)$, $\Delta a_\mu$ and relic data. With these choices of the couplings, we can easily overcome the present constraints on flavour changing $b\to s$ processes, like $B_s-\bar{B_s}$ mixing, $\mathcal{B}(B\to X_s\gamma)$ etc. We will explicitly show this for $B_s-\bar{B_s}$ mixing. We have checked that in our model within the chosen benchmark points given in Table \ref{BP}, the new physics contribution to the branching fraction $\mathcal{B}(B\to X_s\gamma)$ will be of order ${\cal O}(10^{-6})$ which is suppressed with respect to the corresponding SM branching fraction by two order of magnitude. 

\begin{figure}[htb!]
\centering
\subfloat[]
{\includegraphics[width=7.5cm]{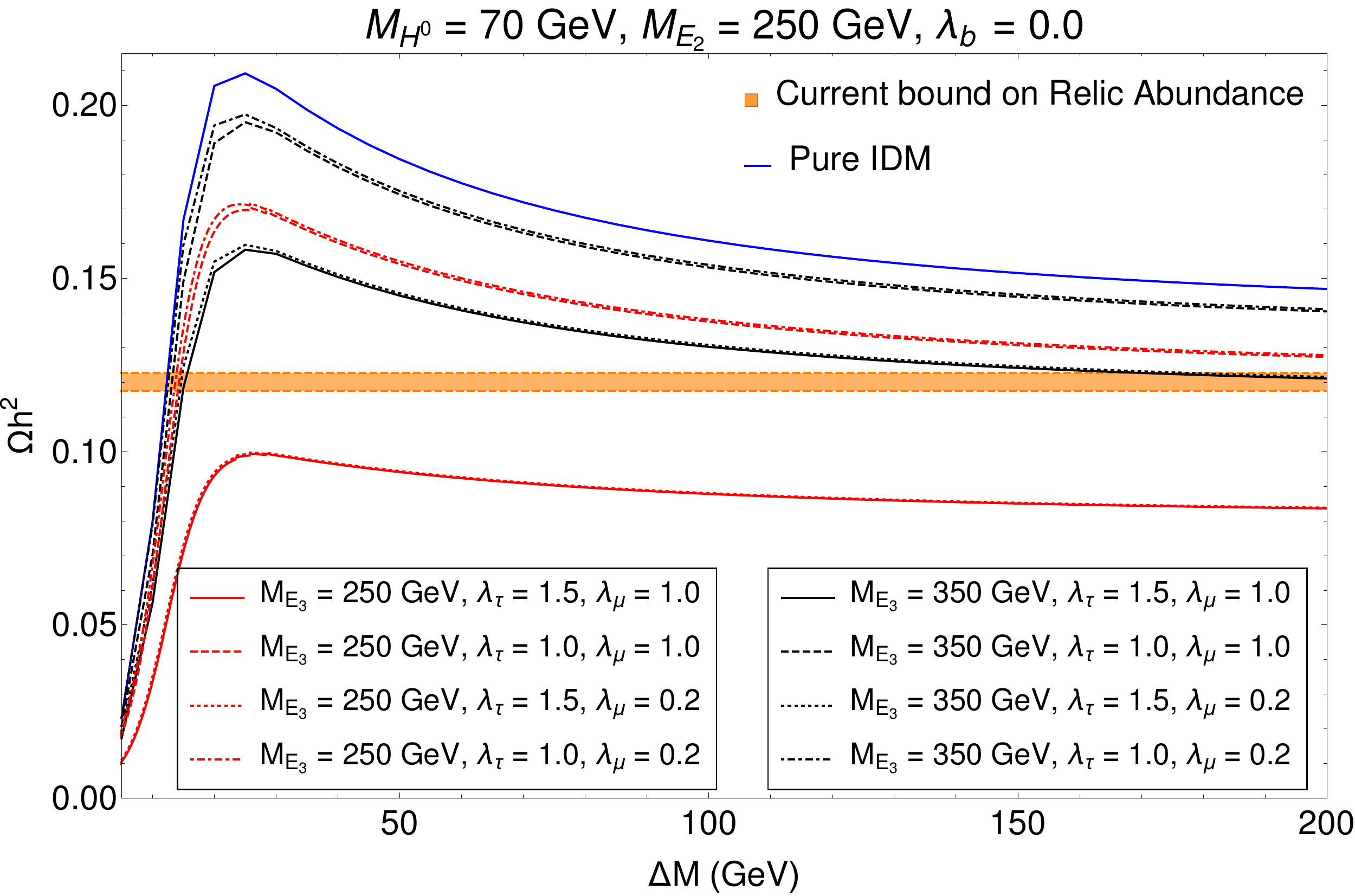}\label{lambdamu}}~~~
\subfloat[ ]
{\includegraphics[width=7.5cm]{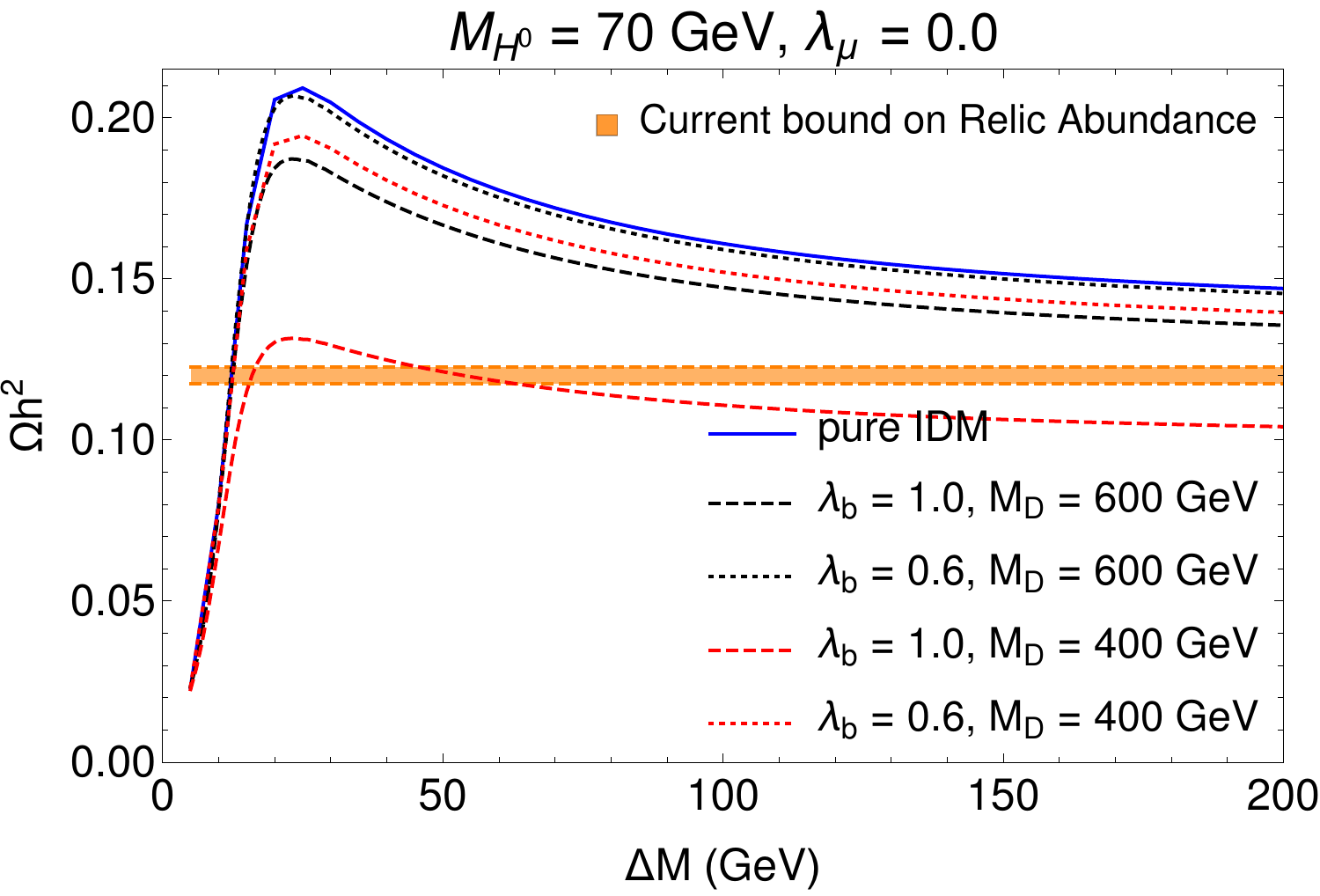}\label{lambdabt}}\\
\subfloat[ ]
{\includegraphics[width=7.5cm]{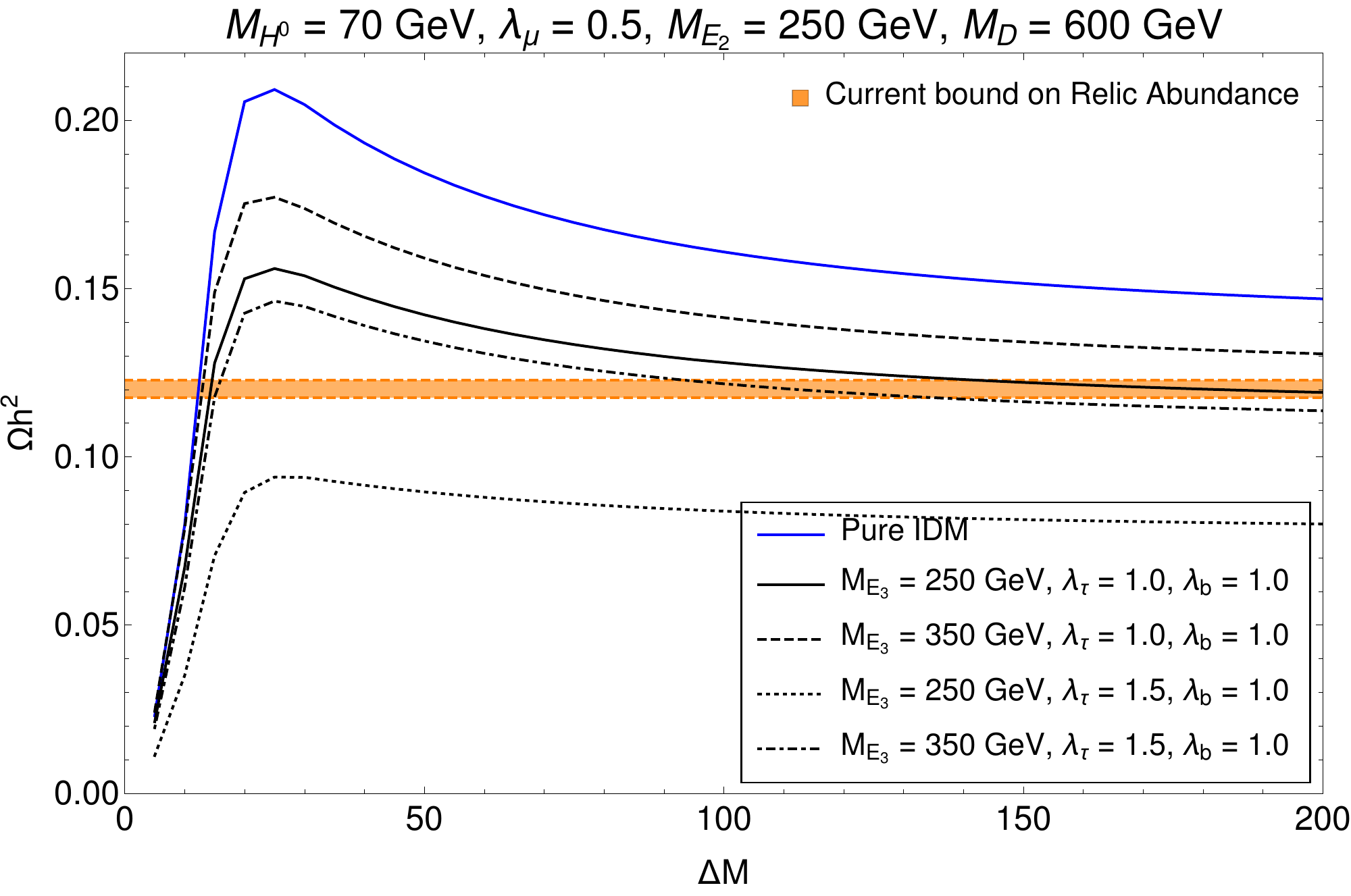}\label{lambdabt05}}~~
\subfloat[ ]
{\includegraphics[width=7.5cm]{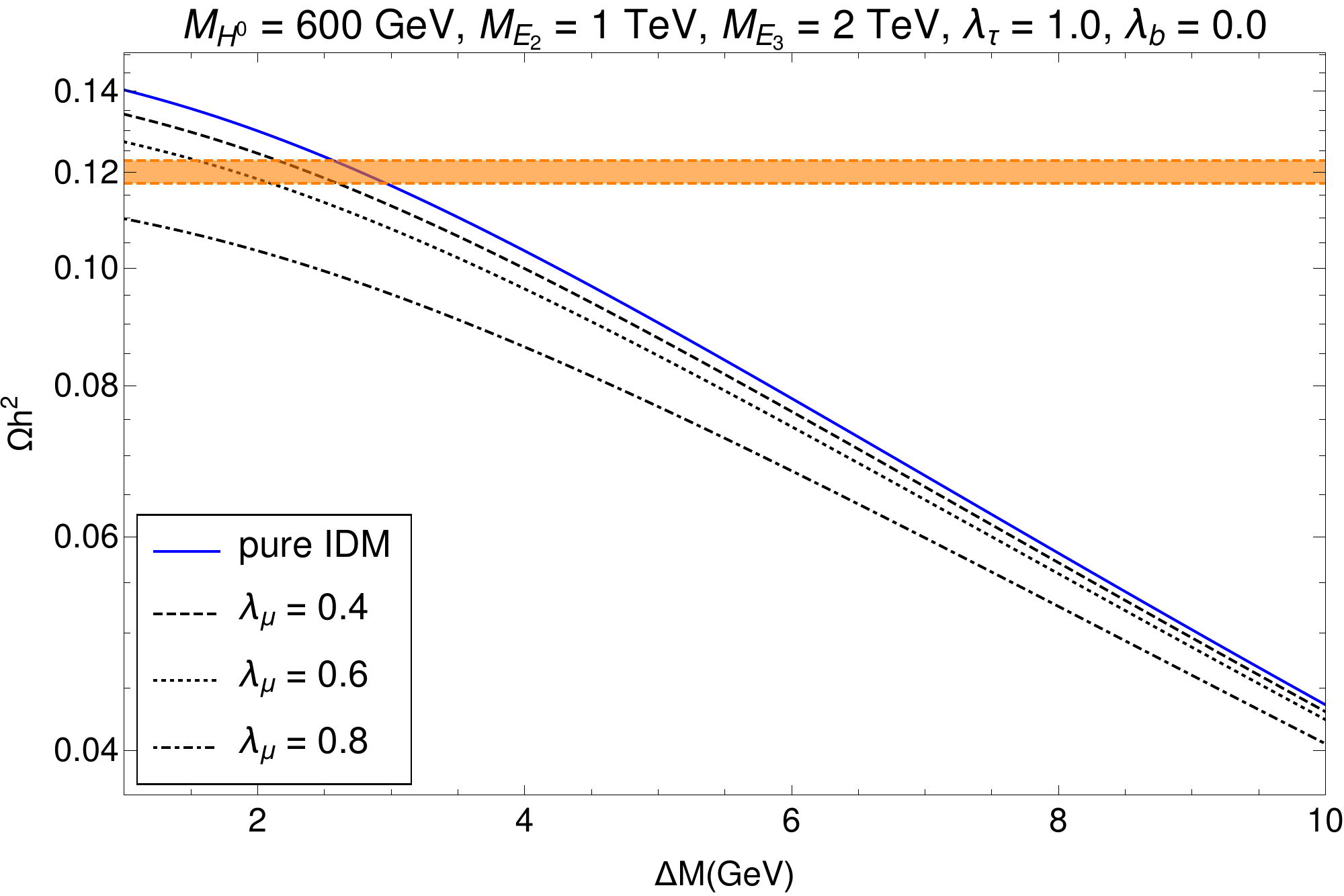}\label{mh0high2}}\\
\captionsetup{style=singlelineraggedleft}
\caption{The plots in the top panel (\ref{lambdamu},\ref{lambdabt}) and the bottom left plot (\ref{lambdabt05}) shows the variations of relic abundance with the mass splitting $\Delta M = M_{H^{\pm}} - M_{H^0} = M_{A^0} - M_{H^0}$ for the DM mass $M_{H^0} = 70$ {\rm GeV}. Here, different benchmark points are chosen for the other new parameters. The similar correlations in the high DM mass region ($M_{H^0} = 600$ GeV) is shown in the bottom right plot (\ref{mh0high2})}
\label{vardeltam}
\end{figure}

First, we will discuss the effects of different parameters of our model on DM relic abundance. As mentioned earlier, in pure IDM, there exists two distinct regions of DM mass which satisfy the relic abundance criterion. In  Fig.~\ref{mh0} we have shown the variations of DM relic abundance with $M_{H^0}$ in two different DM mass regions (low and high) for different new couplings and masses of the exotic vector like fermions. In Fig.~\ref{mh0low}, we have kept $\Delta M = 60$ GeV and varied $M_{H^0}$ between 50 GeV and 75 GeV (low DM mass region). In this region, with the variation of our new parameters, the allowed values of $M_{H^0}$ do not change significantly from that obtained in pure IDM case. In Fig.~\ref{mh0high}, we have kept $\Delta M=2~\rm GeV$ and $M_{H^0}> 500~\rm GeV$ (high DM mass region). Here, the same mass splitting between different components of the inert scalar doublet namely, $\Delta M = M_{H^{\pm}} - M_{H^0} = M_{A^0} - M_{H^0}$ are considered. In this region, the deviation from pure IDM scenario is significant.As expected, for the fixed values of the masses of the vector like fermions, the new couplings and the associated allowed values of $M_{H^0}$ are positively correlated. For the pure IDM scenario, in the low mass region, the allowed values of $M_{H^0}$ is not strongly correlated with the choice of $\Delta M$, while in the high mass region the relic abundance is only satisfied when $\Delta M$ is very small, or in other words when the inert scalars are nearly degenerate. Also, in Fig. \ref{mh0low}, the black-dashed and red-dashed lines overlap with each other. This shows that the DM mostly annihilates to $\tau^+\tau^-$ pairs through vector-like lepton $E_3$ and the annihilation to muon pair is sub-dominant since the coupling $\lambda_\tau$ is much larger than $\lambda_\mu$. Hence small changes in $\lambda_\mu$ does not affect the relic abundance when $\lambda_\tau $ is large for universal vector-like lepton masses.

For simplicity, in the low DM mass region, we have fixed $M_{H^0}$ at 70 GeV for the rest of our analysis. In Fig. \ref{vardeltam}, we have shown the variations of the relic abundance with the mass splitting $\Delta M $ for different benchmark values of the new couplings and masses. From Fig.~\ref{lambdamu}, \ref{lambdabt} and \ref{lambdabt05} we note that, as the new parameters are switched on, the relic abundance decreases compared to pure IDM scenario due to the increase in annihilation cross section. Also, the required mass splitting $\Delta M$ will be less in our model compared to that in pure IDM. Since we have assumed $\lambda_{\tau} > \lambda_{\mu}$ the dominant contributions to the relic abundance will come from the annihilation to $\tau^+\tau^-$, which can be seen from Fig.~\ref{lambdamu} where the variations are almost independent of the choices of $\lambda_{\mu}$. The sensitivity of the relic abundance to the mass splitting in the high mass region is shown in Fig.~\ref{mh0high2}. With the increase in $\lambda_{\mu}$, the mass degeneracies are becoming tighter compared to pure IDM scenario.  Similar trend is also expected with the variation of $\lambda_b$ as well.

\begin{figure}[t]
\centering
\subfloat[]
{\includegraphics[height=5.2cm,width=7.5cm]{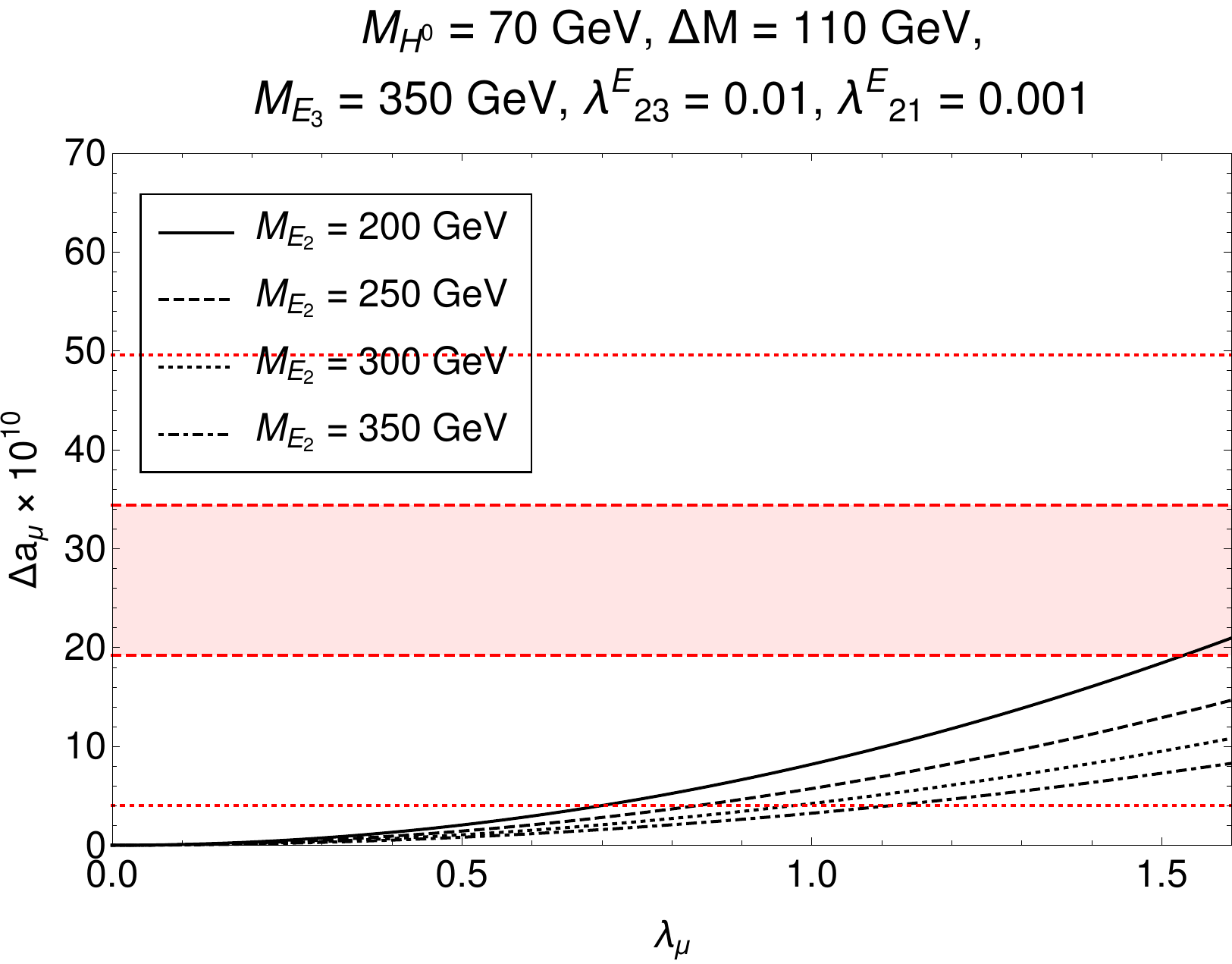}\label{muong2}}~~~~~~\
\subfloat[]
{\includegraphics[height=5.2cm,width=7.5cm]{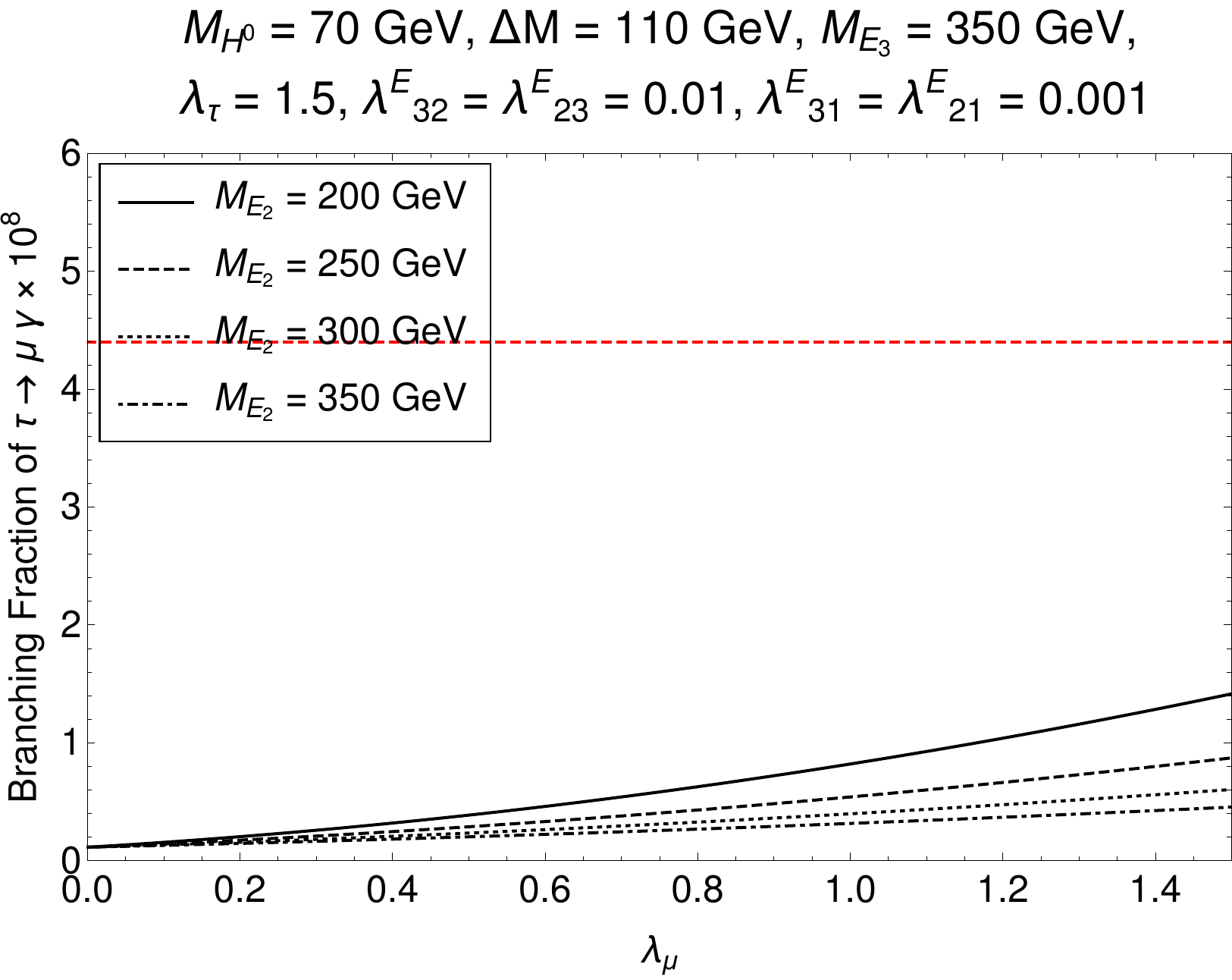}\label{lfv}}~~
\captionsetup{style=singlelineraggedleft}
\caption{The left plot shows the variations of $\Delta a_{\mu}$ (muon $(g-2)$) with the coupling $\lambda_{\mu}$ for different values of $M_{E_2}$. In these plots $M_{H^0}$ has been taken as 70 {\rm GeV}. The red dashed and dotted lines represent the 1-$\sigma$ and 3-$\sigma$ bands of the $\Delta a_{\mu} $, respectively. The right plot shows that with the same benchmark values of the NP parameters the decay width for $\tau\to\mu\gamma$ is well below the present experimental limit \cite{pdg2018}}
\label{lfvg2}
\end{figure}

In section \ref{sec:muong2}, we have discussed various diagrams and their contributions to muon $(g-2)$ and LFV decays $\ell_i\to \ell_j \gamma$. There will be contributions from penguin diagrams with vector like leptons $E_3$, $E_2$ or $E_1$ in the loop. However, since we are assuming hierarchical structure for the couplings: $\lambda^E_{21} < \lambda^E_{23} << \lambda^E_{22} (= \lambda_{\mu}) \lsim \lambda^E_{33}(\equiv \lambda_{\tau})$, the diagrams with $E_3$ and $E_1$ in the loop will not give significant contributions to $\Delta a_{\mu}$. However for completeness, we consider all those contributions in our analysis. The dominant contribution will come from the penguin diagram with $E_2$ in the loop. The variations of $\Delta a_{\mu}$ with the new coupling $\lambda_{\mu}$ for different values of $M_{E_2}$ are shown in Fig.~\ref{muong2}. We note that if we restrict ourselves to the values of $\lambda_{\mu} < 1.5$, then it will be difficult to explain the excess in muon $(g-2)$ within its 1-$\sigma$ range, unless we consider vector like lepton mass $\lsim 200$ GeV. However, the excess can be succesfully explained within its 3$-\sigma$ range for values of $\lambda_\mu \lsim 1.5$ and we restrict ourselves to this limit only. Especially, higher masses prefer higher values of the coupling $\lambda_{\mu}$. In future with more precision measurements, the tension between the predicted and measured values may reduce i.e the data may become more SM-like. In such situations, the data on $a_{\mu}$ will not put strong constraints on our model parameters. In the opposite situations, one may need to consider values of $\lambda_{\mu} \gsim 1.5$. On the other hand, the contribution from all the vector like fermions will be relevant for the LFV decays. However, since in our framework the off-diagonal elements are small compared to the diagonal elements, the contribution to the branching fraction will not be significantly large. As an example, we have chosen $\lambda^E_{32} = \lambda^E_{23} \approx 0.01$. With this choice, the branching fraction $\tau\to \mu\gamma$ will be much below the current experimental limit, even if we choose  $\lambda_{\mu}$ or $\lambda_{\tau}$ roughly $\sim{\cal O}$(1) (Fig.~\ref{lfv}). Here, we have not discussed the LFV $\tau^- \to \mu^-\mu^+\mu^- $ decay. In our model, the leading diagram for this decay is same as $\tau\to\mu \gamma$ , with a virtual photon converting into a muon pair \footnote{There will be one additional box diagram, the contribution of which will be suppressed compared to that of the penguin diagram.}. It is expected that for the same set of NP parameters the branching fraction $\mathcal{B}(\tau\to \mu^-\mu^+\mu^-)$ will be small compared to $\mathcal{B}(\tau\to \mu\gamma)$; as an example see \cite{Bhattacharyya:2009}. Therefore, our NP parameters will be safe with respect to present limit $\mathcal{B}(\tau \to\mu\mu\mu) (\approx {\cal O}(10^{-8})$) \cite{pdg2018}. The LFV decays and $\Delta a_{\mu}$ are insensitive to the coupling $\lambda_{b}$. However, observables like $R(K^{(*)})$ and DM relic abundance are sensitive to all the relevant couplings and masses of the model. 

\begin{figure}[t]
\centering
\includegraphics[scale=0.45]{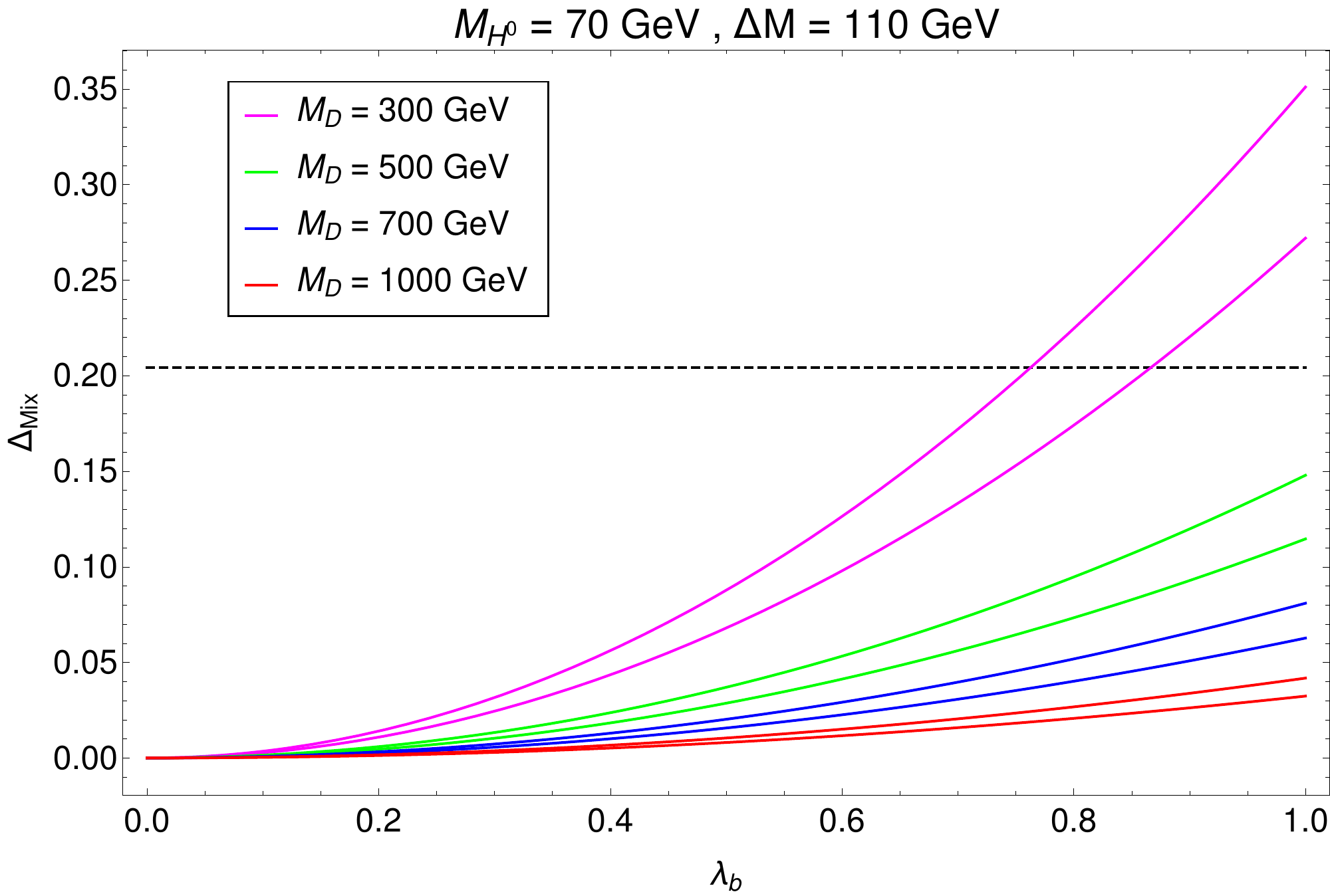} 
\captionsetup{style=singlelineraggedleft}
\caption{ Variation of $\Delta_{\text{Mix}}$ with $\lambda_b$ for four different values of the vector-like quark mass $M_D (=M_{D_3})$, since the dominant contribution will come from $D=D_3$. The black dashed line indicates the maximum allowed value of $\Delta_{\text{Mix}}$ if we take all the inputs in Eq. \ref{delta} within their respective $1\sigma$ confidence interval}
\label{fig:Mixplt}
\end{figure}

In the case of $B_s-\bar{B_s}$ mixing, in Fig.~\ref{fig:Mixplt} we have shown the variations of $\Delta_{\text{Mix}}$ with $\lambda_b$ for different values of the masses of $M_{D_3}$. The black dashed line indicates the $15\%$ allowed range in $\Delta_{\text{Mix}}$, the solid lines represent our model predictions for different values of the vector-like quark mass. The other relevant parameters are fixed as before. We note that for $M_{D_3} \ge 500$ GeV, the allowed value of  $\lambda_b $ could be as big as 1. However, for lower values of $M_{D_3}$ higher values of $\lambda_b$ will be disfavoured, as an example we can see that for $M_{D_3} = 300 $ GeV, $\lambda_b \gsim 0.7$ will not be allowed. Here, we would like to point out that the major uncertainties in the theory prediction is associated with the decay constant. Therefore, if we consider the errors within their 1$\sigma$ CL ranges, then that will give us a conservative estimate of the allowed NP. We would like to stress that data will still allow a NP contribution up to 30-40\% at the $3\sigma$ CL \cite{Charles:2015gya}. 

As mentioned earlier, the branching fraction for the rare decay $B_s\to \mu^+\mu^-$ is consistent with its SM prediction within 1$\sigma$ confidence level. Therefore, the data on $\mathcal{B}(B_s\to\mu^+\mu^-)$ is expected to put tighter constraints on the parameters of any NP model in the decay $b\to s\mu\mu$. More importantly if we see the current data, the measured value is below that of SM prediction and our model has the potential to accommodate it. Although we are not considering it seriously, we have to wait for more precise data and lattice inputs to conclude it further, but at the moment one can not rule out this possibility.
The main source of error in $B_s \to \mu \mu$ is the decay constant $f_{B_s}$ whose different lattice predictions have different errors (for detail see Ref.~\cite{FLAG2019}). Therefore, in order to be conservative, the errors in the measured value have been taken in their 2$\sigma$ confidence level allowed ranges to constrain the NP parameters.

From phenomenology point of view according to the low and high DM mass regions, we divide our analysis into two parts: in one part, we choose $M_{H^0} = 70$ GeV (low DM mass), and in the other we have considered $M_{H^0} = 600$ GeV (high DM mass). These will be discussed in the following sections.

\begin{figure}[h!!!]
\centering
\subfloat[]
{\includegraphics[height=6cm,width=8cm]{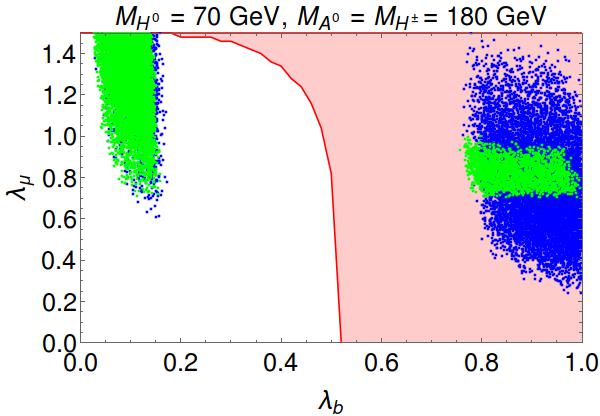}
\label{corrA}}
\subfloat[]
{\includegraphics[height=6cm,width=8cm]{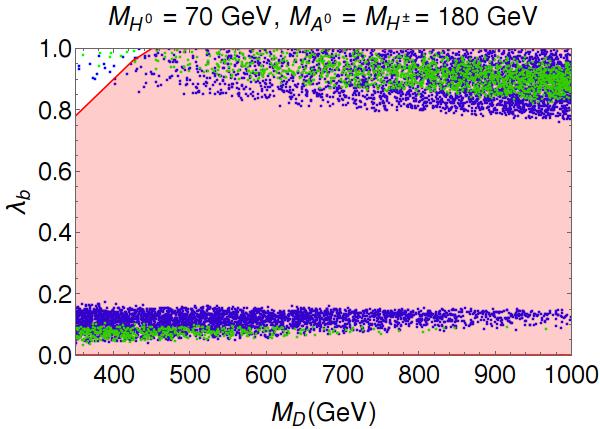}
\label{corrB}}\\
\subfloat[]
{\includegraphics[height=6cm,width=8cm]{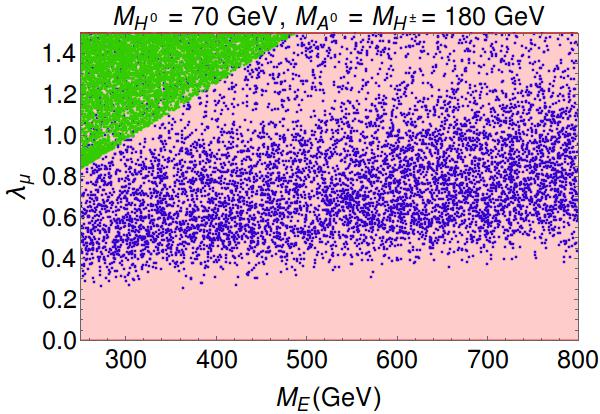}
\label{corrC}}
\subfloat[]
{\includegraphics[height=6cm,width=8cm]{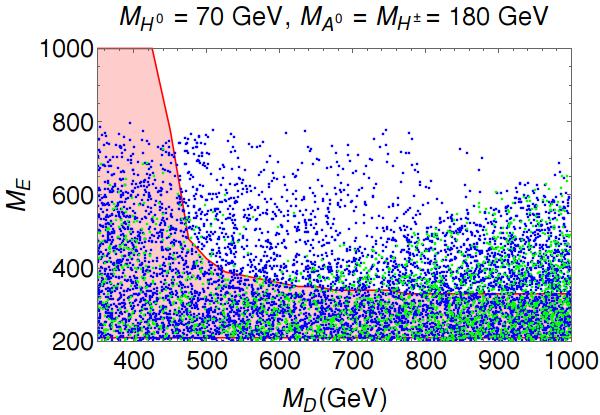}
\label{corrD}}
\captionsetup{style=singlelineraggedleft}
\caption{Correlation between different NP parameters for a Low Mass DM at $M_{H^0} = 70$ GeV. The blue points satisfy all relevant flavour constraints within their  $2\sigma$ confidence intervals except muon (g-2) anomaly. When we further apply the muon anomalous magnetic moment bounds then the allowed region shrinks as depicted by the green points(see text). The red region satisfy the relic and direct search constraints when we consider degenerate vector-like fermion masses}
\label{correlations3}
\end{figure}

\subsection{Low mass DM}
\label{sec:lowmassdm}

\begin{figure}[h!!]
\centering
\subfloat[]
{\includegraphics[height=6cm,width=8cm]{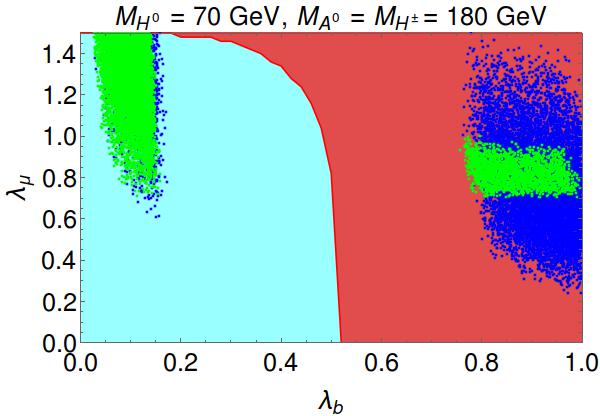}
\label{corrE}}
\subfloat[]
{\includegraphics[height=6cm,width=8cm]{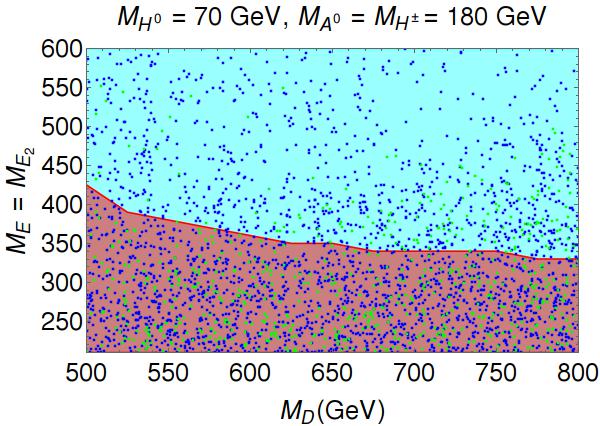}
\label{corrF}}
\captionsetup{style=singlelinecentered}%
\caption{Similar plots as given in \ref{corrA} and \ref{corrD}. In addition, we have considered the case $M_{E_2}$ $\ne$ $M_{E_3}$ which is represented by cyan region. This region satisfies the relic and direct detection constraints when we assume the non-degenerate vector-like lepton masses}
\label{correlations3nd}
\end{figure}

In this section we show our main results for a light dark matter of mass $70$ GeV and mass splitting $\Delta M = 110$ GeV. 
Keeping in mind all the relevant correlations between the different parameters shown above, we do a multi-parameter scan to find out the common parameter space that satisfies all relevant flavour constraints i.e. $R(K^{(*)})$, $\mathcal{B}(B_s \to \mu^+ \mu^-)$, muon magnetic moment anomaly in $\Delta a_\mu$ as well as the correct relic abundance and direct detection bounds of dark matter as shown in Fig.~\ref{correlations3}. To generate these plots we assumed, for simplicity, that all generations of the vector-like leptons have the same mass i.e $M_{E_1}=M_{E_2}=M_{E_3}=M_{E}$. More regions on the parameter spaces of $\lambda_{\mu}$ and $M_{E_2}$ will be allowed if we relax this mass degeneracy. We allow both $M_E$ and $M_D$ to vary between 200 and 1000 GeV instead of keeping them fixed. In these plots $\lambda_\tau$ has been varied between $1.0-1.5$, while we have kept $\lambda_{\mu} < 1.5$ and $\lambda_b < 1$. As mentioned earlier, the major constraints on the new parameters are mainly coming from the flavour data, in particular from $\mathcal{B}(B_s\to \mu\mu)$. 
The blue scattered points satisfy the data on $R(K)$, $R(K^*)$ and $\mathcal{B}(B_s\to \mu\mu)$ in their respective 2$\sigma$ confidence intervals which shrink to the green points when we also consider $\Delta a_{\mu}$ as a constraint. Here, we have considered the excess in $\Delta a_{\mu}$ within its 3-$\sigma$ range. The red regions represent the bound on the NP parameters from the relic density and direct detection cross section of DM. An interesting feature here is the presence of two distinct regions in the parameter spaces of $\lambda_\mu$,  $\lambda_b$ and $M_D$ (Fig.~\ref{corrA} and \ref{corrB}). These two regions correspond to high and low values of $\lambda_b$, respectively. In both the allowed regions, $\lambda_\mu$ can take moderate values $ > 0.5$. However, its magnitude can not be very high ($>>1$) when $\lambda_b \gsim 0.7$. On the other hand when $\lambda_b$ is small, the common parameter spaces are obtained in regions where $\lambda_{\mu} \approx 1.5$, which could be relaxed and the values $\lambda_{\mu} \lsim 1.5$ will be allowed if we lift the mass degeneracies of the vector like leptons, for example see Fig.~\ref{corrE}\footnote{If we lift the mass degeneracies in the vector like leptons than for higher masses of $M_{E_3}$ the DM will mostly annihilate to $\tau^+\tau^-$ and this channel will contribute maximally to the relic abundance. Hence, their won't be any strong constraints on the parameter spaces of $M_{E_2}$ and $\lambda_{\mu}$ from relic abundance which can also be understood from the observations made in Fig.~\ref{lambdamu} and \ref{lambdabt05}.}. Hence, for completeness we have analysed both these regions in the collider searches as will be discussed later. Also, we note that $\lambda_\mu > 0.4$ for $M_{E} \lsim 800$ GeV and $\lambda_b > 0.7$ in the whole range of $M_{D}$ are allowed by all data. Note that muon $(g-2)$ is not sensitive to $\lambda_b$ or $M_{D}$. Also, we see a nice correlation between $M_{E}$ and $M_D$, for the higher values of $M_{D}$ ($\ge 500$ GeV) the relic density prefer $M_{E} \le 400$ GeV. We have checked that this constraint can be relaxed if we assume non degenerate vector like fermion masses for all the generations, for example see Fig.~\ref{corrF}.

Following the above discussions, from the allowed parameter spaces we have chosen seven benchmark points (BPs), listed in Table.~\ref{BP}, for collider analysis elaborated in Sec.~\ref{sec:colliderpheno}. Also, we have checked that the values of the Wilson coefficients $C_9^{NP}$ and $C_{10}^{NP}$ in all these benchmark scenarios are consistent with those obtained from the global fit \cite{Aebischer:2019mlg}, in particular the scenario with $C_9^{NP} = - C_{10}^{NP}$. We have focussed on both the allowed regions of couplings as shown in Fig.~\ref{corrA}, so that we can phenomenologically distinguish them from each other. BP1 to BP5 have the characteristics of high $\lambda_b$ and intermediate $\lambda_\mu$ while a low $\lambda_b$ and high $\lambda_\mu$ characterize BP6 and BP7. Here we would like to mention that there is an existing lower limit on pair-produced charged heavy vector-like leptons from LEP \cite{Achard:2001qw} : $m_{L^\pm} \gsim 101.2$ GeV. Our benchmark points satisfy this limit.

\begin{table}[htb!]
\centering
\renewcommand{\arraystretch}{1.8}
	\begin{tabular}{|c|c|c|c|c|c|c|c|c|c|c|c|c|}
	\hline
	$BP$ &  $M_{D}$ (GeV) & $\lambda_b$ & $M_{E_2}$ (GeV) &$\lambda_\mu$ & $M_{E_3}$ (GeV) & $\lambda_\tau$ & $\Omega h^2$ & R(K) & R(K*) & $\Delta a_\mu \times 10^{10}$ & $\mathcal{B}(\tau \to \mu \gamma) \times 10^9$ & $\mathcal{B}(B_s \to \mu \mu)\times 10^9 $ \\
	\hline
	1. &  500 & 0.9 & 150 & 0.5 & 350 & 1.5 & 0.120 & 0.769 & 0.760 & 3.15 & 5.98 & 1.95\\
	\hline
	2. &  750 & 0.9 & 250 & 0.7 & 250 & 1.2 & 0.117 & 0.777 & 0.769 & 2.8 & 5.71 & 2.49\\
	\hline
	3. &  750 & 0.9 & 200 & 0.7 & 350 & 1.5 & 0.122 & 0.758 & 0.749 & 4.01 & 5.39 & 2.55\\
	\hline
	4. &  850 & 0.9 & 250 & 0.7 & 250 & 1.2 & 0.118 & 0.808 & 0.800 & 2.81 & 5.71 & 2.59 \\
	\hline
	5. &  900 & 0.9 & 350 & 0.8 & 350 & 1.5 & 0.119 & 0.803 & 0.796 & 2.07 & 2.67 & 2.69 \\
	\hline
	6. &  800 & 0.1 & 300 & 1.5 & 350 & 1.5 & 0.121 & 0.908 & 0.903 &9.5 & 6.03 & 1.89 \\
	\hline
	7. &  800 & 0.1 & 180 & 1.5 & 500 & 1.5 & 0.121 & 0.885 & 0.880 & 4.62 & 5.35 & 1.84\\
	\hline
	\end{tabular}
	\bigskip
	\captionsetup{style=singlelinecentered}
	\caption{Model prediction of different relevant observables corresponding to our chosen benchmark points}
	\label{BP}
\end{table}
\bigskip

In pure IDM, the electroweak precision observables (EWPO) like $S$ and $T$, play an important role in constraining the mass splitting $\Delta M$ between the inert scalars \cite{Barbieri:2006dq}. We have already taken care of this constraint while scanning the new parameter spaces. Our model contains singlet vector fermions which do not mix. Hence, there will not be any additional significant contributions in $S$, $T$ and $U$ parameters, although there will be diagrams that contribute to $Z\to \mu\bar{\mu}$ and $Z\to b\bar{b}$ decays at one loop level. However, we have checked that within our chosen model parameters those contributions are highly suppressed. Therefore, the EWPO will not put any stringent constraint on our model parameters.

\begin{figure}[t]
\centering
\includegraphics[scale=0.33]{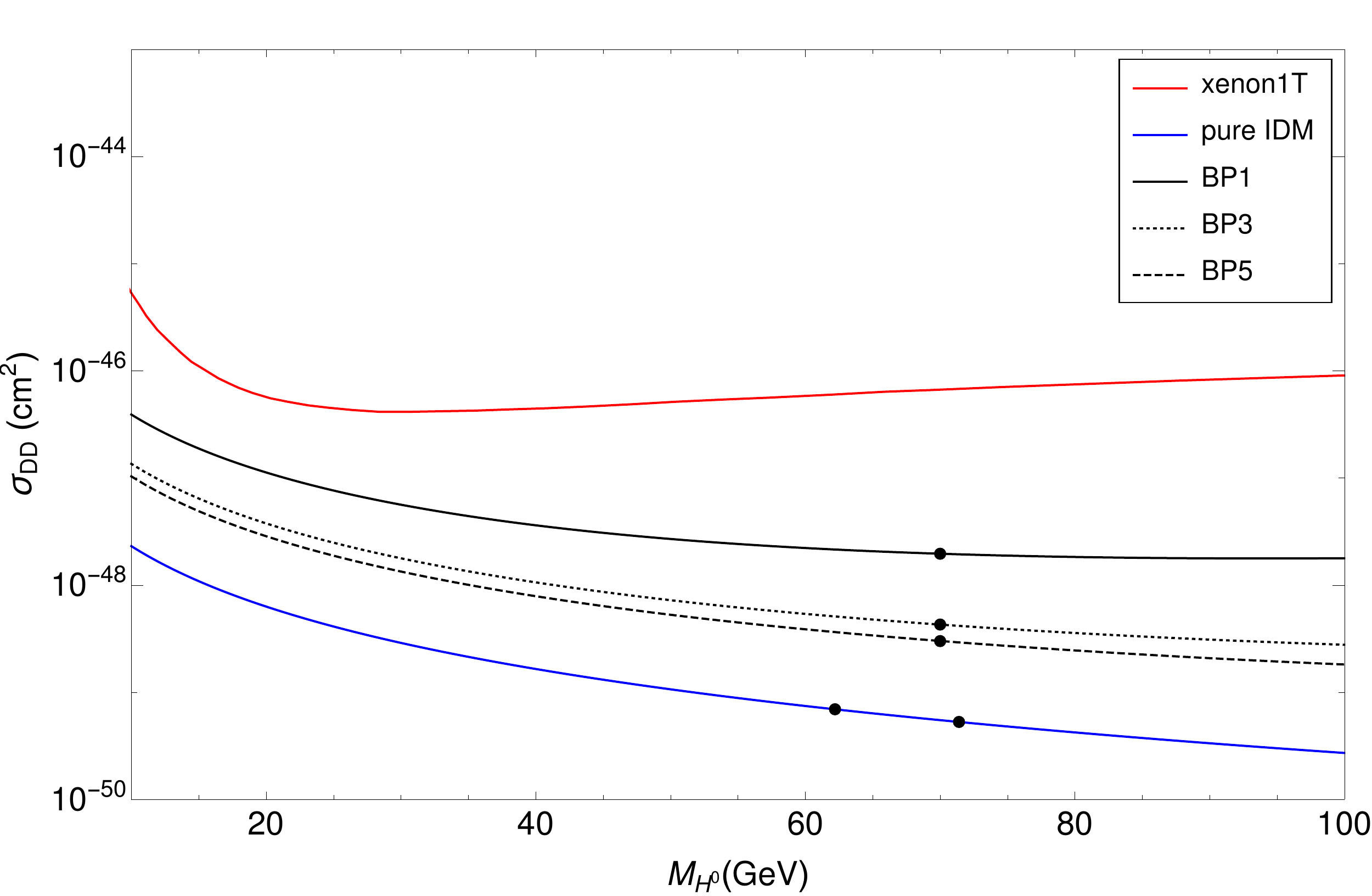} 
\captionsetup{style=singlelineraggedleft}
\caption{The plot shows the variation direct detection cross section with DM mass in direct search plane. The black lines correspond to different BPs, where the DM mass has been varied, while the dots correspond to specific choices of the DM mass (see text for details), the red line is the exclusion limit from recent XENON-1T data and the blue line is the direct search limit from pure IDM case}
\label{fig:ddxsection}
\end{figure}

Before moving on to the collider analysis, we show, for illustrative purpose, the variation of spin-independent direct search cross section (per nucleon) with the DM mass in Fig.~\ref{fig:ddxsection} for some of the chosen BPs in Table.~\ref{BP}. For a comparison, the similar correlation for pure IDM is presented in the same figure. The solid red line is the exclusion limit from recent XENON-1T data~\cite{Aprile:2017iyp}. The black dots on each black line refer to particular point corresponding to a fixed $M_{H^{0}}$, satisfying constraints from relic density, direct search (as they lie below the experimental exclusion limit) and flavour bounds (which we have discussed in Sec.~\ref{sec:muong2} and Sec.~\ref{sec:bsll}).  
As pointed out earlier, and can now be seen from these plots, the presence of exotic quarks increases the direct detection rates compared to the pure IDM keeping it more promising for observing at ongoing direct search experiments.

\subsection{High Mass DM}

\begin{figure}[h!!!]
\centering
\subfloat[]
{\includegraphics[scale=0.38]{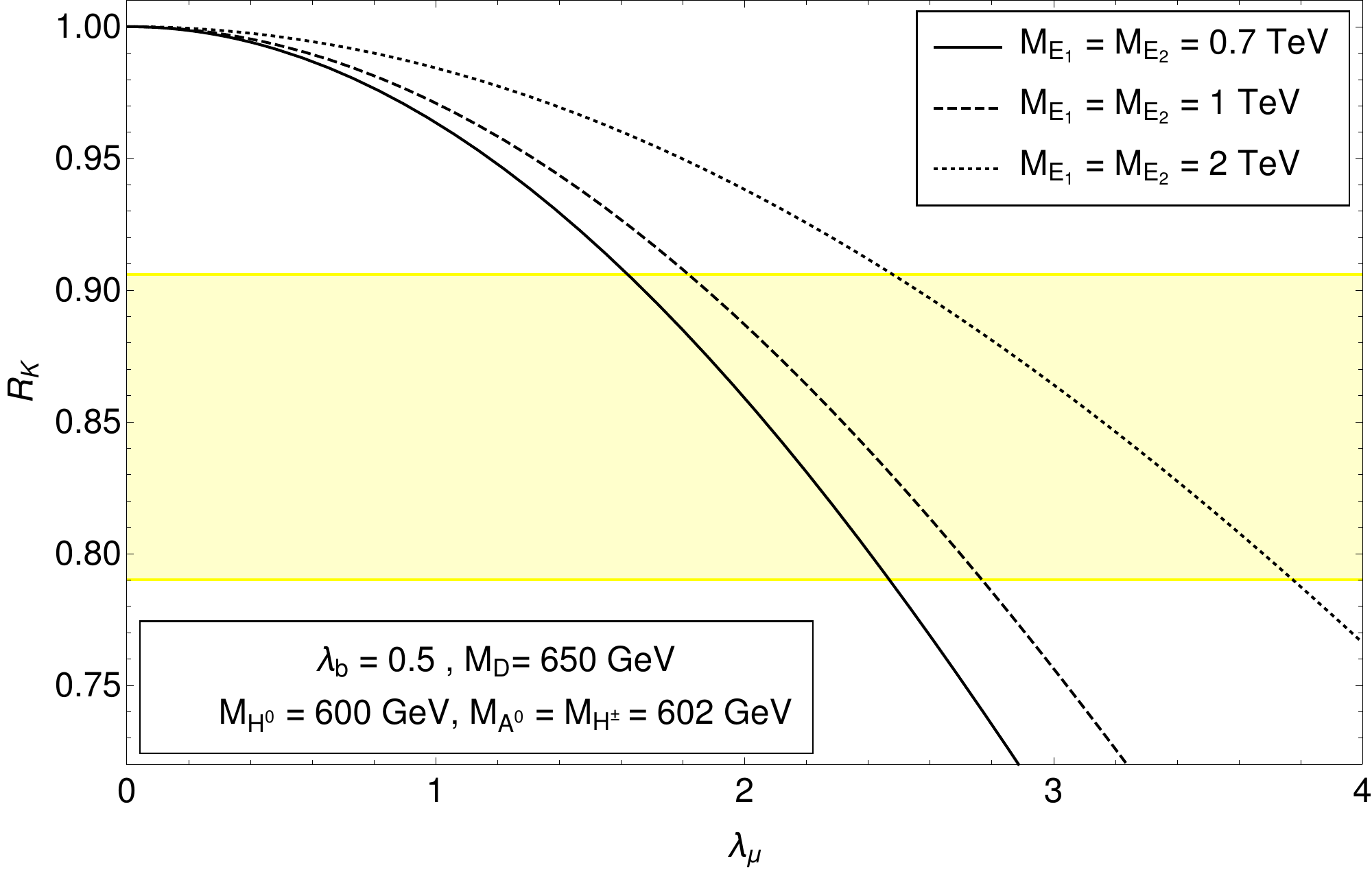}
\label{high1}}
\subfloat[]
{\includegraphics[scale=0.38]{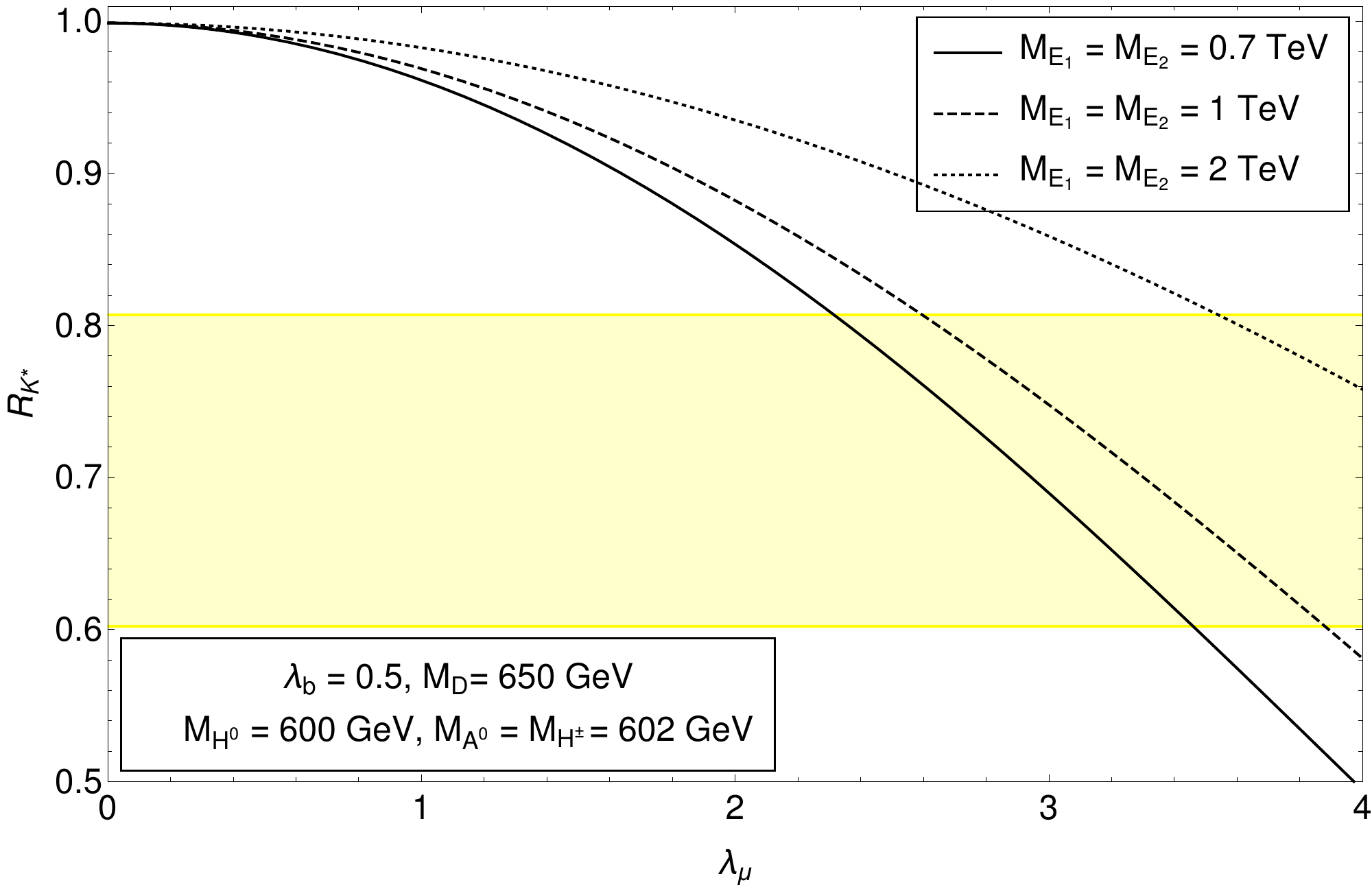}
\label{high2}}\\
\subfloat[]
{\includegraphics[scale=0.41]{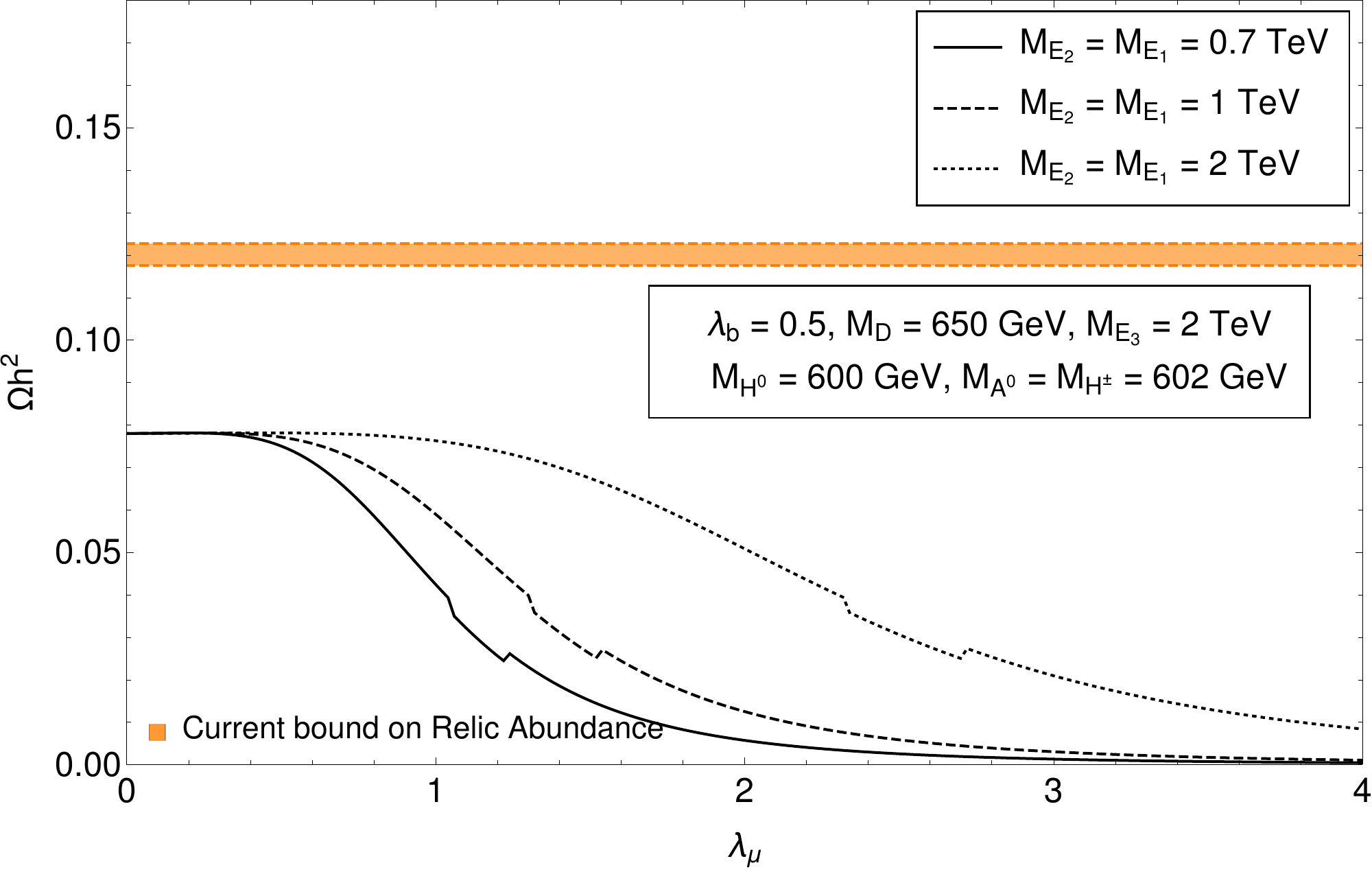} \label{high3}} \\
\captionsetup{style=singlelineraggedleft}
\caption{Top Left (\ref{high1}): Variation of $R(K)$ with $\lambda_{\mu}$ for fixed values of DM mass and $\lambda_b$, for three different values of $M_{E_2}$. The yellow band shows the 1$\sigma$ experimental range of $R(K)$. Top Right (\ref{high2}): Same for 1$\sigma$ experimental range of $R(K^{(*)})$. Middle (\ref{high3}): Variation of relic abundance with $\lambda_{\mu}$  for the same chosen parameters as in~\ref{high1} and~\ref{high2}, the orange band shows the Planck-observed relic density bound} 
\end{figure}

We analyze the high DM mass region of the IDM in context of our extended framework. As discussed earlier, we need to consider degenerate masses for the IDM scalars (as we need to resort on co-annihilation channels in order to satisfy relic density) and also tune $\lambda_L (= \lambda_3 + \lambda_4 + \lambda_5)$, which involves the DM-Higgs interaction, to an appropriate value. So the masses of vector-like leptons ($E_i$) and the vector-like bottom partner ($D$) will have to be greater than the masses of $A^{0}$ or $H^{\pm}$ to maintain the stability of the DM. We consider a mass splitting of 2 GeV between the inert scalars and set $\lambda_L$ to 0.0001.

The parameter space which are allowed by flavour data are shown in Fig.~\ref{high1} and Fig.~\ref{high2}. We have kept the value of the DM mass fixed at $M_{H^0} = 600$ GeV. It is interesting to note that, in this case,  we will be able to explain the $R(K^{(*)})$ anomaly only for higher values of $\lambda_{\mu}$ ($\approx 3$ or $4$). However, for the same masses, such high values of $\lambda_{\mu}$ will not allow us to achieve right relic abundance (see Fig.~\ref{high3}). Hence, it is not possible to obtain a common parameter space that satisfies both relic abundance  and flavour constraints simultaneously. When we add new interactions, new annihilation channels open up and they make the DM under-abundant. So in order to make the effects of NP minimal, we require the couplings to be small but masses to be large (Fig~.\ref{mh0high}). However, if we also want to explain the flavour anomalies for such high values of the vector like fermions masses, we need very high values of couplings as well ($\lambda_\mu \gsim 3$) . So it is impossible to achieve solution in this region of DM mass and hence we discard further investigations for this case.

\section{Collider Phenomenology}
\label{sec:colliderpheno}
Our goal is to investigate the implications of our model on collider searches at the LHC. As mentioned earlier, we have expanded the contact interaction of the DM with the SM and included the vector like fermions (mediators) as propagating degrees of freedom of the theory. Also, it is clear from the above discussions that the mediators have decay channels to the SM fermions. In this section, we will analyze the prospects for detecting our model at the LHC through various channels. Due to the presence of the exotic vector like leptons and quarks, the model gives rise to several tantalizing collider signatures. Here, we have discussed a few of them:
\begin{figure}[h!!]
\centering
\includegraphics[scale=0.5]{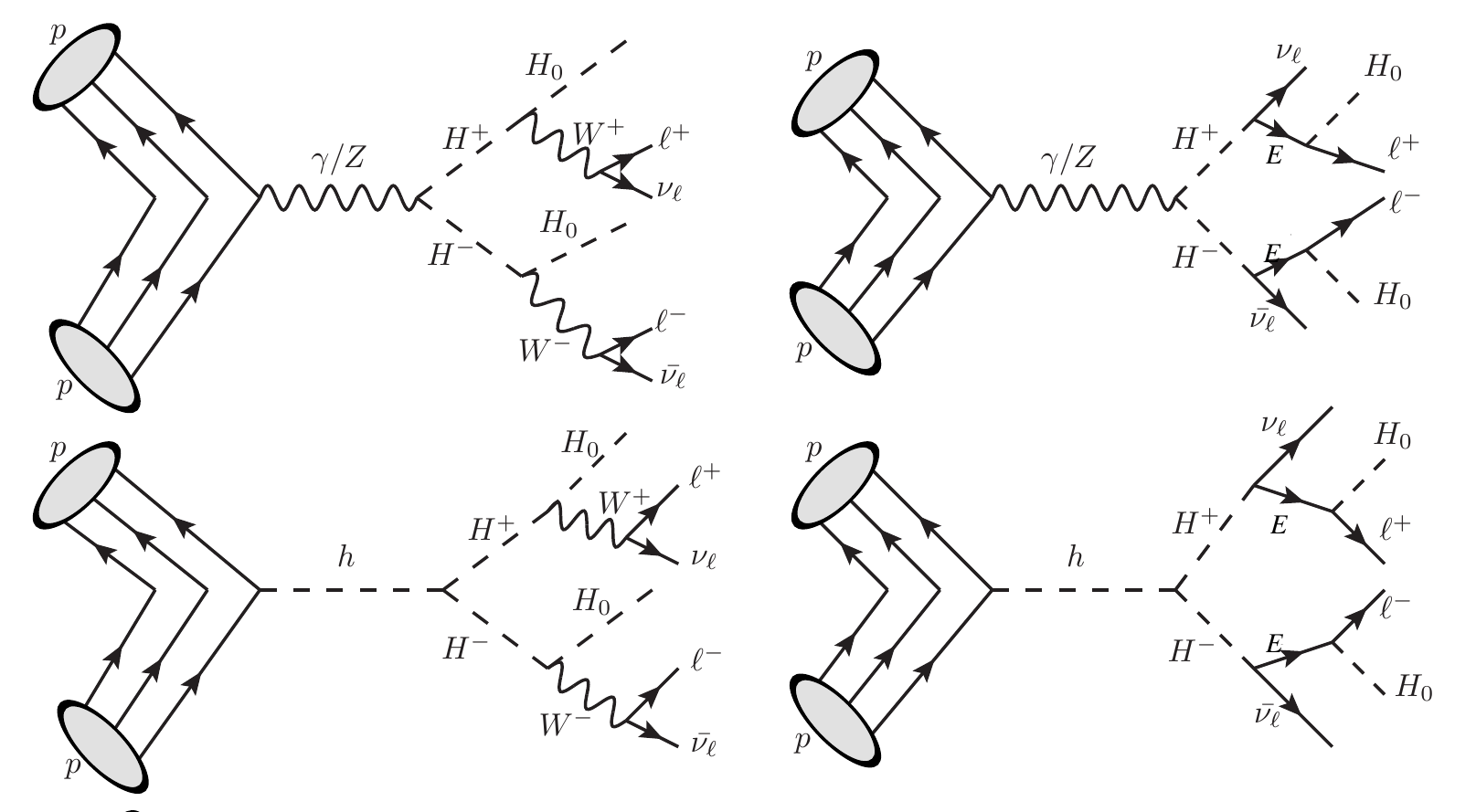}
\includegraphics[scale=0.45]{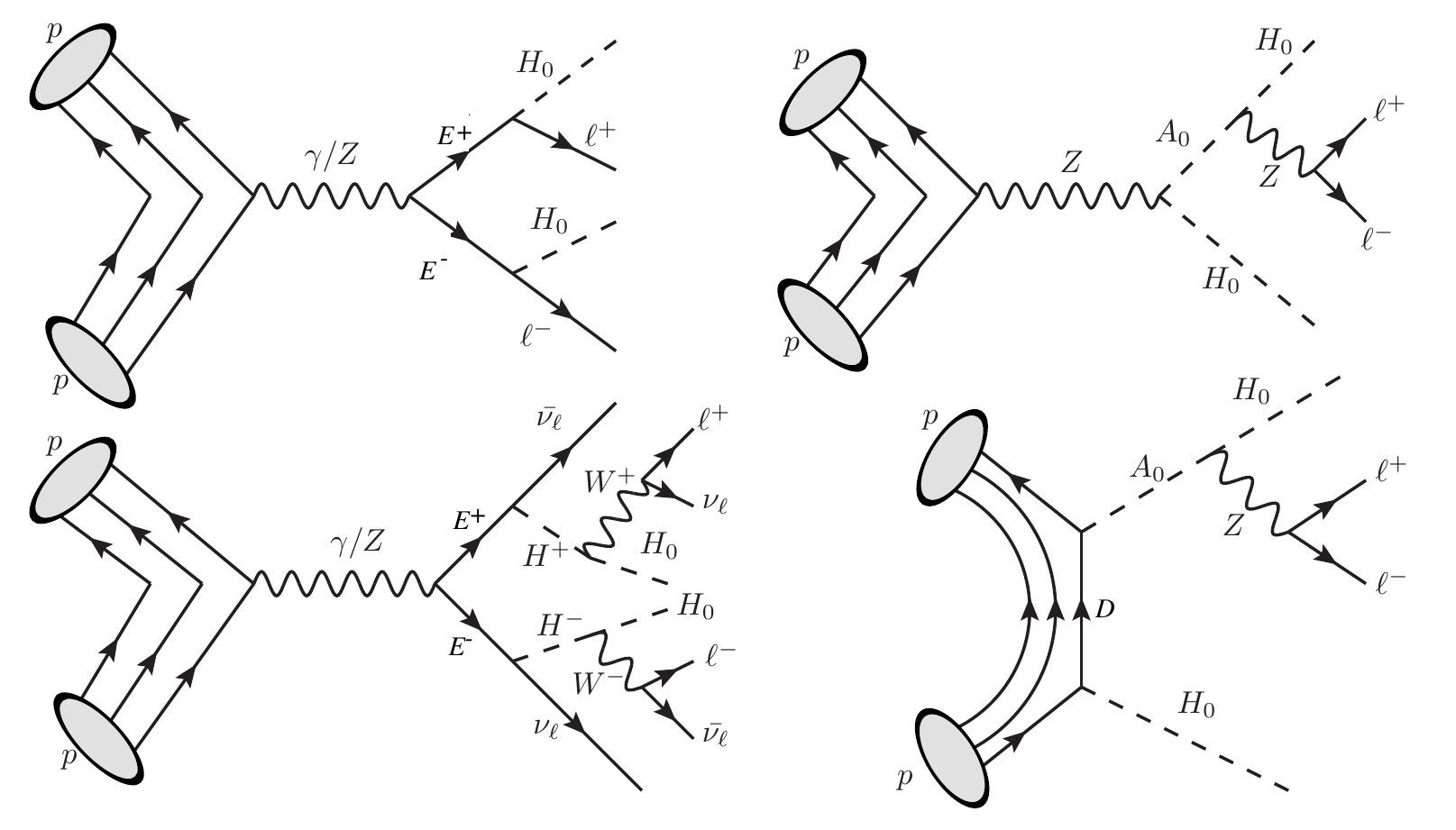}\\
\includegraphics[scale=0.52]{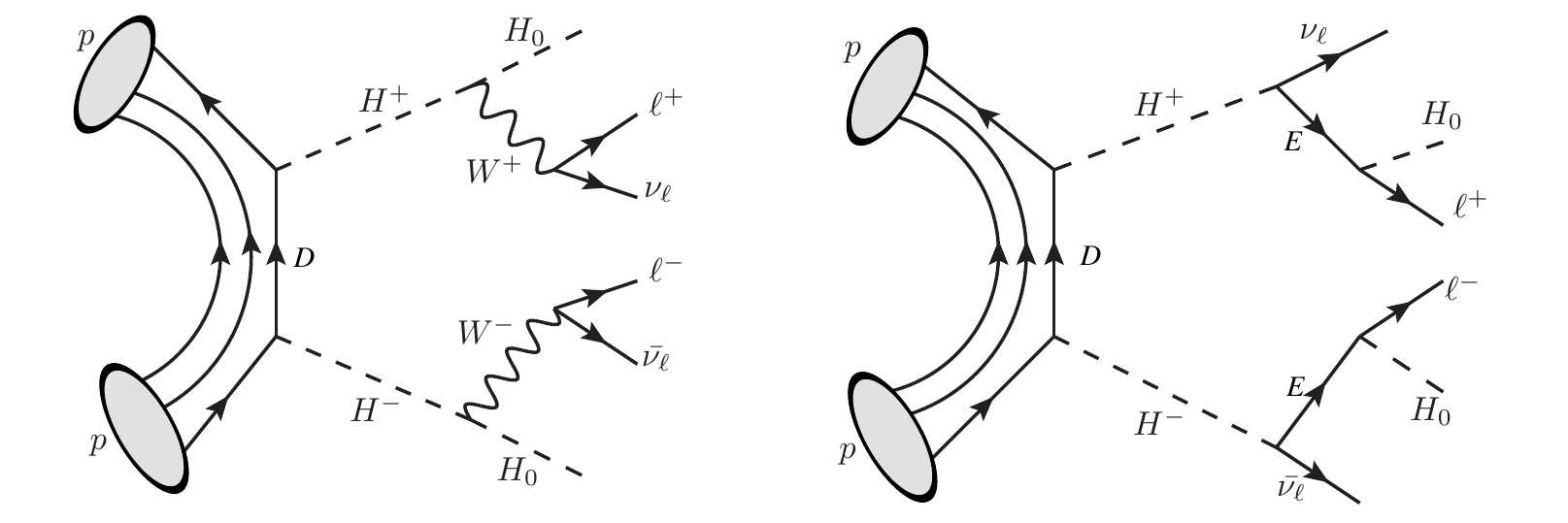}
\caption{Feynman diagrams for the production of $(\ell^+\ell^-+\slashed{E_T})$ final state at a hadron collider like LHC}
\label{fig:dilep_diag}
\end{figure}

\begin{figure}[h!!!]
\centering
\includegraphics[scale=0.42]{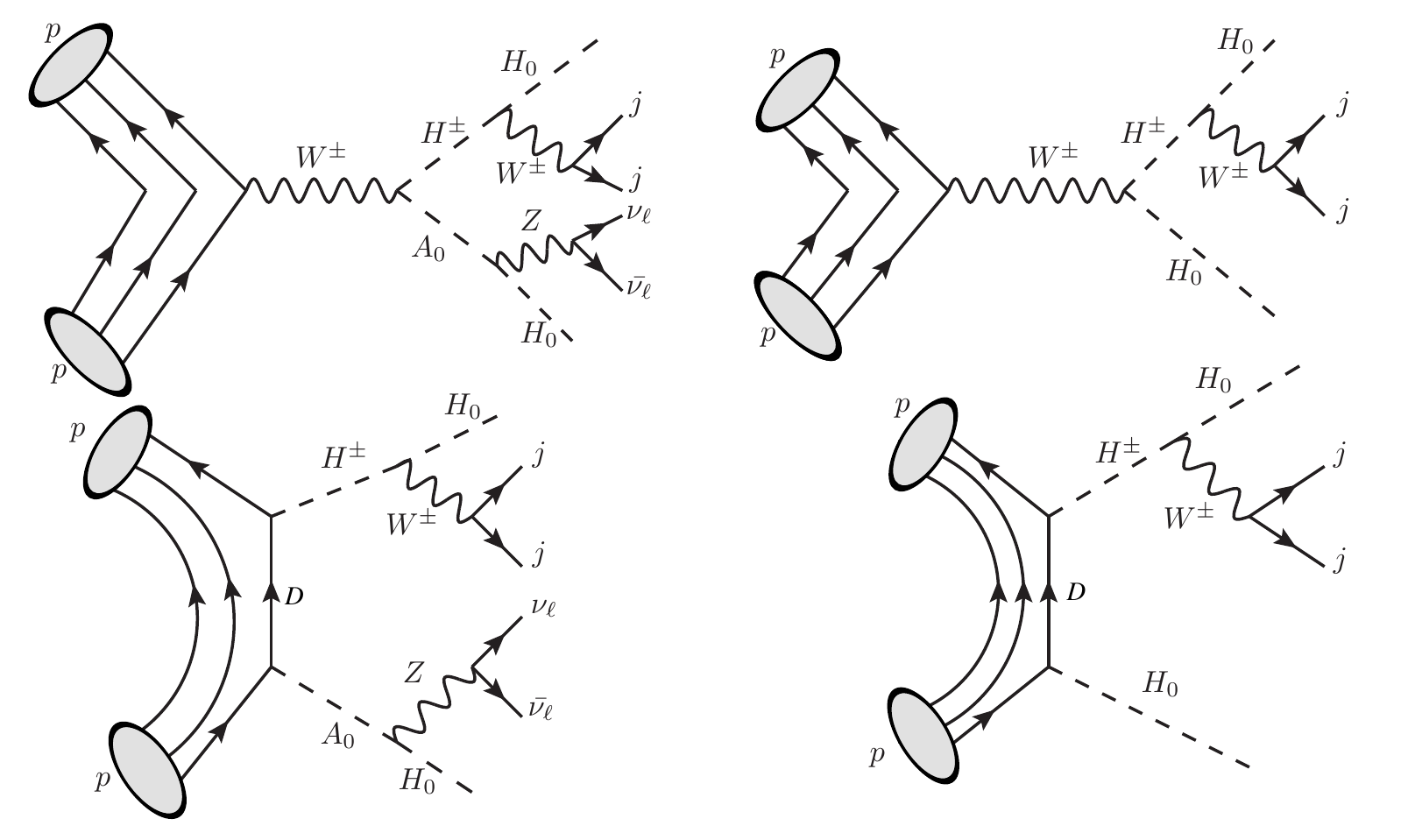}
\includegraphics[scale=0.42]{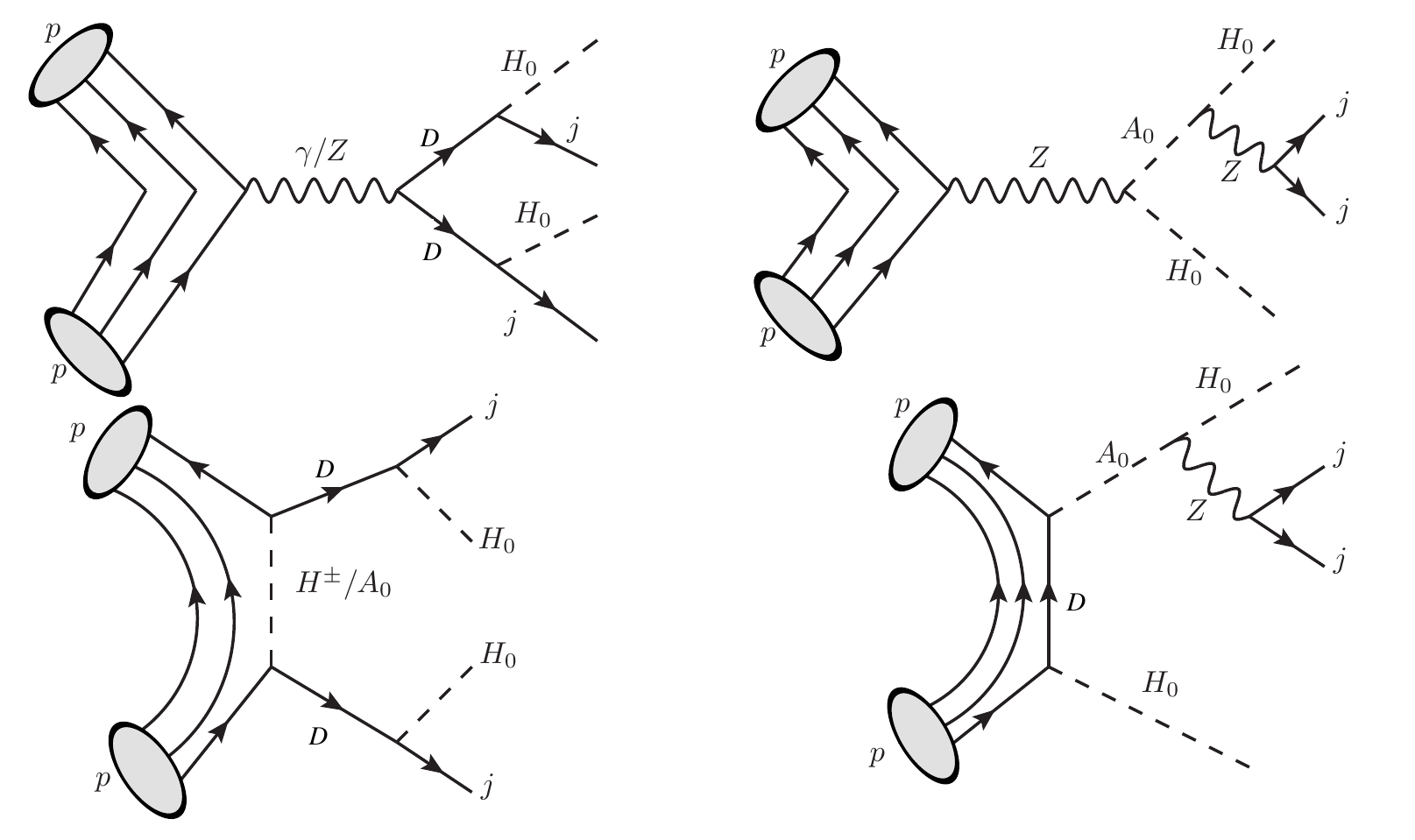}
\includegraphics[scale=0.28]{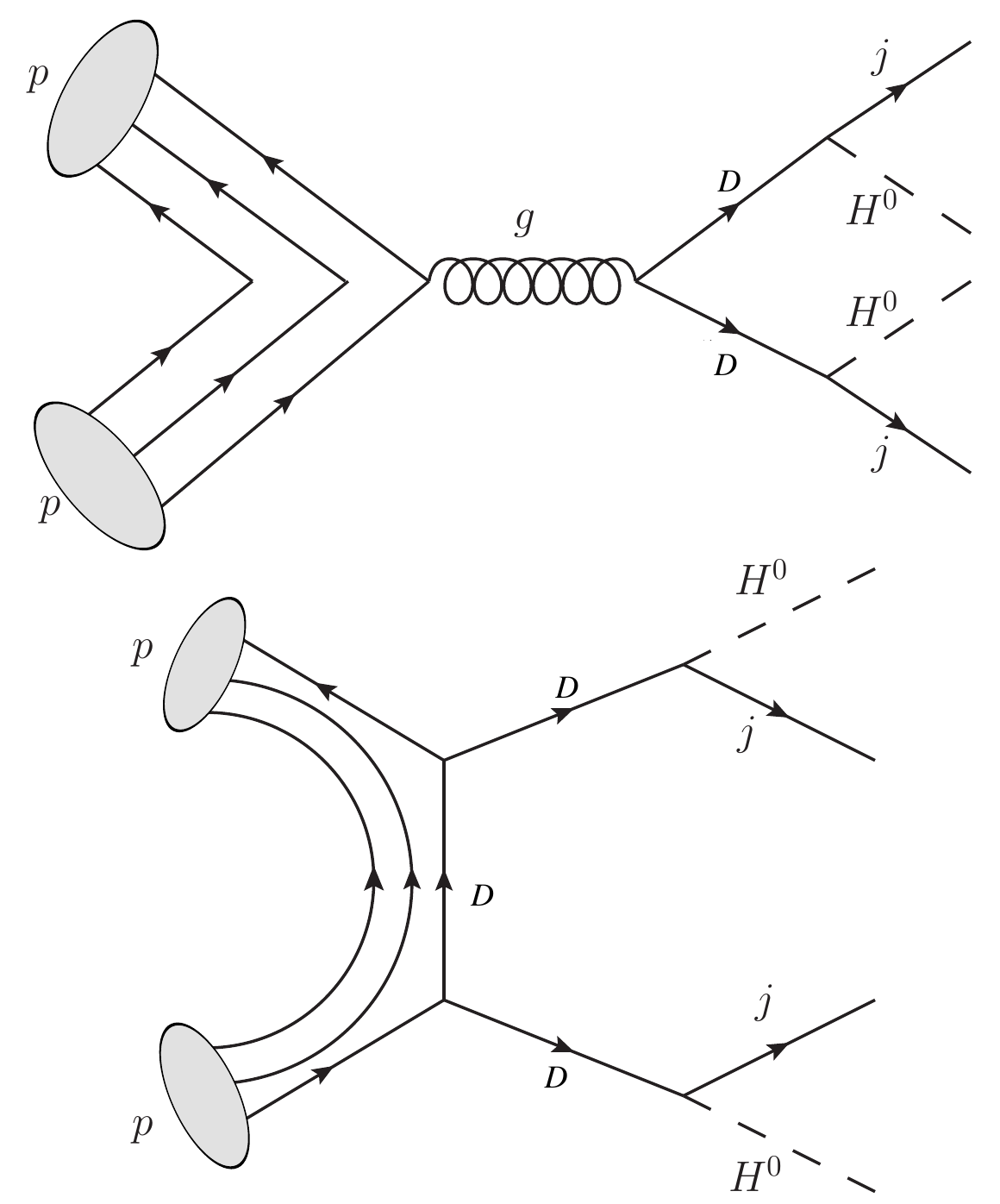}
\caption{Feynman diagrams for channels contributing to $jj +\slashed{E_T}$ final state. The gluon fusion diagram has also been considered in this case}
\label{fig:dijet_diag}
\end{figure}

\begin{figure}[h!!!]
\centering
\includegraphics[scale=0.5]{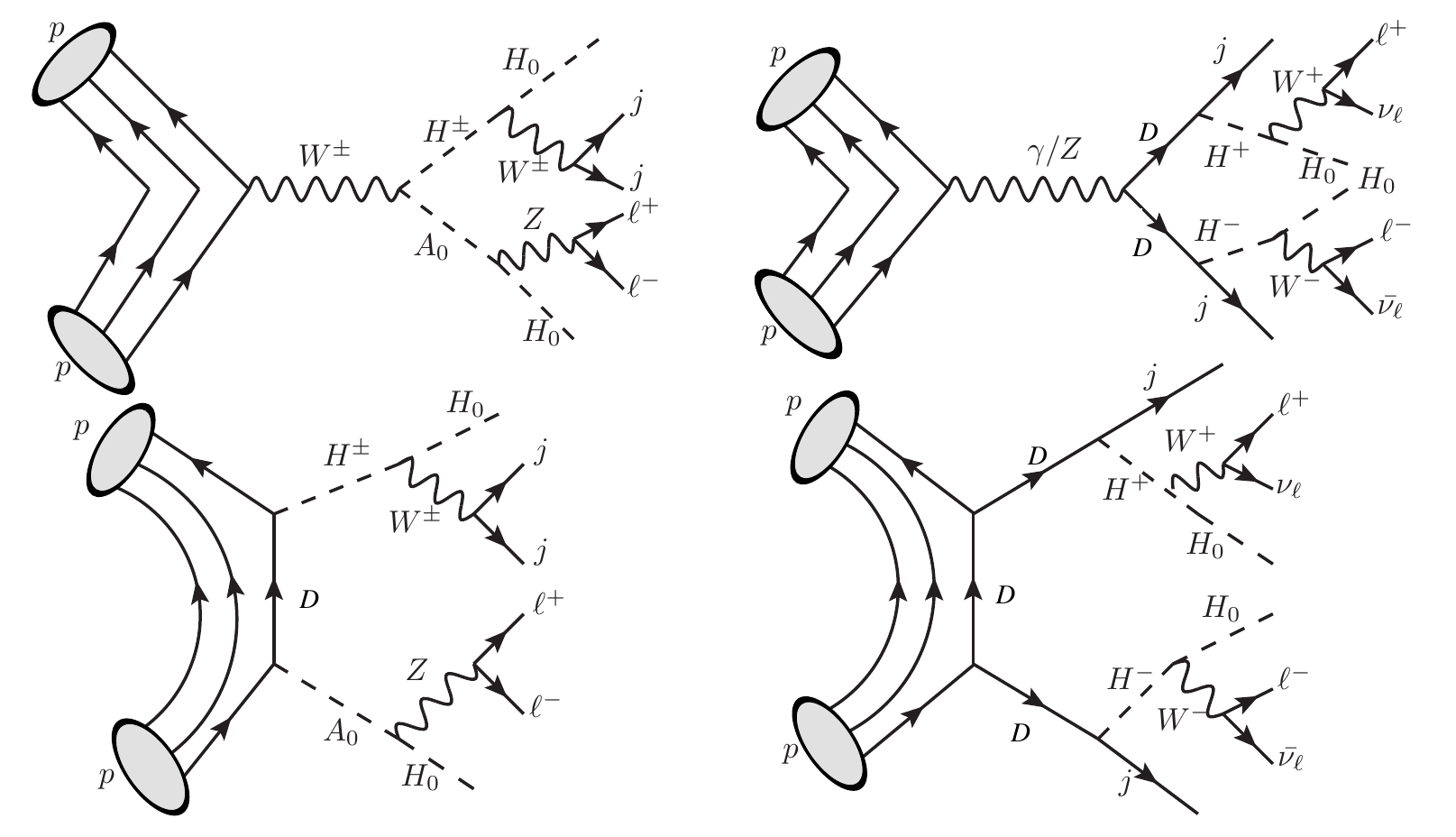}
\includegraphics[scale=0.25]{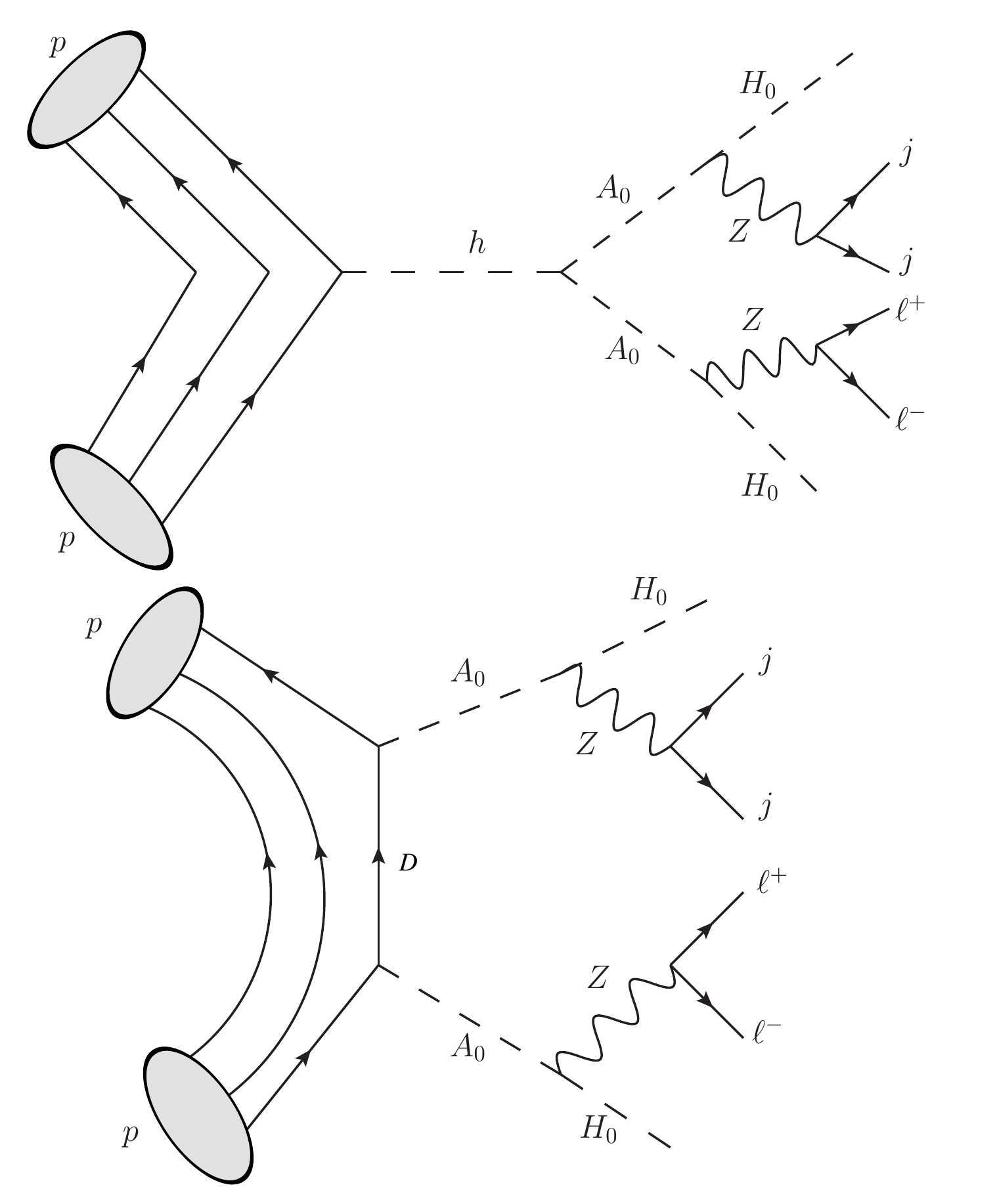}
\captionsetup{style=singlelinecentered}
\caption{Feynman diagrams for channels contributing to $\ell^+ \ell^- + jj+\slashed{E_T}$ final state}
\label{fig:dijet_dilep_diag}
\end{figure}

\begin{itemize}
 \item Opposite sign dilepton (OSD) with missing energy $(\ell^+\ell^-+\slashed{E_T})$\footnote{Hadronically quiet dilepton with missing energy}.
 \item Dijet with missing energy $(jj+\slashed{E_T})$, there are cases where the b-jets have been tagged separately.
 \item Dilepton with dijet and missing energy $(\ell^+\ell^-+jj+\slashed{E_T})$.
\end{itemize}

All these three final states could be tested at the LHC. The corresponding Feynman diagrams for the final states $(\ell^+\ell^-+\slashed{E_T})$, $(jj+\slashed{E_T})$ and $(\ell^+\ell^-+jj+\slashed{E_T})$ are given in figures \ref{fig:dilep_diag}, \ref{fig:dijet_diag} and \ref{fig:dijet_dilep_diag}, respectively. In all these diagrams, the proton ($p$) is considered a multiparticle composed of both quarks and gluons. Hence, all possible initial states with quarks and gluons have been taken into account. We note that amongst all the diagrams, some will appear only in pure IDM case. Therefore, it is important to test whether it is possible to discriminate the signatures of our model from that of pure IDM at the LHC.  


As mentioned earlier, from the common parameter space satisfying flavour constraints, relic density of DM and direct detection bound, we have chosen our benchmark points (BP), which are given in Table~\ref{BP}. All other parameters are fixed at the values:
$M_{H^0}=70$ GeV, $\Delta M = 110$ GeV, $m_h = 125$ GeV, $\lambda_L = 0.0001$, $\lambda_2 = 0.1$, $\lambda_{\text{d}} = 0.001$, $\lambda_{\text{s}} = 0.01$, $\lambda_{\text{e}} = 0.001$.

\begin{table}[h!!!]
\centering
\renewcommand{\arraystretch}{1.8}
	\begin{tabular}{|c|c|c|c|}
	\hline
	$BP$ &  $\sigma_{\ell^+ \ell^- + \slashed{E_T}}$ (fb) & $\sigma_{jj + \slashed{E_T}}$ (fb) & $ \sigma_{\ell^+ \ell^- + jj + \slashed{E_T}}$ (fb) \\
	\hline
	1. &  588 & 278 & 0.31 \\
	\hline
	2. &  54 & 181 & 2.91 \\
	\hline
	3. &  173 & 177 & 2.73 \\
	\hline
	4. &  54 & 172 & 2.75 \\
	\hline
	5. &  17 & 173 & 2.98 \\
	\hline
	6. &  20 & 174 & 2.88 \\
	\hline
	7. &  307 & 174 & 1.00 \\
	\hline
	\end{tabular}
	\bigskip
	\captionsetup{style=singlelinecentered}
	\caption{Production cross-sections for the different signals corresponding to the chosen benchmark points}
	\label{Cross-sec}
\end{table}

\begin{figure*}[htb!]
\centering
\subfloat[]
{\includegraphics[scale=0.35]{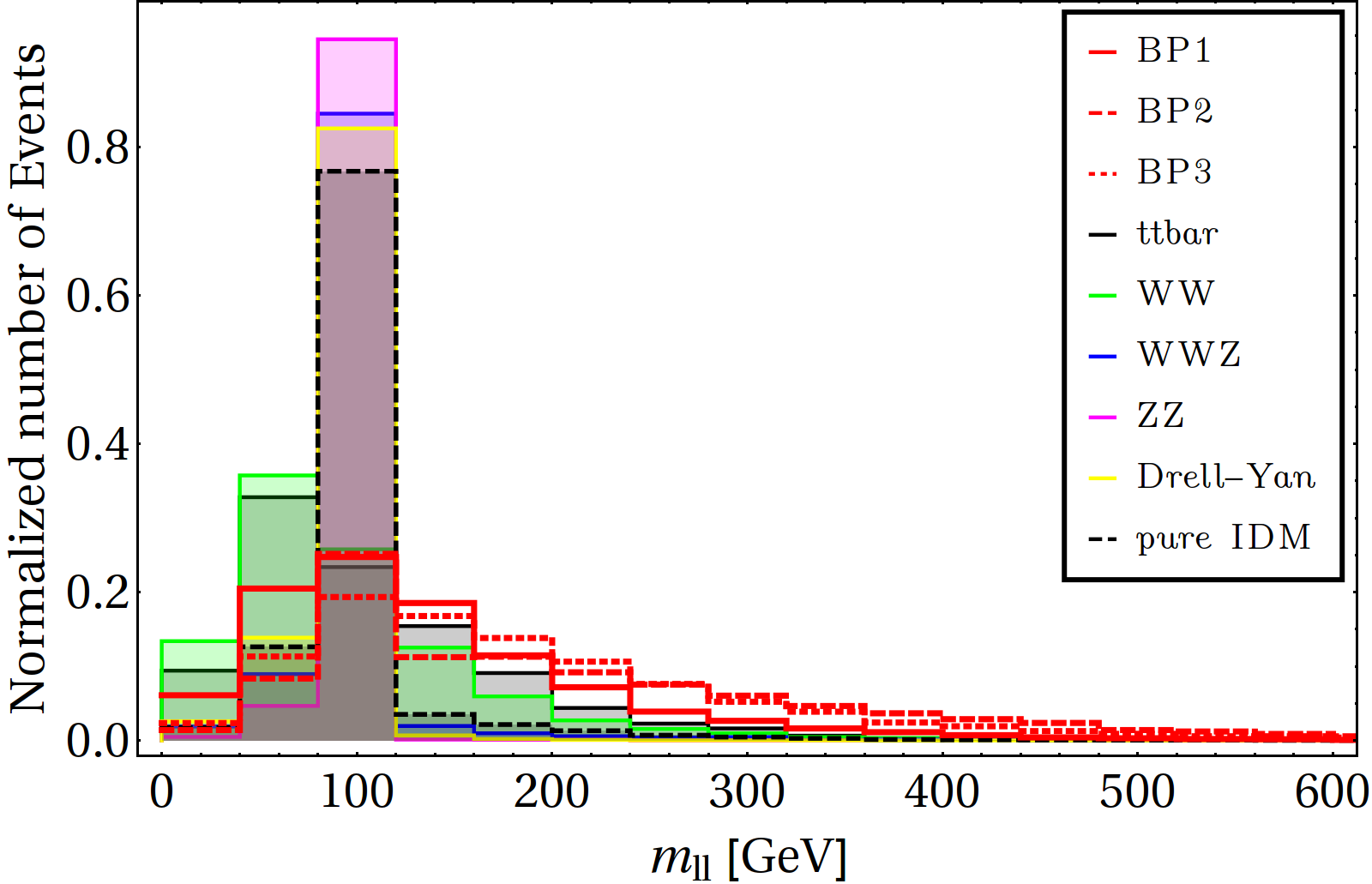}
\label{mlldilepBP1}}~~
\subfloat[]
{\includegraphics[scale=0.35]{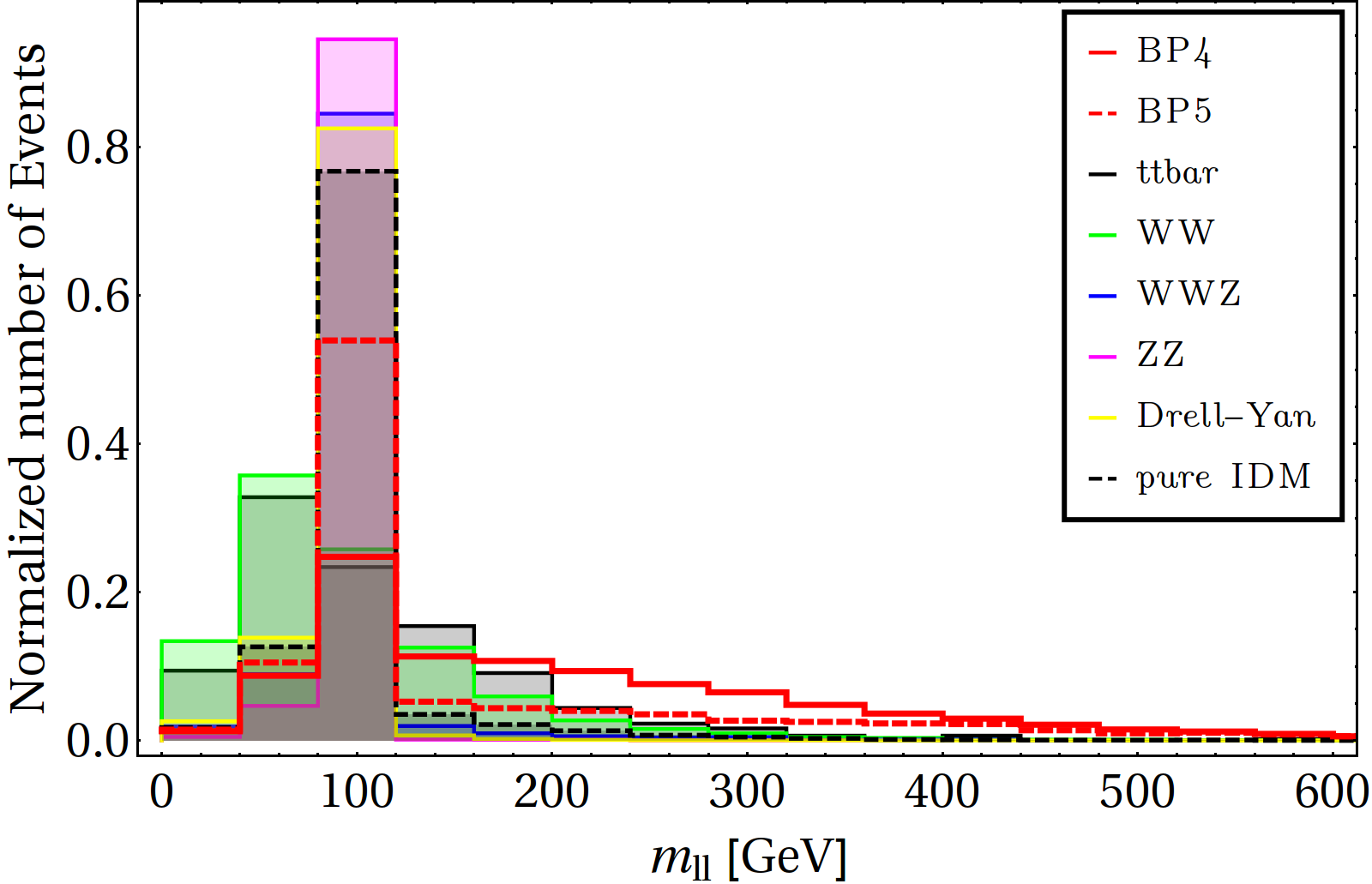}
\label{mlldilepBP}}\\
\captionsetup{style=singlelinecentered}
\caption{Invariant dilepton mass ($M_{\ell\ell}$) distribution for the signals (in red), with SM backgrounds for $(\ell^+\ell^-+\slashed{E_T})$ channel}
\label{fig:mlldilep}
\end{figure*}
\subsection{Object reconstruction and simulation strategy}
\label{sec:simul}


We implemented the model in {\tt FeynRule}~\cite{Alloul:2013bka}. The parton level events are generated in {\tt MADGRAPH}~\cite{Alwall:2014hca}, which are further showered through {\tt PYTHIA}~\cite{Sjostrand:2006za}. All the events are generated at $\sqrt{s}=14$ TeV using {\tt CTEQ6l}~\cite{Placakyte:2011az} as the parton distribution function. All the leptons and jets are reconstructed in order to mimic the LHC environment using the following criteria:

\begin{itemize}
 \item {\it Lepton ($l=e,\mu$):} Leptons are identified with a minimum transverse momentum $p_T>20$ GeV and pseudorapidity $|\eta|<2.5$ such that they are in the central part of the detector. Two leptons are distinguished as isolated objects if their mutual distance in the $\eta-\phi$ plane is $\Delta R=\sqrt{\left(\Delta\eta\right)^2+\left(\Delta\phi\right)^2}\ge 0.2$, while that separation between an isolated lepton and a jet is given by $\Delta R\ge 0.4$.
 
 \item {\it Jets ($j$):} The cone jet algorithm {\tt PYCELL} has been used to build jets inside {\tt PYTHIA}. All the partons within $\Delta R=0.4$ from the jet initiator cell are included to form the jets. We require $p_T>20$ GeV for a clustered object to be considered as jet. Jets are isolated from unclustered objects for $\Delta R>0.4$.   
 
 \item {\it Unclustered Objects:}  All the final state objects which are neither clustered to form jets, nor identified as  isolated leptons, belong to this category. All particles with $0.5<p_T<20$ GeV and $|\eta|<5$, are considered as unclustered.
 
 \item {\it Missing Energy ($\slashed{E}_T$):} The transverse momentum of all the missing particles (those are not registered in the detector) can be estimated from momentum imbalance of the visible particles in the transverse direction. Thus, missing energy (MET) is defined as:
 
 \bea
 \slashed{E}_T = -\sqrt{(\sum_{\ell,j} p_x)^2+(\sum_{\ell,j} p_y)^2},
 \eea
 where the sum runs over all visible objects that include the leptons and jets, and the unclustered components. 
 
 \item { $H_T$: We have used another observable for collider searches which is the scalar sum of all isolated lepton/jet transverse momentum: 
 
 \bea
 H_T = \sum_{\ell, j} p_T 	
 \eea
 }
\end{itemize}

The dominant SM backgrounds have been generated in {\tt MADGRAPH} and then showered through {\tt PYTHIA}. Also appropriate $K$-factors were used to match them with the Next-to-Leading order (NLO) cross section. We have identified dominant SM backgrounds as: $t\bar{t}$, $W^{+}W^{-}$, $W^{\pm}Z$, $ZZ$, $Wj$, $Zj$ and $Drell-Yan$ for the chosen signal regions. The discovery potential of the the model, in terms of signal significance, are shown for only those cases where the signal can be clearly distinguished from the SM background. In each case, we have also shown the status of pure IDM scenario for comparison purpose.  Once again we would like to remind the readers that the purpose of the collider phenomenology in our context is not to search for signals of the VLFs, but to look for the signature of the model itself. In that sense, our choice of observables and cuts are different from that of dedicated experimental searches done at the LHC or LEP.

\begin{figure*}[htb!]
\centering
\subfloat[]
{\includegraphics[scale=0.35]{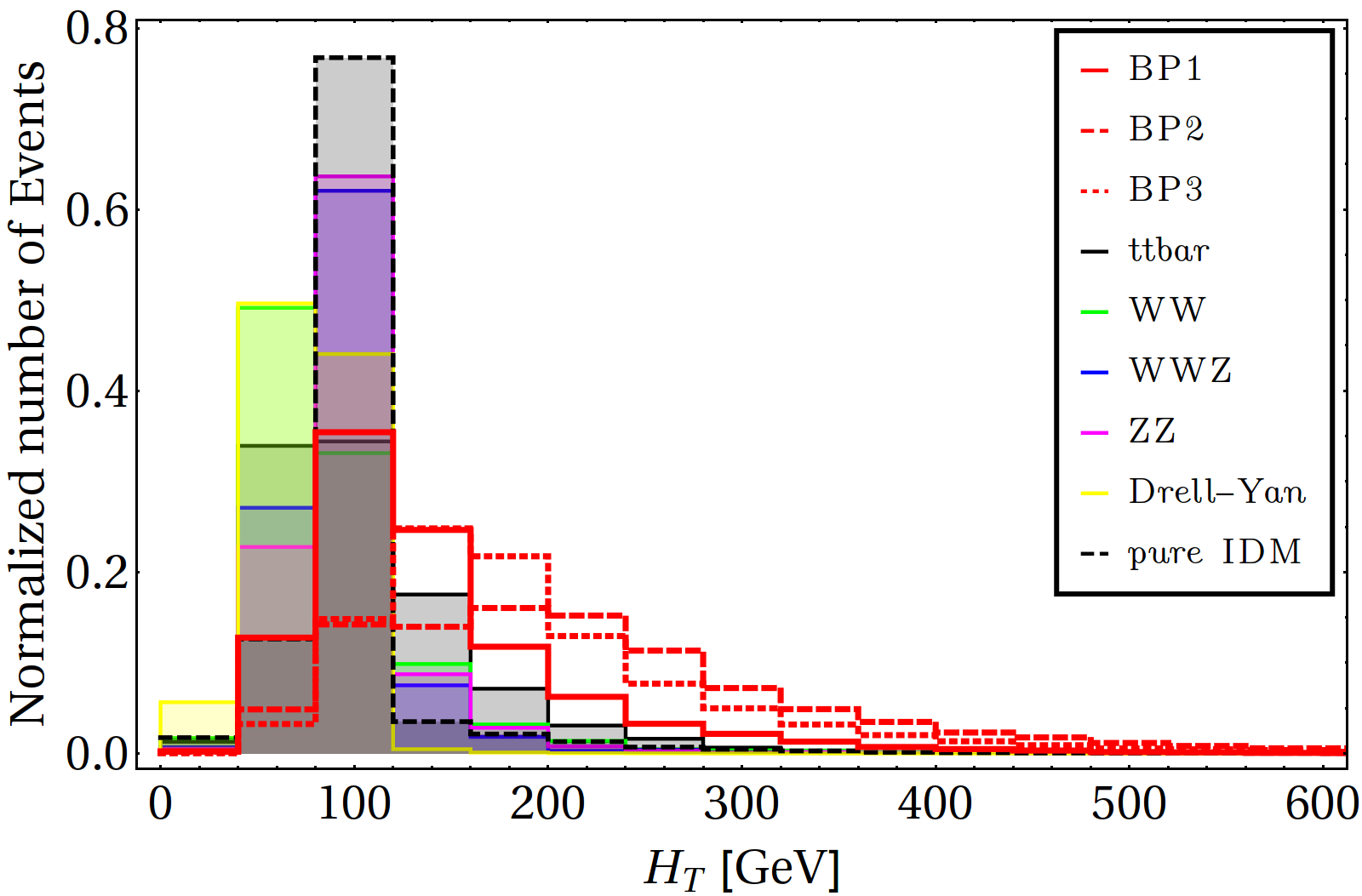}
\label{htdilepBP1}}~~
\subfloat[]
{\includegraphics[scale=0.35]{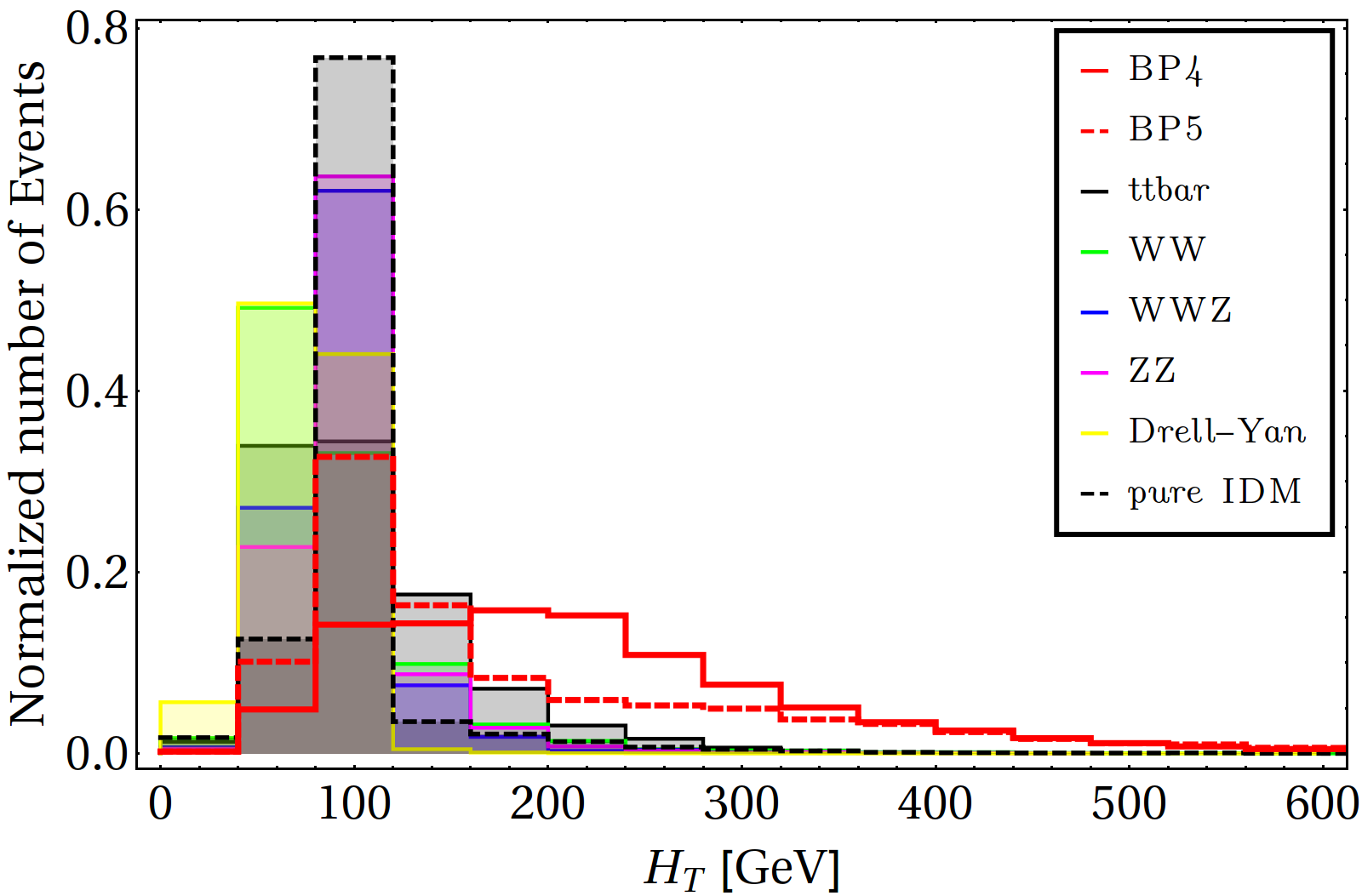}
\label{htdilepBP2}}\\
\captionsetup{style=singlelinecentered}
\caption{$H_T$ distribution for the signals (in red), with SM backgrounds $(\ell^+\ell^-+\slashed{E_T})$ channel}
\label{fig:htdilep}
\end{figure*}

\subsection{ Dilepton with missing energy final state}
\label{sec:osdmet}



In Fig.~\ref{fig:mlldilep} and~\ref{fig:htdilep}, we have shown respectively the dilepton invariant mass ($M_{\ell\ell}$) and $H_T$ distributions for the signal, in comparison with the relevant SM background for final states containing opposite sign dilepton and MET. As one can see from Fig.~\ref{fig:mlldilep}, the primary difference in the $M_{\ell\ell}$ distribution between the signal and the background lies in the fact that we observe a long tail in case of signal, which is not present in the background. The reason being, in case of signal, the source of the leptons in the final state are the heavy new particles apart from the SM gauge bosons. Since these particles are heavier than the SM gauge bosons, hence the leptons in this case are much boosted than those coming only from the SM gauge bosons. As a result, we see a flatter distribution for $M_{\ell\ell}$ for the signal compared to the SM background (Fig.~\ref{fig:mlldilep}). As an example in Fig.~\ref{fig:dilep_diag}, we see that for signal, the leptons in the final state arise not only from SM $Z$ or $W$ but also from the decay of heavy new particles {\it e.g,} $E^+\to H^0 \ell^+$. The same arguement holds for $H_T$, where a flatter signal distribution (Fig.~\ref{fig:htdilep}) arises because of the less boosted leptons in the final state.




From $M_{\ell\ell}$ distributions shown in Fig.~\ref{fig:mlldilep} we can see that a cut: $M_{\ell\ell}\gsim300~\rm GeV$ can help us to get rid of the SM background completely for all BPs except BP1, retaining the signal intact. For BP1, on the other hand, such a cut kills the background along with the signal as well. Similar kind of trend is also observed in the $H_T$ distribution. In order to determine the signal significance We have used $M_{\ell\ell}>200~\rm GeV$ and $H_T> 280~\rm GeV$ cuts for BP1-5.

\begin{table}[htb!]
\begin{center}
\begin{tabular}{|c|c|c|c|c|c|c|c|c|c|}
\hline
Benchmark & $\sigma^{\rm production}$&  & $\sigma^{\rm OSD}$  \\ [0.5ex] 
Points     & (pb)  &  & (fb)    \\ [0.5ex] 
\hline\hline

BP1 & 0.58 && 11.98 \\
\cline{4-4}
\cline{1-2}
BP2 & 0.05 &$M_{\ell\ell}>$ 200 GeV& 3.05 \\
\cline{4-4}
\cline{1-2}
BP3 & 0.16 &$H_T>$ 280 GeV& 8.77 \\
\cline{4-4}
\cline{1-2}
BP4 & 0.05 && 3.04 \\
\cline{4-4}
\cline{1-2}
BP5 & 0.01 && 0.45 \\
\cline{4-4}
\cline{1-2}
BP6 & 0.02 && 0.48 \\
\cline{4-4}
\cline{1-2}
BP7 & 0.30 && 0.85\\
\cline{4-4}
\cline{1-2}
Pure IDM & 0.16 && 0.03 \\
\hline\hline
\end{tabular}
\end{center}
\caption {Final state signal cross-section with $M_{\ell\ell}>200~\rm GeV$ and $H_T>280~\rm GeV$ for $OSD+\slashed{E_T}$ final state. All simulations are done at $\sqrt{s}=14~\rm TeV$.} 
\label{tab:osd-sig}
\end{table}

\begin{table}[htb!]
\begin{center}
\begin{tabular}{|c|c|c|c|c|c|c|c|c|c|}
\hline
Benchmark & $\sigma^{\rm production}$&  & $\sigma^{\rm OSD}$  \\ [0.5ex] 
Points     & (pb)  &  & (fb)    \\ [0.5ex] 
\hline\hline

$t\bar t$ & 81.64 && 48.78 \\
\cline{4-4}
\cline{1-2}
$WW$ & 99.98 &$M_{\ell\ell}>$ 200 GeV& 36.99 \\
\cline{4-4}
\cline{1-2}
$WWZ$ & 0.15 &$H_T>$ 280 GeV& 0.06 \\
\cline{4-4}
\cline{1-2}
$ZZ$ & 14.01 && 0.14 \\
\cline{4-4}
\cline{1-2}
Drell-Yan & 2272.80 && 0.14 \\
\cline{4-4}
\cline{1-2}
\hline\hline
\end{tabular}
\end{center}
\caption {Final state SM background cross-section with $M_{\ell\ell}>200~\rm GeV$ and $H_T>280~\rm GeV$ for $OSD+\slashed{E_T}$ final state. All simulations are done at $\sqrt{s}=14~\rm TeV$.} 
\label{tab:osd-bckh}
\end{table} 

For all the benchmarks, final state cross-section with all the above mentioned cuts imposed are tabulated in Table~\ref{tab:osd-sig}. Corresponding cross-sections for dominant SM backgrounds are also shown in Table~\ref{tab:osd-bckh}. Inspite of the fact that such a cut kills most of the signals for BP1, due to its high production cross section (as shown in Table~\ref{Cross-sec}) it shows a very high significance along with BP3. Similarly, $\slashed{E_T}>200~\rm GeV$ and $H_T> 320~\rm GeV$ have been used to compute the significance of BP6 and BP7 and a similar excess is seen in case of BP7 due to its high production cross section. Therefore,the benchmarks BP1, BP3 and BP7 are prone to be eliminated in ongoing LHC runs, if no excess is being found in those channels.

\begin{figure*}[h!!]
\centering
\subfloat[]
{\includegraphics[scale=0.35]{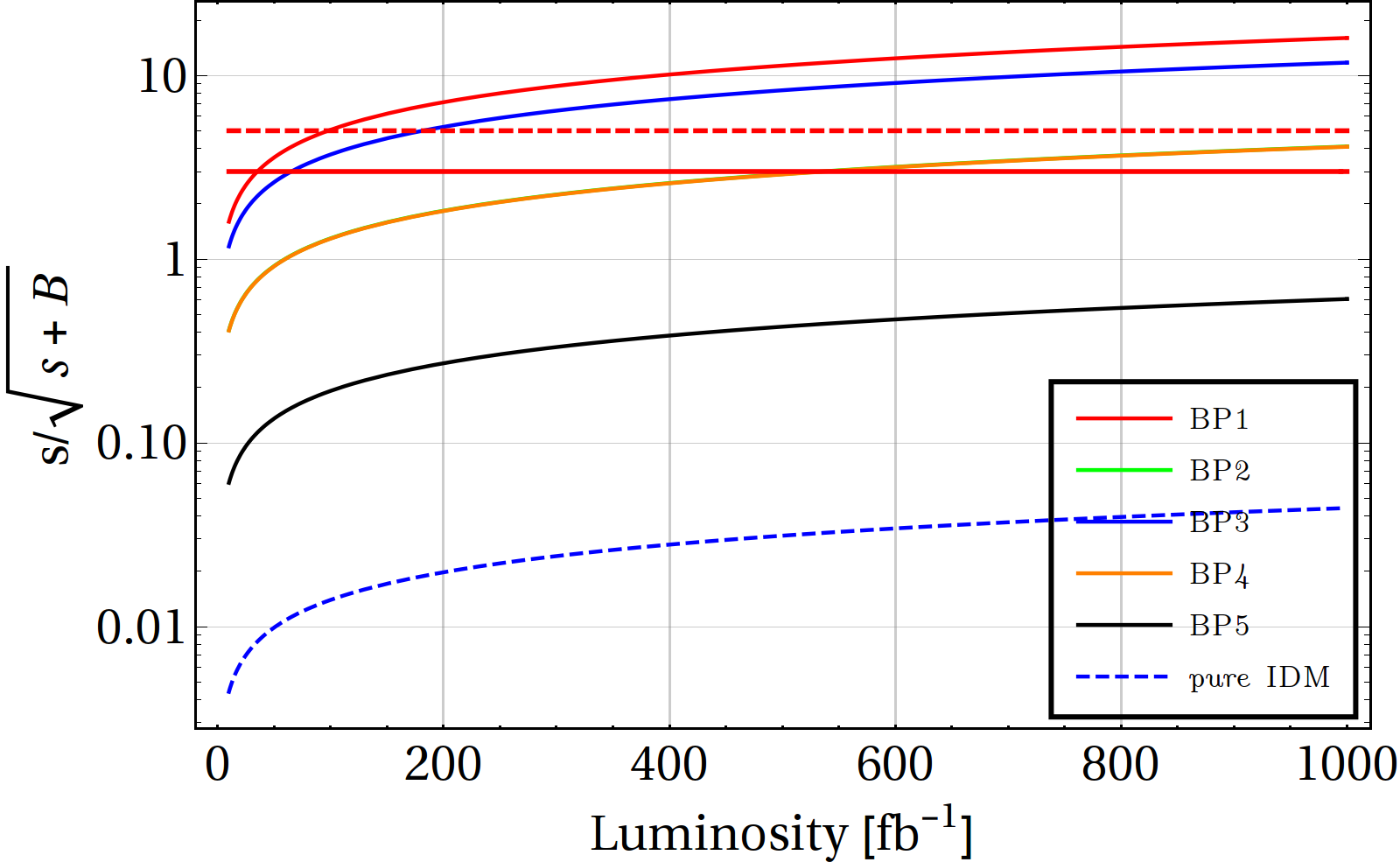}
\label{dilepsig1-6}}
\subfloat[]
{\includegraphics[scale=0.35]{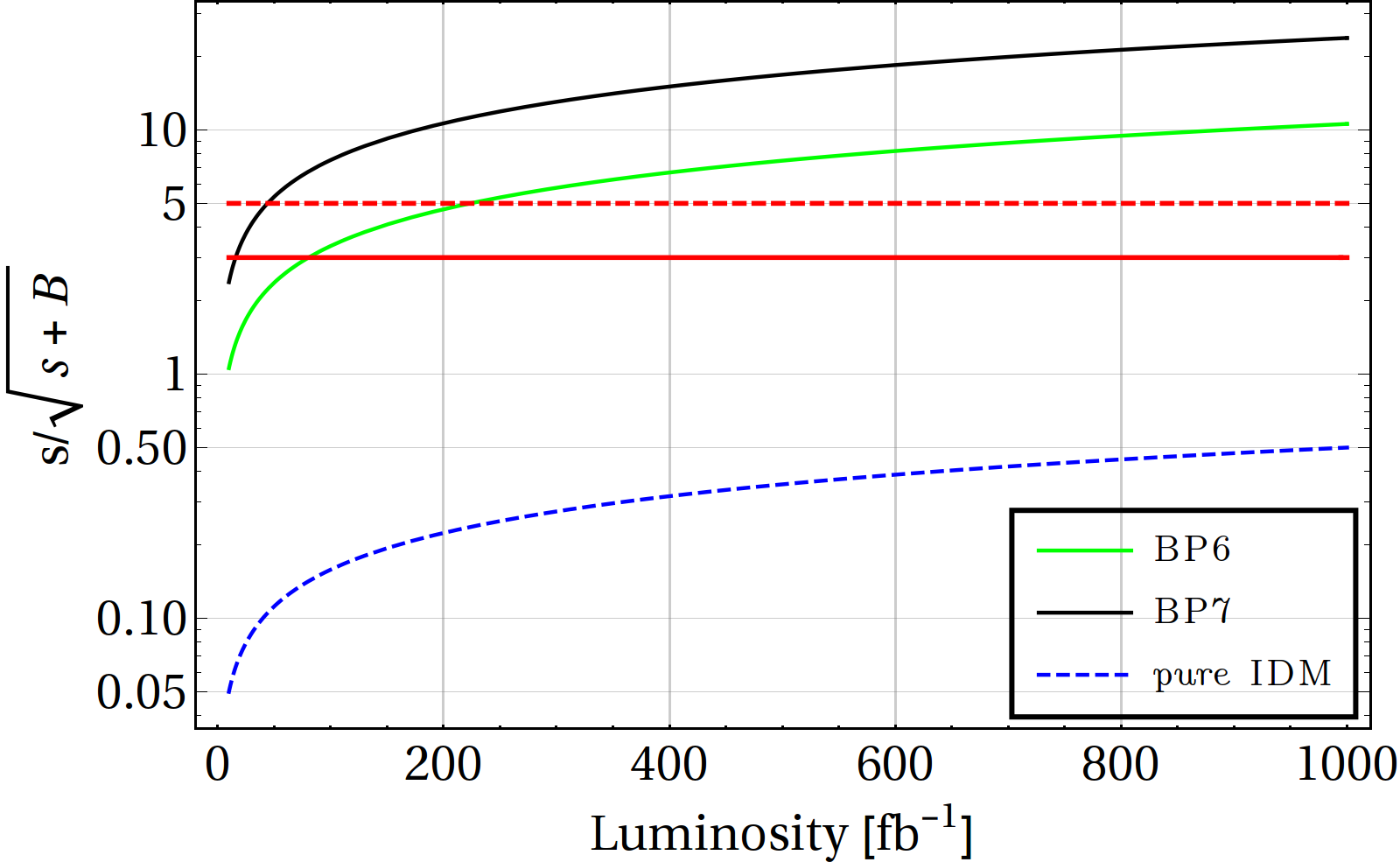}
\label{dilepsig67}}
\captionsetup{style=singlelineraggedleft}
\caption{Significance plot for dilepton plus $\slashed{E_T}$ final state for the different BPs. The dashed blue line shows the significance for the pure IDM scenario for comparison. The thick red and the dashed red lines are respectively showing the $3\sigma$ and $5\sigma$ confidences. Note that we have used $M_{\ell\ell}>200~\rm GeV$ and $H_T> 280~\rm GeV$ cuts to compute the significance of BP1 to BP5 (left plot) while $\slashed{E_T}>200~\rm GeV$ and $H_T> 320~\rm GeV$ have been used to compute the significance of BP6 and BP7 (right plot)}
\label{fig:dilepsignific}
\end{figure*}

\begin{figure*}[htb!]
\centering
\subfloat[]
{\includegraphics[scale=0.35]{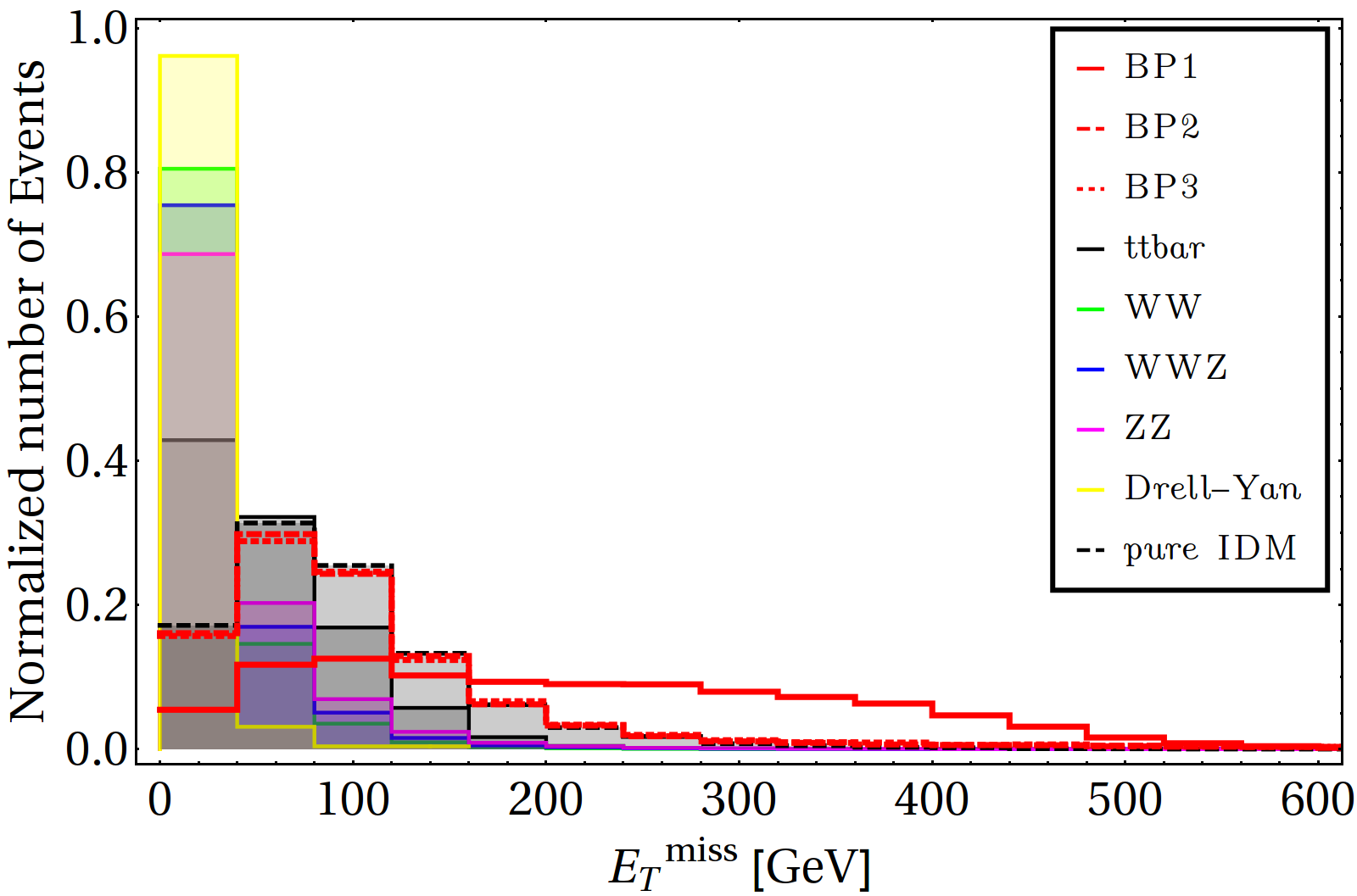}}~~
\subfloat[]
{\includegraphics[scale=0.35]{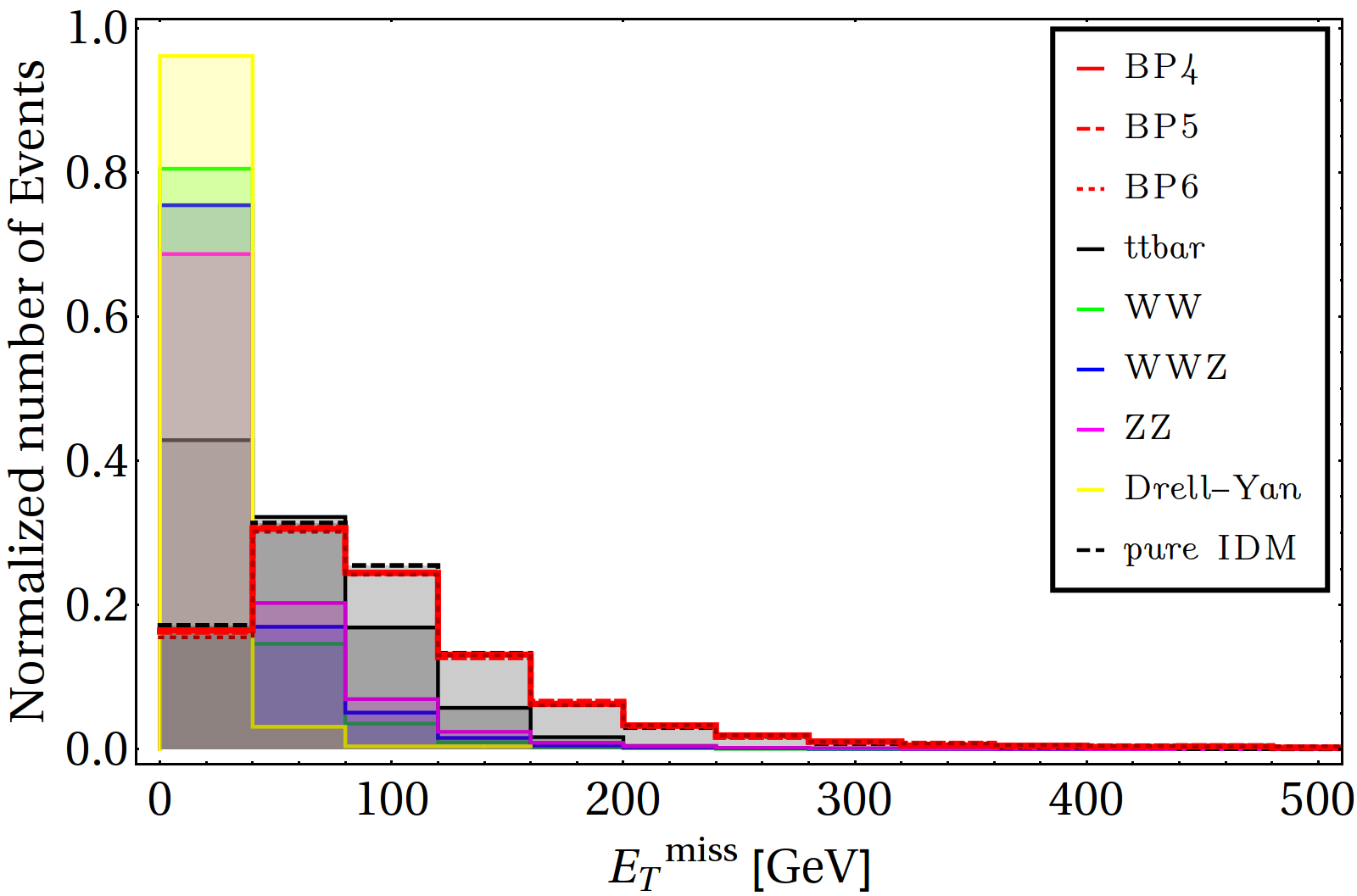}}\\
\captionsetup{style=singlelinecentered}
\caption{Missing energy ($\slashed{E_T}$) distribution for the signals (in red), with SM backgrounds for dijet+$\slashed{E_T}$ channel. We note that in the left plot, it is hard to distinguish the distributions of the benchmark scenarios BP2 and BP3 from each other. Similarly in the right plot all the signal distributions almost overlap with each other}
\label{fig:METdijet}
\end{figure*}

\begin{figure*}[t]
\centering
\subfloat[]
{\includegraphics[scale=0.35]{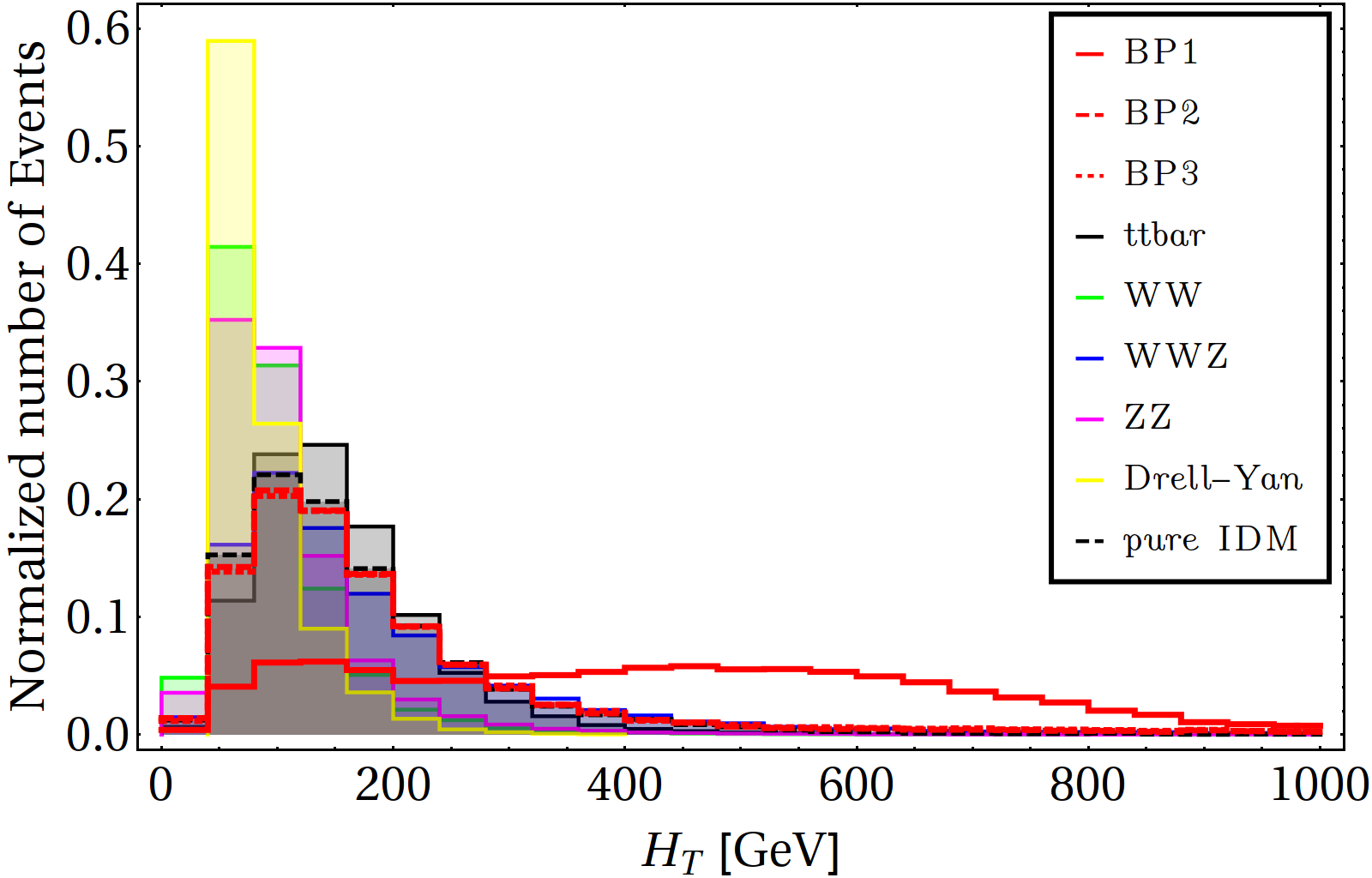}}~~
\subfloat[]
{\includegraphics[scale=0.35]{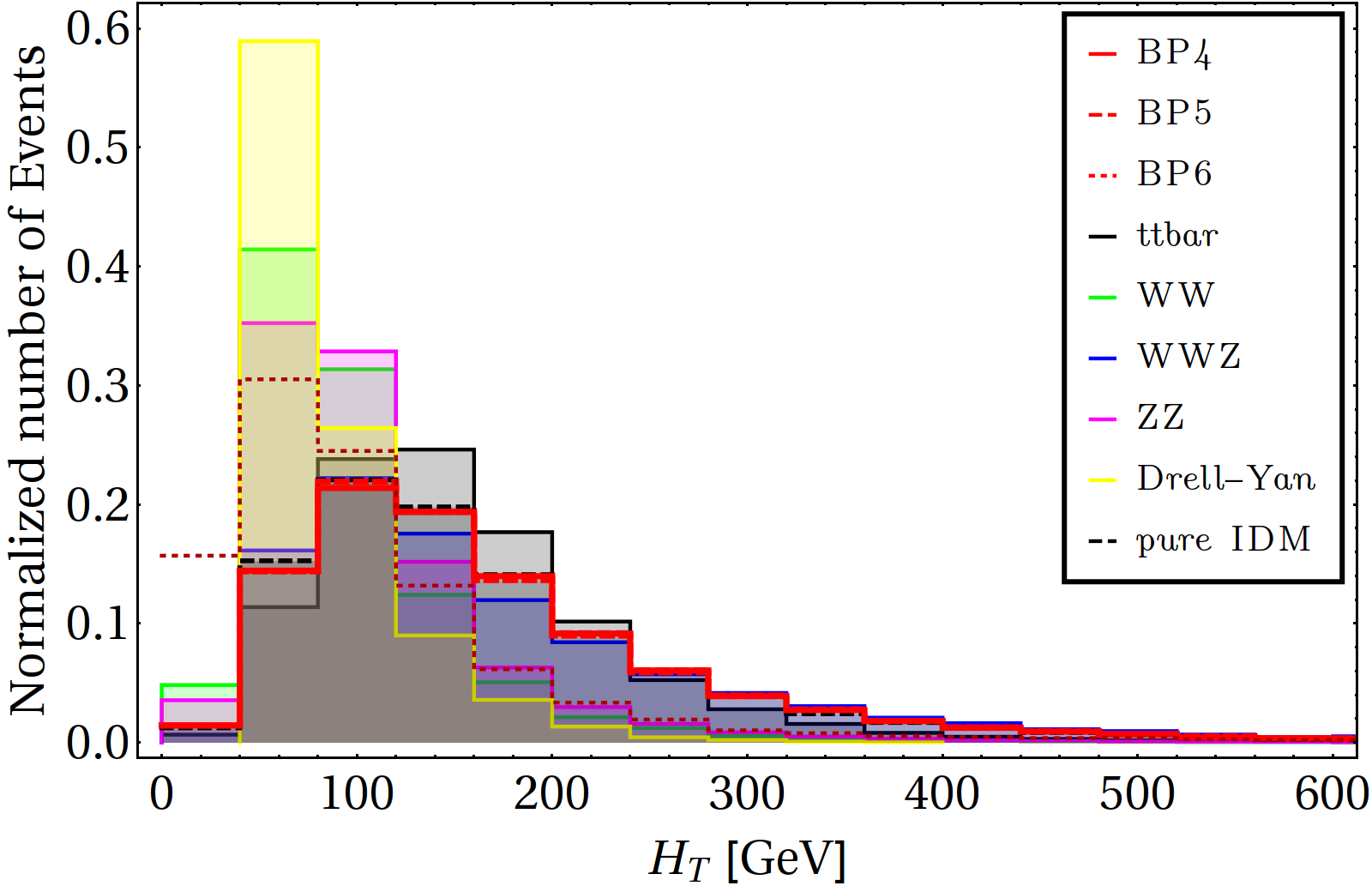}}\\
\captionsetup{style=singlelinecentered}
\caption{$H_T$ distribution for the signals (in red), with SM backgrounds for dijet+$\slashed{E_T}$ channel. We note that in the left plot, it is hard to distinguish the distributions of the benchmark scenarios BP2 and BP3 from each other. Similarly in the right plot all the signal distributions for BP4 and BP5 almost overlap with each other}
\label{fig:HTdijet}
\end{figure*}

\begin{figure*}[htb!]
\centering
\subfloat[]
{\includegraphics[scale=0.35]{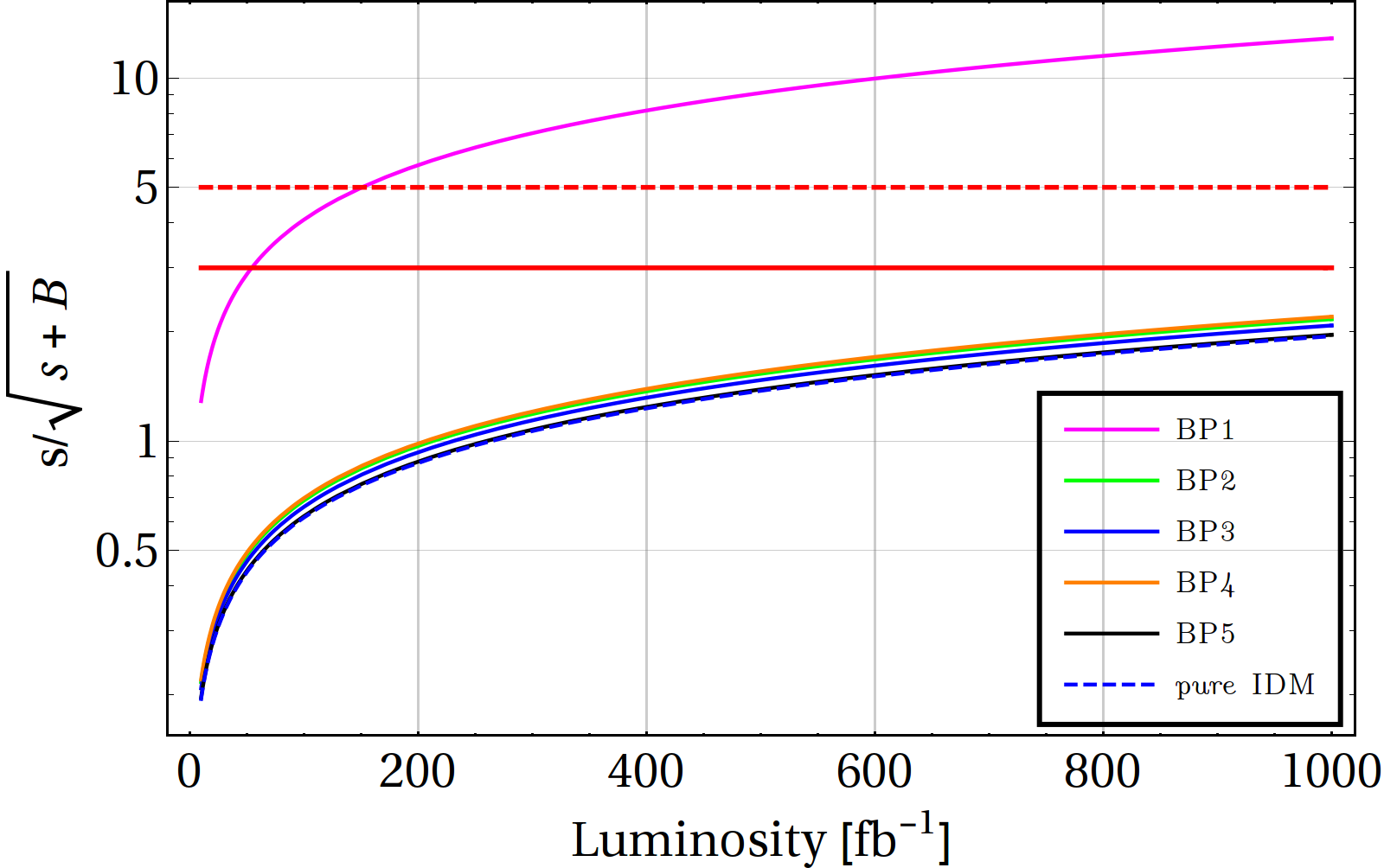}}~~
\subfloat[]
{\includegraphics[scale=0.35]{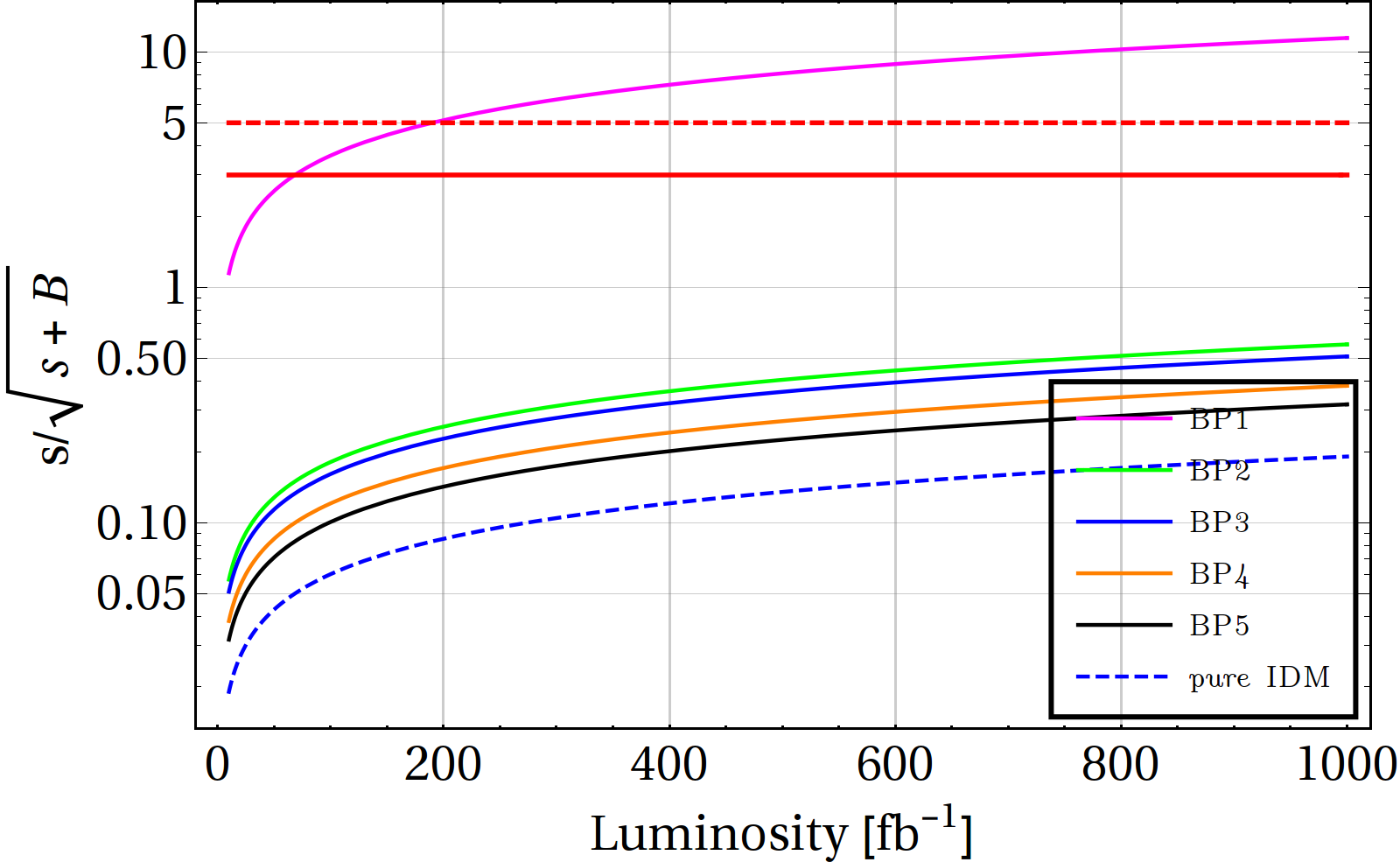}}~
\captionsetup{style=singlelineraggedleft}
\caption{Left: Significance plot for dijet+$\slashed{E_T}$ final state for different BPs. Right: Same when only $b$-jets are considered. In both the plots, the dashed blue line shows the significance for pure IDM scenario; the
thick red and the dashed red lines are respectively showing the $3\sigma$ and $5\sigma$ confidences. We have refrained from showing the significance plots for BP6 and BP7 because they were similar and very close to BP3-5, making the plot look messy}
\label{fig:dijetsignific}
\end{figure*}

The significance plots for $OSD+\slashed{E_T}$ channels are shown in Fig.~\ref{fig:dilepsignific}. From the  plot we can infer that:

\begin{itemize}
\item So far at LHC, we have not seen any excess in $\ell^+\ell^- + \slashed{E_T}$ channels, therefore, our model parameters in BP1 and BP7 are most likely to be ruled out by LHC. 
\item There is still a possibility that BP3 and BP6 might be probed in the future run of LHC at a luminosity $\mathcal{L}\sim 60~\rm fb^{-1}$ and $200~\rm fb^{-1}$ respectively.
\item Due to identical masses of the vector like leptons (and hence same production cross section for dilepton final state), the significance of BP2 and BP4 are exactly the same. Both of them reach a discovery potential of 5$\sigma$ at very high luminosity ($\mathcal{L}\sim 700~\rm fb^{-1}$).
\item It will be very hard to distinguish BP5 from pure IDM and seems almost impossible to be probed within the future limit of LHC luminosity.  
\end{itemize}

Therefore, from the collider searches of the $(\ell^+\ell^- + \slashed{E_T})$ final state, it is hard to rule out the model parameters entirely for some specific choices of BPs. What we see is that, vector like leptons with masses $\lsim 200~\rm GeV $ are prone to be ruled out by LHC at present luminosity. But vector like leptons with masses of $\gsim$ 250 GeV and above are yet to be probed by the future high luminosity runs. In the future runs the non-observance of any excess in the data will help us to rule out higher mass regions of the vector like leptons.

\subsection{ Dijet plus $\slashed{E_T}$ final state with and without $b$-tagging}
\label{sec:dijetmet}


We have used $\slashed{E_T}$ and $H_T$ as observables in order to distinguish the jet final state signal from that of the SM backgrounds. For the SM, the only source of $\slashed{E_T}$ are the SM neutrinos, which can be approximately taken to be massless at the colliders. As a result, the peak of the $\slashed{E_T}$ distribution for the background lies at a very low value of $\slashed{E_T}$. For signal, on the other hand, we can see from Fig.~\ref{fig:METdijet} and Fig.~\ref{fig:HTdijet} that except for BP1, in all the other benchmark scenarios it is extremely difficult to separate the signal of our model from the SM background and also from pure IDM signal. This is due to fact that in our model the dominant contributions to missing energy are coming from the diagrams in Fig.~\ref{fig:dijet_diag}, where final states contain both SM neutrinos and the DM. Hence the shape of the distribution is not just dictated by the massive DM, but the massless neutrinos as well. The final state cross-sections are listed in Table~\ref{tab:dijet-sig} for the signal where we can see that the cross-section for BP1 is significantly large, while for other benchmarks they are almost the same. For the SM backgrounds corresponding cross-sections are also tabulated in Table~\ref{tab:dijet-bck}. Note that, due to hadronic final state, the backgrounds are more vigorous here, making the signal less significant than that of leptonic final state.

 
\begin{table}[htb!]
\begin{center}
\begin{tabular}{|c|c|c|c|c|c|c|c|c|c|}
\hline
Benchmark & $\sigma^{\rm production}$&  & $\sigma^{\rm jj}(\sigma^{\rm bb})$  \\ [0.5ex] 
Points     & (pb)  &  & (fb)    \\ [0.5ex] 
\hline\hline

BP1 & 0.27 && 7.62 (1.87) \\
\cline{4-4}
\cline{1-2}
BP2 & 0.18 && 1.27 (0.09) \\
\cline{4-4}
\cline{1-2}
BP3 & 0.16 &$\slashed{E_T}:$ 240-280 GeV& 1.22 (0.08) \\
\cline{4-4}
\cline{1-2}
BP4 & 0.17 && 1.29 (0.06)\\
\cline{4-4}
\cline{1-2}
BP5 & 0.17 && 1.15 (0.05) \\
\cline{4-4}
\cline{1-2}
Pure IDM & 0.16 && 1.14 (0.03) \\
\hline\hline
\end{tabular}
\end{center}
\caption {Final state signal cross-section with  $\slashed{E_T}:240-280~\rm GeV$ for $dijet+\slashed{E_T}$ final state. The numbers in the parenthesis is the corresponding cross-section for exclusive $b$-jet final state . All simulations are done at $\sqrt{s}=14~\rm TeV$.} 
\label{tab:dijet-sig}
\end{table}

\begin{table}[htb!]
\begin{center}
\begin{tabular}{|c|c|c|c|c|c|c|c|c|c|}
\hline
Benchmark & $\sigma^{\rm production}$&  & $\sigma^{\rm jj}(\sigma^{\rm bb})$  \\ [0.5ex] 
Points     & (pb)  &  & (fb)    \\ [0.5ex] 
\hline\hline

$t\bar t$ & 814.64 && 317.71 (24.43) \\
\cline{4-4}
\cline{1-2}
$WW$ & 99.98 && 12.99 (0.00) \\
\cline{4-4}
\cline{1-2}
$WWZ$ & 0.15 &$\slashed{E_T}:$ 240-280 GeV& 0.14 (0.00) \\
\cline{4-4}
\cline{1-2}
$ZZ$ & 14.01 && 11.76 (0.28)\\
\cline{4-4}
\cline{1-2}
\hline\hline
\end{tabular}
\end{center}
\caption {Final state signal cross-section with  $\slashed{E_T}:240-280~\rm GeV$ for $dijet+\slashed{E_T}$ final state. The numbers in the parenthesis is the corresponding cross-section for exclusive $b$-jet final state . All simulations are done at $\sqrt{s}=14~\rm TeV$.} 
\label{tab:dijet-bck}
\end{table}

In Fig.~\ref{fig:dijetsignific}, we have shown the significance of different BPs for dijet final state with (left) and without (right) $b$ tagging. For BP1, a significant excess can be seen in all the distributions and it shows a very high significance for present LHC luminosity, which are very likely to be ruled out by LHC data. So far, no excess has been found in the present LHC run~\cite{Sirunyan:2018vjp}.
For inclusive dijet search, all other BPs reach a 5$\sigma$ significance for a luminosity $\mathcal{L}\sim 600~\rm fb^{-1}$, as shown in the LHS of Fig.~\ref{fig:dijetsignific}. It is also impossible to distinguish BP5 from pure IDM scenario, while all other BPs (except BP1) lie very close to each other because of comparable production cross sections. On the other hand, for exclusive $b$-tagged final states, although the BPs can be distinguished from pure IDM case, but none of them reach a discovery limit even at very high luminosity. This makes the model impossible to be probed at the LHC for final states containing $b$-jets. Thus, we can infer that, at the present LHC luminosity the non-observation of any excess rules out the vector like quark with masses $\le 500~\rm GeV$ (BP1). However, 
$ M_{D}$ in between 700 GeV and 900 GeV are still allowed which can further be constrained if we do not see any excess in the dijet plus $\slashed{E_T}$ signal at high luminosity LHC runs. We have also noted that $ M_{D} \gsim 900$ GeV can not be ruled out at the LHC even at very high luminosity run. This makes the model difficult to be probed in dijet plus $\slashed{E_T}$ final state even at high luminosity LHC runs.

In passing we would like to mention that dilepton plus dijet with missing energy final state has production cross-section lower than that of other final state signatures we discussed so far (as tabulated in Table.~\ref{Cross-sec}) which will result in very low signal significance over background. Therefore, we refrain from elaborating the fate of such final states at the LHC.

\section{Renormalization Group Equation (RGE) Running of the Couplings}
\label{sec:rge}
Any finite coupling is expected to hit a Landau pole at some scale. The Landau pole is the scale where the couplings become infinite. We check the high scale validity of our model by solving the RGEs using PyR@TE 2~\cite{Lyonnet:2016xiz}. The RG equations for pure IDM has been studied extensively before~\cite{Chakrabarty:2015yia}, but the addition of vector-like particles will modify the gauge, quartic and Yukawa couplings as given below. We have considered the running of the SM Yukawas and gauge couplings at two loop level. The running of all the new couplings are considered at one loop level.

The RG equations can be in general written as :
\begin{equation}
\frac{d\lambda}{d\mu} \equiv \beta_\lambda = \beta_{\lambda}^{I} + \beta_{\lambda}^{II}
\end{equation}
where $\lambda$ is any arbitrary coupling, $\mu$ is the energy scale and $\beta^{I(II)}$ is the one(two)-loop beta function.
In this section we present the one-loop RG evolution equations for the gauge, quartic and Yukawa couplings.

The evolution of the gauge couplings are given by :

\begin{equation}
(4\pi)^2 \beta_{g_s}^{I} =  -6 g_s^3 ,  \ \ \  (4\pi)^2 \beta_{g}^{I} = -3 g^3 , \ \ \ (4\pi)^2 \beta_{g_1}^{I} = +\frac{35}{3} g_1^3 
\end{equation}
where, $g_s,g$ and $g_1$ are the SU(3), SU(2) and U(1) gauge couplings respectively.

The RG equations for the quartic couplings are given by :
\begin{equation}
\centering
\begin{split}
(4\pi)^2 \beta_{\lambda_1}^{I} &=-24 Y_t^2 Y_b^2 -12 Y_b^4 -12 Y_t^4 +4\lambda_3^2 +2\lambda_4^2 +2 \lambda_5^2 +12\lambda_1^2 +4\lambda_3\lambda_4 -3\lambda_1 g_1^2 \\ & -9\lambda_1 g^2 + \frac{3g_1^4}{4} + \frac{9g^4}{4} + \frac{3}{2}g^2 g_1^2 -4Y_\tau^4 +4\lambda_1 Y_\tau^2 \\ 
(4\pi)^2 \beta_{\lambda_2}^{I} &=  - 12\lambda_b^2 +2\lambda_4^2 +2 \lambda_5^2 -4\lambda_\mu^4 -4\lambda_\tau^4 -8\lambda_\mu^2 \lambda_\tau^2 +12\lambda_2^2 + 4\lambda_3^2 + 4 \lambda_3 \lambda_4 +4\lambda_2 \lambda_\mu^2 \\ & + 4\lambda_2 \lambda_\tau^2 + \frac{3g_1^4}{4}  + \frac{9g^4}{4} + \frac{3}{2}g^2 g_1^2  -3\lambda_2 g_1^2  -9 \lambda_2 g^2 + 12 \lambda_2 \lambda_b^2\\
(4\pi)^2 \beta_{\lambda_3}^{I} &= +4\lambda_3^2 +2\lambda_4^2 +2 \lambda_5^2  +2\lambda_1\lambda_4 +2\lambda_2\lambda_4 +6\lambda_1\lambda_3 +6\lambda_2\lambda_3 + \frac{3g_1^4}{4} + \frac{9g^4}{4} -\frac{3}{2}g^2 g_1^2 \\& - 9 \lambda_3 g^2  - 3 \lambda_3 g_1^2  + 2 \lambda_3 Y_\tau^2 + 2 \lambda_3 \lambda_\mu^2 + 2 \lambda_3 \lambda_\tau^2 + 6\lambda_3 Y_b^2 + 6 \lambda_3 Y_t^2 + 6\lambda_3 \lambda_b^2 
\\ (4\pi)^2 \beta_{\lambda_5}^{I} & =+2\lambda_1\lambda_5 +2\lambda_2\lambda_5 +8\lambda_3\lambda_5 +12\lambda_3\lambda_5 - 9 \lambda_5 g^2  - 3 \lambda_5 g_1^2 + 2 \lambda_5 Y_\tau^2 + 2 \lambda_5 \lambda_\mu^2 \\& + 2 \lambda_5 \lambda_\tau^2 + 6\lambda_5 Y_b^2 + 6 \lambda_5 Y_t^2 + 6\lambda_5 \lambda_b^2
\\(4\pi)^2 \beta_{\lambda_{45}}^{I} & = -12 \lambda_b^2 Y_b^2 - 12 \lambda_b^2 Y_t^2 + 4\lambda_4^2 + 8\lambda_5^2 - 4\lambda_\mu^2 Y_\tau^2 - 4\lambda_\tau^2 Y_\tau^2 + 2 \lambda_1 \lambda_4 +2 \lambda_1 \lambda_5- 9 \lambda_4 g^2  \\& - 3 \lambda_4 g_1^2  - 9 \lambda_5 g^2  - 3 \lambda_5 g_1^2 +2\lambda_2\lambda_4 +2\lambda_2 \lambda_5 +2\lambda_4 Y_\tau^2 +2\lambda_4 \lambda_\mu^2 +2\lambda_4 \lambda_\tau^2 +2\lambda_5 Y_\tau^2 \\& +2\lambda_5 \lambda_\mu^2   +2\lambda_5 \lambda_\tau^2 + 3 g_1^2 g^2 +6\lambda_4 Y_t^2  +6\lambda_4 Y_b^2  +6\lambda_4 \lambda_b^2+6\lambda_5 Y_t^2  +6\lambda_5 Y_b^2 +6\lambda_5 \lambda_b^2 \\& + 8\lambda_3 \lambda_4 +8\lambda_3\lambda_5 + 12\lambda_4 \lambda_5
\end{split}
\end{equation}
where $\lambda_{45} = \lambda_4 +\lambda_5$.

The RG equations for the Yukawa couplings are given by :

\begin{eqnarray}
(4\pi)^2 \beta_{Y_t}^{I} &=& Y_t\bigg(\frac{1}{2}\lambda_b^2 +\frac{9}{2}Y_b^2 +\frac{9}{2}Y_t^2 + Y_\tau ^2 - 8g_s^2 -\frac{9}{4}g^2 -\frac{5}{12}g_1^2\bigg) \nonumber \\
(4\pi)^2 \beta_{Y_b}^{I} &=& Y_b\bigg(\frac{1}{2}\lambda_b^2 +\frac{9}{2}Y_b^2 +\frac{9}{2}Y_t^2 + Y_\tau ^2 - 8g_s^2 -\frac{9}{4}g^2 -\frac{17}{12}g_1^2\bigg) \nonumber \\
(4\pi)^2 \beta_{Y_\tau}^{I} &=& Y_\tau\bigg(\frac{1}{2}\lambda_\mu^2 + \frac{1}{2}\lambda_\tau^2 +3Y_b^2+ 3Y_t^2 + \frac{5}{2}Y_\tau ^2  -\frac{9}{4}g^2 -\frac{15}{4}g_1^2\bigg)\nonumber \\
(4\pi)^2 \beta_{\lambda_b}^{I} &=& \lambda_b\bigg(\frac{9}{2}\lambda_b^2 +\frac{1}{2}Y_b^2 +\frac{1}{2}Y_t^2 + \lambda_\mu^2+ \lambda_\tau ^2 - 8g_s^2 -\frac{9}{4}g^2 -\frac{5}{12}g_1^2\bigg) \nonumber \\
(4\pi)^2 \beta_{\lambda_\mu}^{I} &=& \lambda_\mu \bigg(\frac{5}{2}\lambda_\mu^2 + \frac{5}{2}\lambda_\tau^2 +3\lambda_b^2 + \frac{1}{2}Y_\tau ^2 -\frac{9}{4}g^2 -\frac{15}{4}g_1^2\bigg) \nonumber\\
(4\pi)^2 \beta_{\lambda_\tau}^{I} &=& \lambda_\tau \bigg(\frac{5}{2}\lambda_\mu^2 + \frac{5}{2}\lambda_\tau^2 +3\lambda_b^2 + \frac{1}{2}Y_\tau ^2 -\frac{9}{4}g^2 -\frac{15}{4}g_1^2\bigg)
\end{eqnarray}

\begin{figure}[htb!]
\centering
	\includegraphics[scale=0.45]{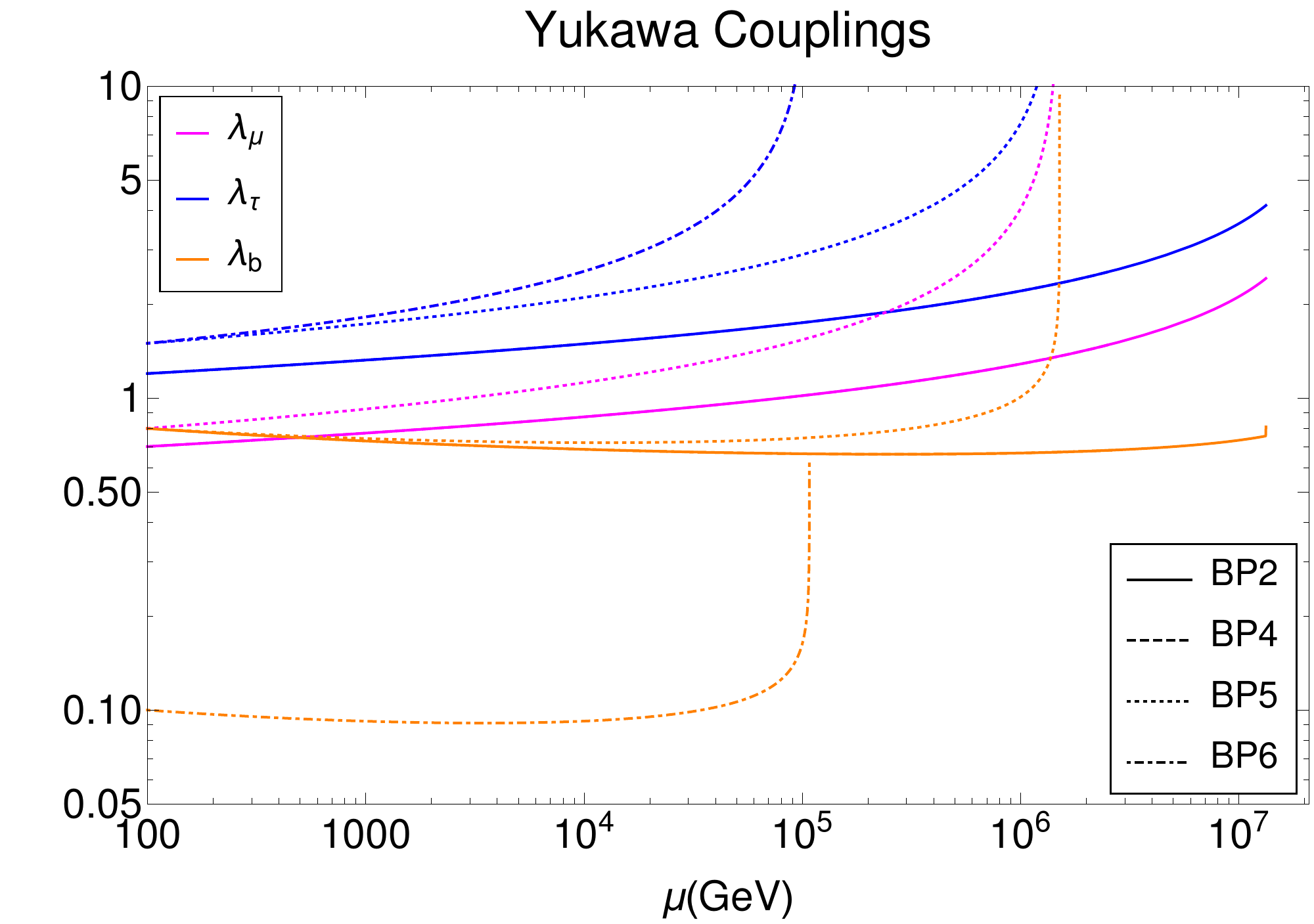}
	\captionsetup{style=singlelineraggedleft}
	\caption{Running of the NP Yukawa couplings of our model with energy scale $\mu$ (in GeV) is shown here. The magenta lines represent the evolution of $\lambda_\mu$, the blue ones represent that of $\lambda_\tau$ while the orange ones are for $\lambda_b$. The solid, dashed, dotted and dotdashed styles for the legends are used to represent the running for benchmarks points 2, 4, 5 and 6 respectively. Note that the solid and dashed lines overlap (due to similar variation of the couplings) and hence are indistinguishable. Also the magenta and blue dot-dashed lines for BP6 overlap due to similar coupling strengths}
	\label{RG}
\end{figure}

In Fig.~\ref{RG}, we have shown the variation of the Yukawa couplings in our model for the allowed benchmark scenarios. We note that, the Landau pole is reached at $\sim 10^7$ GeV for both BP2 and BP4, for BP5 the Landau pole is reached at$\gsim 10^6 $ GeV, while for BP6, it is reached at $10^5$ GeV. We note that the cut-off scale is dependent on the initial value of our NP couplings. The cut-off scale is higher if any one of the  couplings are small as in BP2 (here the smaller one being $\lambda_\mu = 0.7$ while the largest one being $\lambda_\tau = 1.2$) while for BP6 in which both $\lambda_\mu$ and $\lambda_\tau $ equal $1.5$ and $\lambda_b = 0.1$, the Landau pole is reached at $10^5$ GeV. Thus the Landau pole is mainly driven by the smaller coupling and since we have chosen $\lambda_\tau$ to be high in all the cases, the perturbativity of our theory is mostly determined by the value either $\lambda_\mu$ or $\lambda_b$, whichever is smaller. 

\section{Summary}
\label{sec:concl}

In this paper we have studied an extension of the inert Higgs doublet model with vector like fermions singlet under $SU(2)_L$ gauge symmetry. The model offers a DM candidate same as that in pure IDM, and the vector like fermions act as mediators between the dark sector and the visible sector, apart from the usual SM Higgs portal interactions. Due to the presence of the Yukawa interaction and family-dependent couplings, we can now have interesting FCNC and LFV processes, which can explain anomalies in $R(K^{(*)})$ data and also have the potential to incorporate the observed data on muon $(g-2)$. At the moment, satisfying all the other constrain, our model can comfortably explain the lower limit of the observed discrepancy in muon $(g-2)$. We have to wait for more precise estimate of $a_{\mu}$ for the final verdict. We have studied the parameter space of the model in detail considering bounds from DM relic density, direct detection, as well as from flavour data for both low  and high mass region of the DM. Apart from change in allowed DM mass values from relic density requirements, due to the existence of new interactions mediated by vector like fermions, we also find more promising direct detection rates compared to pure IDM, for chosen benchmark points.

From the resulting constrained parameter space, we have chosen a set of benchmark points for further collider studies. Because of the presence of the exotic particles, this model gives rise to several interesting signals in collider consisting of: hadronically quiet dilepton channel ($\ell^{+}\ell^{-}+\slashed{E_T}$), dijet channel ($jj+\slashed{E_T}$) and dilepton plus dijet channel ($\ell^{+}\ell^{-}+jj+\slashed{E_T}$), along with missing energy. Final states containing two leptons with two jets plus missing energy provide a very small production cross-section, and we refrain from analyzing such signals in our work. 
Of the other two channels, hadronically quiet dilepton final state shows a 5$\sigma$ significance in few of our chosen benchmark scenarios (BP2 and BP4) with $M_{E_2}$ and $M_{E_3} \gsim 200$ GeV for $M_{\ell\ell} > 200$ GeV and $H_T > 280$ GeV, and the required luminosity is $\sim 700~\rm fb^{-1}$. BP6 is also likely to be probed at the future run of the LHC when the luminosity reaches $\sim 200~\rm fb^{-1}$. We note that the vector like lepton masses $\lsim 200$ GeV are very likely ruled out by LHC data since they have not seen any excess in the above mentioned final states at the present luminosity.  For inclusive dijet plus $\slashed{E_T}$ final state, on the other hand, a 5$\sigma$ discovery can be claimed in couple of benchmark scenarios  with $M_{D} \gsim 600$ GeV for luminosity $\ge 600~\rm fb^{-1}$.

To summarize, the vector like fermion extension of IDM is capable of explaining anomalous results like $R(K^{(*)})$ and muon $(g-2)$; the required new parameter spaces are allowed by other flavour data like the rare and radiative $B_q~(q=d,s)$ decays, $B_q-{\bar B_q}$ mixing and the LFV decays like $\tau \to \mu \gamma$, $\mu \to e \gamma$ etc. The DM of the model satisfies Planck-observed relic density, obeying bounds from recent direct search data. The model can also be probed in the LHC experiment for a higher luminosity for some particular final states satisfying all the constraints mentioned above. We also check the perturbative unitarity of the model and find that for the chosen benchmark points the model can remain perturbative up to an energy scale $10^5-10^7$ GeV.          

\section{Acknowledgement}

BB and LM would like to thank Subhadeep Mondal and Subhaditya Bhattacharya for useful discussions on collider analysis. DB acknowledges the support from IIT Guwahati start-up grant (reference number: xPHYSUGIITG01152xxDB001). We would like to thank Tanmoy Mondal for his useful comments. SN acknowledges the support from research grant by the Science and Engineering Research Board, Govt. of India, under the grant CRG/2018/001260

\begin{appendices}
\section{Two Loop RG Equations}
\label{sec:RG-2Loop}

The two loop RG equation for the SM gauge couplings is given by:
\begin{eqnarray}
\beta_{g_s}^{II} &=& \frac{g_s^3}{(4\pi)^4}\bigg(-7 g_s^2 -2 Y_b^2 -2 Y_t^2  + \frac{13}{6}g_1^2 + \frac{9}{2} g^2  - 2 \lambda_b^2 \bigg) \nonumber \\
\beta_{g}^{II} &=& \frac{g^3}{(4\pi)^4}\bigg(8g^2 + 2 g_1^2  + 12 g_s^2 - \frac{3}{2}Y_b^2 -\frac{3}{2} Y_t^2 - \frac{3}{2} \lambda_b^2 - \frac{1}{2} Y_\tau^2 -\frac{1}{2} \lambda_\mu^2 -\frac{1}{2} \lambda_\tau^2 \bigg) \\
\beta_{g_1}^{II} &=& \frac{g_1^3}{(4\pi)^4}\bigg(\frac{214}{9} g_1^2 + 6 g^2 + \frac{52}{3} g_s^2 - \frac{17}{6} Y_b^2 - \frac{5}{6} Y_t^2 - \frac{5}{2} Y_\tau^2 -\frac{5}{2} \lambda_\mu^2 -\frac{5}{2} \lambda_\tau^2 -\frac{5}{6} \lambda_b^2\bigg) \nonumber
\label{gauge_2loop}
\end{eqnarray}

The two loop RG equations for the SM Yukawa couplings is given by :

\begin{equation}
\centering
\begin{split}
\beta_{Y_t}^{II} = \frac{Y_t}{(4\pi)^4}\bigg( &-12 Y_t^4 -12 Y_b^4 -24 Y_b^2 Y_t^2 -\frac{5}{2}Y_b^2 \lambda_b^2 -\frac{5}{2}Y_t^2 \lambda_b^2 -\frac{5}{2}\lambda_b^4 +\lambda_3^2 + \lambda_4^2 +\lambda_3\lambda_4 +\frac{3}{2} \lambda_1^2 +\frac{3}{2} \lambda_5^2 -6\lambda_1 Y_b^2 \\& -6\lambda_1 Y_t^2  -4\lambda_4 \lambda_b^2 -2\lambda_3 \lambda_b^2 +16g_s^2 Y_b^2 +16g_s^2 Y_t^2 -\frac{304}{3}g_s^4 -\frac{47}{72}g_1^4 -\frac{21}{4}g^4 -\frac{9}{4}Y_\tau^2 Y_b^2 -\frac{9}{4}Y_\tau^2 Y_t^2 \\& -\frac{9}{4}Y_\tau^4 -\frac{3}{4}\lambda_\mu^2 \lambda_b^2 -\frac{3}{4}\lambda_\tau^2 \lambda_b^2 -\frac{3}{4}Y_\tau^2 \lambda_\mu^2 -\frac{3}{4}Y_\tau^2 \lambda_\tau^2 +\frac{16}{3}g_s^2 \lambda_b^2 +\frac{29}{16}g_1^2 Y_b^2 +\frac{33}{16}g^2 \lambda_b^2 +\frac{135}{16}g^2Y_b^2 \\& + \frac{135}{16}g^2 Y_t^2  +\frac{187}{48}g_1^2 Y_t^2 +\frac{247}{144}g_1^2 \lambda_b^2 +9g^2 g_s^2 +20g_s^2 Y_b^2 +20g_s^2 Y_t^2 -\frac{9}{4}g_1^2 g^2 +\frac{15}{8}g^2 Y_\tau^2 +\frac{25}{8}g_1^2 Y_\tau^2 \\& +\frac{25}{24}g_1^2 Y_t^2 +\frac{31}{9}g_1^2 g_s^2 +\frac{45}{8}g^2 Y_b^2 +\frac{45}{8}g^2 Y_t^2 +\frac{85}{24}g_1^2 Y_b^2 \bigg)
\end{split} 
\end{equation}

\begin{equation}
\centering
\begin{split}
\beta_{Y_b}^{II} = \frac{Y_b}{(4\pi)^4}\bigg(& -12Y_b^4 -12 Y_t^4 -24Y_b^2 Y_t^2 -\frac{5}{2}\lambda_b^2 Y_b^2 -\frac{5}{2}\lambda_b^2 Y_t^2 -\frac{5}{2}\lambda_b^4 +\lambda_3^2 +\lambda_4^2 +\lambda_3\lambda_4 +\frac{3}{2}\lambda_1^2 +\frac{3}{2}\lambda_5^2 -6\lambda_1 Y_b^2 \\& -6\lambda_1 Y_t^2 -4\lambda_4 \lambda_b^2 -2\lambda_3 \lambda_b^2 +16g_s^2 Y_b^2 +16g_s^2 Y_t^2 -\frac{304}{3}g_s^4 -\frac{41}{144}g_1^2 \lambda_b^2 -\frac{21}{4}g^4 -\frac{9}{4}Y_\tau^2 Y_b^2 -\frac{9}{4}Y_\tau^2 Y_t^2 \\& -\frac{9}{4}Y_\tau^4 -\frac{3}{4}\lambda_\mu^2 \lambda_b^2 -\frac{3}{4}\lambda_\tau^2 \lambda_b^2 -\frac{3}{4}\lambda_\mu^2 Y_\tau^2 -\frac{3}{4}\lambda_\tau^2 Y_\tau^2 +\frac{16}{3} g_s^2 \lambda_b^2 +\frac{33}{16}g^2 \lambda_b^2 +\frac{35}{48}g_1^2 Y_t^2  +\frac{135}{16}g^2 Y_b^2  \\&+\frac{135}{16}g^2 Y_t^2 +\frac{223}{48}g_1^2 Y_b^2 +9g^2 g_s^2 +20g_s^2 Y_b^2 +20g_s^2 Y_t^2 +\frac{77}{8}g_1^4 -\frac{3}{4}g_1^2 g^2 +\frac{15}{8}g^2 Y_\tau^2 +\frac{19}{9}g_1^2 g_s^2 \\& +\frac{25}{8}g_1^2 Y_\tau^2 +\frac{25}{24}g_1^2 Y_t^2 +\frac{45}{8}g^2 Y_b^2 +\frac{45}{8}g^2 Y_t^2 +\frac{85}{24}g_1^2 Y_b^2 \bigg)
\end{split}
\end{equation}

\begin{equation}
\centering
\begin{split}
\beta_{Y_\tau}^{II} = \frac{Y_\tau}{(4\pi)^4}\bigg( &-\frac{3}{4}Y_\tau^4 -\lambda_\mu^2 Y_\tau^2 -\lambda_\tau^2 Y_\tau^2 -2\lambda_\mu^2 \lambda_\tau^2 -\lambda_\mu^4 - \lambda_\tau^4 + \lambda_3^2 + \lambda_4^2 + \lambda_3 \lambda_4 + \frac{3}{2} \lambda_1^2 + \frac{3}{2} \lambda_5^2 -6\lambda_1 Y_\tau^2  - 4 \lambda_4 \lambda_\mu^2 \\ &  - 4 \lambda_4 \lambda_\tau^2 - 2 \lambda_3 \lambda_\mu^2 - 2\lambda_3 \lambda_\tau^2 - \frac{27}{2} Y_t^2 Y_b^2  - \frac{27}{4} Y_b^2 Y_\tau^2 - \frac{27}{4} Y_t^2 Y_\tau^2 - \frac{27}{4} Y_t^4 - \frac{27}{4} Y_b^4  \frac{21}{4} g^4 -\frac{9}{4} \lambda_b^2 Y_b^2 \\ & -\frac{9}{4} \lambda_b^2 Y_t^2  + \frac{33}{16} g^2 \lambda_\mu^2 + \frac{33}{16} g^2 \lambda_\tau^2 +\frac{103}{16}g_1^2 \lambda_\mu^2  +\frac{103}{16}g_1^2 \lambda_\tau^2 +\frac{129}{16}g_1^2 Y_\tau^2  + \frac{135}{16}g^2 Y_\tau^2 + \frac{791}{24}g_1^4 \\ & + 20g_s^2 Y_b^2 + 20g_s^2 Y_t^2 + \frac{9}{4}g_1^2 g^2 +\frac{15}{8} g^2 Y_\tau^2 +\frac{25}{8} g_1^2 Y_\tau^2 + \frac{25}{24}g_1^2 Y_t^2 + \frac{45}{8} g^2 Y_b^2 + \frac{45}{8} g^2 Y_t^2 +\frac{85}{24}g_1^2 Y_b^2 \bigg)
\end{split}
\end{equation}

\end{appendices}


\end{document}